\documentclass[12pt, english, singlespacing, headsepline]{MastersDoctoralThesis}

\usepackage{amsmath}
\usepackage{wasysym}
\usepackage{physics}
\usepackage{subcaption}
\usepackage[utf8]{inputenc} 
\usepackage[T1]{fontenc} 

\usepackage{mathpazo} 

\usepackage[style=ieee]{biblatex}

\usepackage[autostyle=true]{csquotes} 

\thesistitle{The  Stationary Worldlines and  Power Distributions of a Point Charge} 
\supervisor{Michael\textsc{ R.R. Good}} 
\examiner{} 
\degree{Master of Physics} 
\author{Maksat \textsc{Temirkhan}} 
\addresses{} 

\subject{Theoretical Physics} 
\university{\href{http://www.nu.edu.kz}{Nazarbayev University}} 
\department{\href{http://sst.nu.edu.kz}{School of Science and Technology}} 
\group{\href{http://sst.nu.edu.kz}{Physics Department}} 
\faculty{\href{http://sst.nu.edu.kz}{Department of Physics}} 

\AtBeginDocument{
\hypersetup{pdftitle=\ttitle} 
\hypersetup{pdfauthor=\authorname} 
\hypersetup{pdfkeywords=\keywordnames} 
}

\begin{document}

\frontmatter 

\pagestyle{plain} 


\begin{titlepage}
\begin{center}

\vspace*{.06\textheight}
{\scshape\LARGE \univname\par}\vspace{1.5cm} 
\textsc{\Large Master Thesis}\\[0.5cm] 

\HRule \\[0.4cm] 
{\huge \bfseries \ttitle\par}\vspace{0.4cm} 
\HRule \\[1.5cm] 
 
\begin{minipage}[t]{0.4\textwidth}
\begin{flushleft} \large
\emph{Author:}\\
{\authorname}
\end{flushleft}
\end{minipage}
\begin{minipage}[t]{0.4\textwidth}
\begin{flushright} \large
\emph{Supervisor:}\\
{\supname}
\end{flushright}
\end{minipage}\\[3cm]
 
\vfill

\large \textit{A thesis submitted in fulfillment of the requirements\\ for the degree of \degreename}\\[0.3cm] 
\textit{in the}\\[0.4cm]
\groupname\\\deptname\\[2cm] 
 
\vfill

{\large \today}\\[4cm] 

\vfill
\end{center}
\end{titlepage}

\let\cleardoublepage\clearpage
\begin{declaration}
\addchaptertocentry{\authorshipname} 
\noindent I, \authorname, declare that this thesis titled, \enquote{\ttitle} and the work presented in it are my own. I confirm that:

\begin{itemize} 
\item This work was done wholly or mainly while in candidature for a degree at this University.
\item Where any part of this thesis has previously been submitted for a degree or any other qualification at this University or any other institution, this has been clearly stated.
\item Where I have consulted the published work of others, this is always clearly attributed.
\item Where I have quoted from the work of others, the source is always given. With the exception of such quotations, this thesis is entirely my own work and that work done jointly with others.
\item I have acknowledged all main sources of help.
\end{itemize}
 
\noindent Signed:\\
\rule[0.5em]{25em}{0.5pt} 
 
\noindent Date:\\
\rule[0.5em]{25em}{0.5pt} 
\end{declaration}

\cleardoublepage





\let\cleardoublepage\clearpage
\begin{abstract}

\addchaptertocentry{\abstractname} 

The uniform acceleration of a point charge moving along a stationary worldline, which  emits constant radiated power was investigated.  A classification of motion of a particle along stationary worldlines into six types is made. The angular distribution of this power is found for all stationary worldlines including those with torsion and hypertorsion and their properties and features are also described. Their proper accelerations, acceleration ratios, minimum velocities and constant power emissions are computed.  The graphs  of emitted radiation  with different speeds are illustrated in two and three dimensional space. Additionally, the maximum angle and Thomas precession of radiation is found.

\end{abstract}



\let\cleardoublepage\clearpage
\tableofcontents 
\let\cleardoublepage\clearpage
\listoffigures 

\mainmatter 

\pagestyle{thesis} 


\newcommand{\aidar}[1]{\textcolor{blue}{[{\bf Maksat}: #1]}} 
\newcommand{\mike}[1]{\textcolor{red}{[{\bf MG}: #1]}} 
\newcommand{\Lagr}{\mathcal{L}}
\newcommand{\Hamil}{\mathcal{H}}
\newcommand{\extdz}[1]{\sideremark{#1}}
\newcommand{\edz}[1]{\sideremark{#1}}
\newcommand{\bo}{\raise-1mm\hbox{\Large$\Box$}} 
\newcommand{\ox}[1]{ \noindent\fbox{\parbox{\columnwidth}{#1 }}}
\newcommand{\e}{\epsilon_0}
\newcommand{\f}[2]{\frac{#1}{#2}}
\newcommand{\pderiv}[2]{\frac{\partial #1}{\partial #2}}
\newcommand{\y}{\gamma}
\newcommand{\bd}{\boldsymbol}
\newcommand{\la}{\langle}
\newcommand{\ra}{\rangle}
\newcommand{\w}{\omega}
\newcommand{\kp}{\kappa}
\newcommand{\be}{\begin{equation}}
\newcommand{\ee}{\end{equation}}
\newcommand{\bea}{\begin{eqnarray}}
\newcommand{\eea}{\end{eqnarray}}
\newcommand{\eqn}[1]{(\ref{#1})}

\newcommand{\oM}{\overline{M}}
\newcommand{\oV}{\overline{V}}
\newcommand{\2}{$^2$}
\newcommand{\3}{$^3$}
\newcommand{\4}{$_4$}
\newcommand{\5}{$_5$}
\newcommand{\x}{arXiv:}
\newtheorem{theorem}{Theorem}
\newtheorem{proposition}{Proposition}
\newtheorem{Lemma}{Lemma}
\newtheorem{proof}{{\em Proof}}
\renewcommand{\theproof}{}
\newtheorem{Cor}{Corollary}[theorem]
\newtheorem{remark}{{\em Remark}}
\renewcommand{\theremark}{}
\definecolor{amaranth}{rgb}{0.9, 0.17, 0.31}
\definecolor{purple(munsell)}{rgb}{0.62, 0.0, 0.77}
\definecolor{americanrose}{rgb}{1.0, 0.01, 0.24}
\definecolor{palatinateblue}{rgb}{0.15, 0.23, 0.89}
\definecolor{royalblue(web)}{rgb}{0.25, 0.41, 0.88}
\definecolor{hanpurple}{rgb}{0.32, 0.09, 0.98}
\definecolor{beaublue}{rgb}{0.74, 0.83, 0.9}
\definecolor{carminered}{rgb}{1.0, 0.0, 0.22}
\definecolor{brightpink}{rgb}{1.0, 0.0, 0.5}
\definecolor{vividviolet}{rgb}{0.62, 0.0, 1.0}
\hypersetup{ linktoc=all,
   linkcolor={palatinateblue},
    citecolor={brightpink}, urlcolor={amaranth}}

\newcommand{\changeurlcolor}[1]{\hypersetup{urlcolor=#1}}    
\def\sideremark#1{\ifvmode\leavevmode\fi\vadjust{\vbox to0pt{\vss
 \hbox to 0pt{\hskip\hsize\hskip1em
 \vbox{\hsize2cm\tiny\raggedright\pretolerance10000
 \noindent #1\hfill}\hss}\vbox to8pt{\vfil}\vss}}}%

\renewcommand{\d}[1]{\ensuremath{\operatorname{d}\!{#1}}}

\def\rcurs{{\mbox{$\resizebox{.09in}{.08in}{\includegraphics[trim= 1em 0 14em 0,clip]{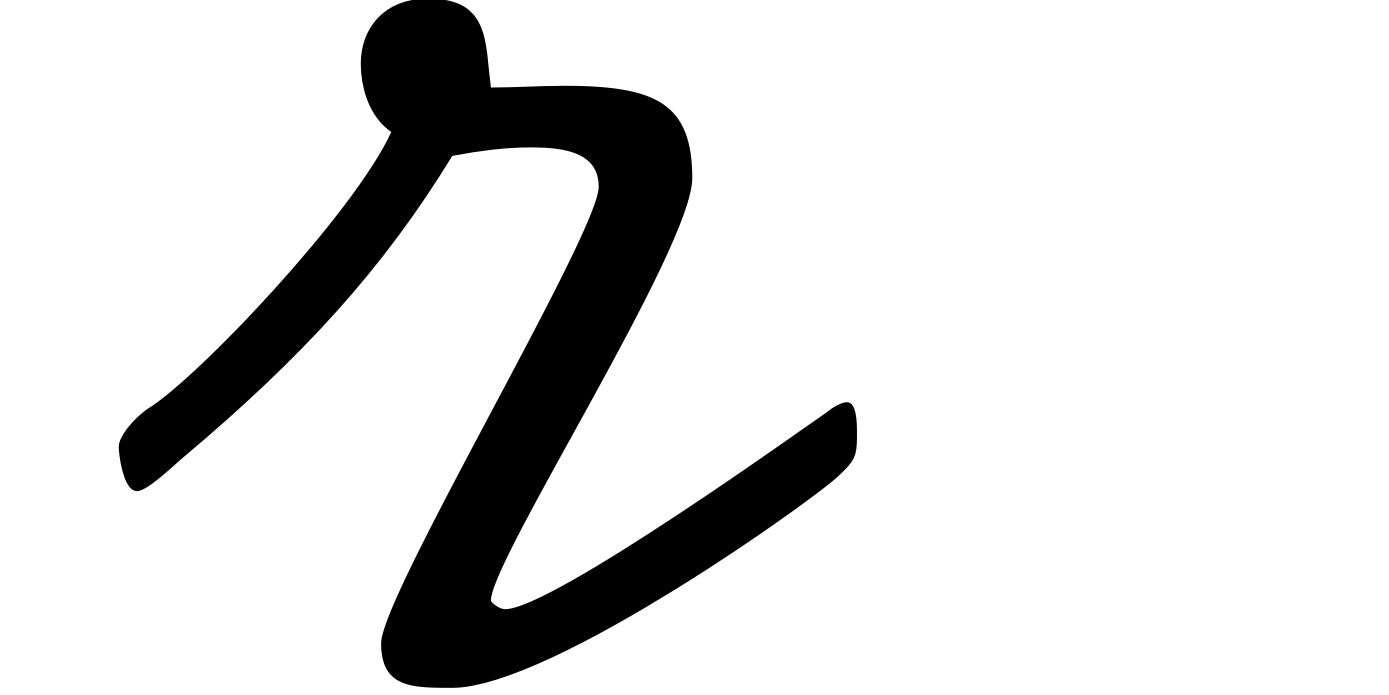}}$}}}
\def\brcurs{{\mbox{$\resizebox{.09in}{.08in}{\includegraphics[trim= 1em 0 14em 0,clip]{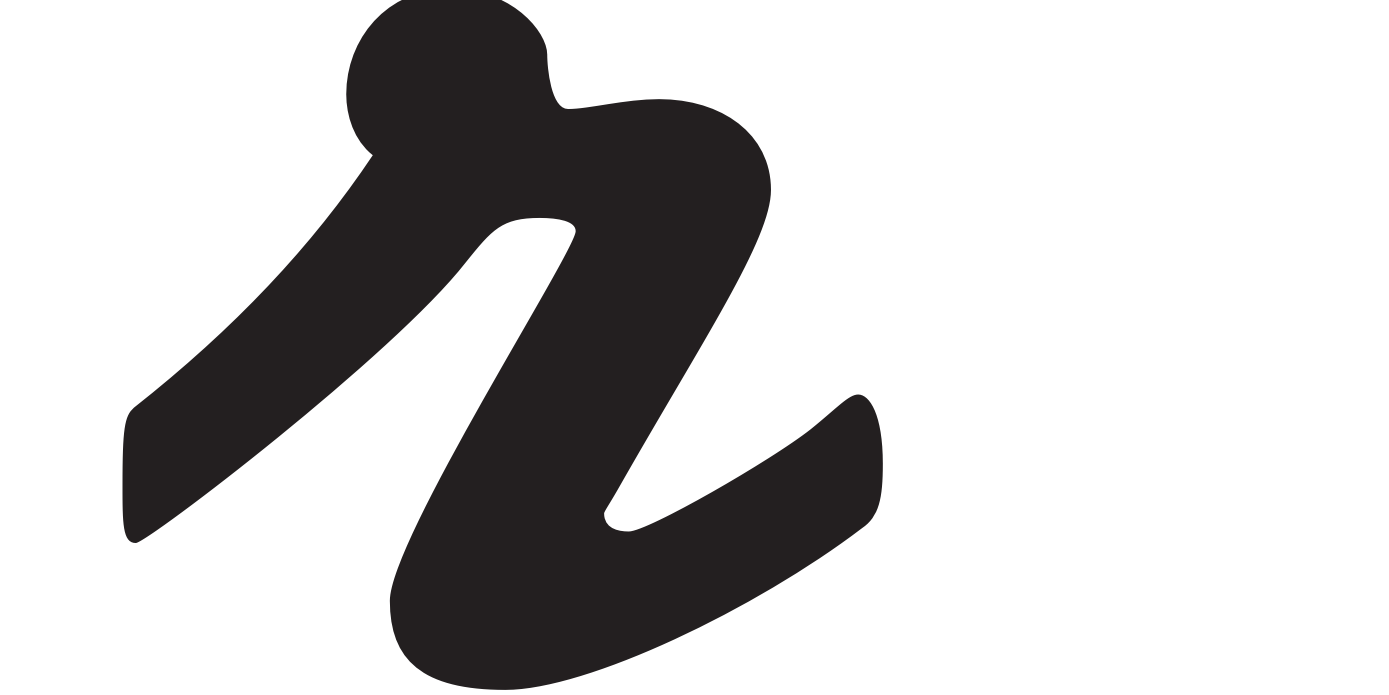}}$}}}

\def\hrcurs{{\mbox{$\hat \brcurs$}}}

\let\cleardoublepage\clearpage

\chapter{Introduction} 

\label{Chapter1} 

\section{History}
In this MS thesis, we will consider the electromagnetic field of an accelerating charge. Accelerated moving charges according to the laws of electrodynamics as described by James Clerk Maxwell \cite{Maxwell61} must emit electromagnetic waves. In 1865 Maxwell derived a system of equations which can now be expressed in differential or integral form describing an electromagnetic field and its connection with electric charges and currents in vacuum and continuum. Together with the expression for the Lorentz force, which specifies the measure of the effect of an electromagnetic field on charged particles, these equations form a complete system of  classical electrodynamics. The equations formulated by  Maxwell on the basis of experimental results accumulated by the middle of the 1800s, played a key role in the development of theoretical physics and had a strong influence not only on all areas of physics directly related to electromagnetism, but also on many later fundamental theories, like  special theory of relativity. 

In 1897 Joseph Larmor \cite{Larmor1897} first calculated the total power radiated by a non-relativistic point charge as it decelerated or accelerated. The Larmor radiation formula describes a uniformly accelerating  single point charge which emits radiation  toward infinity.

In 1955, Gold and Bondi  wrote a paper \cite{BondiGold} where they established the production of the electromagnetic field by a uniformly accelerated charged particle. This was a problem that had  been considered before but without finding the complete solution. Later, a new question was raised : could the equivalence principle apply to such a situation?

The equivalence principle is that used by Albert Einstein in the derivation of the general theory of relativity. Its formulation briefly can be expressed as: `the gravitational and inertial masses of a particle are equal'. 
From the equivalence principle follows interesting predictions about the behavior of light in a gravitational field. Imagine that in an elevator at the moment of accelerated upward movement, you send a light pulse (for example, using a laser pointer) in the horizontal direction. During the time that the light pulse is travelling to a target on the wall, the elevator will accelerate, and the light will flash on the wall below the target. (Of course, on Earth, you will not notice this deviation, so just imagine that you are able to consider the deviation in thousandths of a micron.) Now, returning to the principle of equivalence,  we can conclude that a similar effect of a deflecting light beam should be observed not only in a non-inertial system, but also in a gravitational field. For a light beam, according to the equivalence principle  of gravity and inertial masses, in the number of postulates of the general theory of relativity, the deviation of the light beam of a particular star located on the edge of the solar disk would be deflected radially away from the Sun by about 1.75 angular seconds \cite{DonaldG}. When the measurements carried out by Sir Arthur Eddington during the total solar eclipse in 1919 revealed the deflection of the beam through an angle of 1.6 angular seconds, it was an experimental confirmation of the general theory of relativity. Similarly, it is not difficult to see that the equivalence principle predicts that a light beam directed parallel to the gravitational field should  has a red shift, and this prediction has also  received  experimental confirmation.

Uniform acceleration can  be described by the four - acceleration vector. It is a motion that is  hard to imagine  in a real physical context, because it involves the particle asymptotically approaching the speed of light in the distant future  and  past.  By using the so-called method of retarded potentials (the Lienard–Wiechert retarded potentials) Born \cite{MBorn} obtained the electromagnetic fields that  he supposed would be generated by a charged particle.  These  potentials were developed in part by Alfred-Marie Lienard in 1898 and later independently by Emil Wiechert \cite{Wiechert} in 1900. However, Born's solution is only correct in a certain region of spacetime. The aforementioned Gold and Bondi’s paper is thus concerned with obtaining a proper solution of Maxwell’s equations,  where one must include distributions as Dirac delta functions. Their method involves imagining a physical process and considering the idealistic uniformly accelerated motion as a limit. 

In 1965 Fritz Rohrlich \cite{Rohrlichbook}  showed that a charged particle and a neutral particle fall equally rapidly in a gravitational field. Similarly, Rohrlich \cite{Rohrlich} established that a charged particle that is at rest in a gravitational field does not radiate in its rest system, but does so in the framework of a freely falling observer.

In 1977 Davies and Fulling \cite{Davies:1977yv} \cite{Fulling1976RadiationAnomaly} found radiation from an accelerating perfectly reflecting boundary condition on the quantum field (i.e. a moving mirror) shortly after Hawking's black hole work \cite{Hawking:1974sw}.  They demonstrated that at late times moving mirrors with certain asymptotic motion can be thermal and radiate energy. Here the Planck spectrum is found at late times and mimics a black hole evaporating. Interesting trajectories are included in these recent sources \cite{Good:2017kjr,Good:2018ell,Good:2016yht,Good:2017ddq,Good:2016oey,Good:2016bsq,Good:2015jwa,Anderson:2015iga,Good:2012cp,Good:2014iua,Good2013TimeMirrors,Good2017OnMirror,Good:2018,Myrzakul2018,Good:2018aer}.

In 1976 Bill Unruh \cite{Unruh:1976db} demonstrated, by a uniformly accelerating detector in a vacuum field, that an observer moving linearly with constant acceleration in the vacuum of flat spacetime will radiate. The observer will behave as if they are in contact with scalar particles with distribution according to a Planck spectrum at temperature equal $T = \kappa/ 2\pi$. The same results have been found by Boyer \cite{Boyer} with detectors of electromagnetic radiation in 1980.

The Unruh effect is the quantum field effect of observing thermal radiation in an accelerated reference frame. In other words, an accelerating observer will see radiation with temperature proportional to their acceleration. Unruh showed that the vacuum depends  on  the observer which moves through space-time. If around the stationary observer is only a vacuum, then an accelerating observer will see around many particles in thermodynamic equilibrium i.e. a warm gas.  Experimental confirmation and the existence of the Unruh effect is not fully resolved. Some researchers believe that the Unruh effect has been confirmed experimentally \cite{LuisCrispino}, others believe that the formulation of the problem itself contains erroneous assumptions \cite{Igor}. According to modern definitions, the concept of a vacuum is not the same as empty space, since the entire space is filled with quantized fields.

The vacuum state is the simplest, lowest energy state. The energy levels of any quantized field depend on the Hamiltonian, which in the general case depends on generalized coordinates. Therefore, the Hamiltonian or  the concept of vacuum, depends on the reference system. The number of particles is an eigenvalue of the number operator and depends on the creation and annihilation operators. Before defining the  operators, we need to decompose the field into positive and negative frequency components.  The decomposition will be different in the Minkowski and Rindler coordinates, despite the fact that the operators in them are connected by the Bogolyubov transformation. That is why the number of particles depends on the reference system.

The Unruh effect  explains Hawking radiation, but can not be considered as completely analogues  to it \cite{Mensky}. The difference in the boundary conditions of the problems gives different solutions for these effects. Hawking radiation may occur at the border or `atmosphere' of a black hole and causes it to gradually evaporate, and due to the Unruh effect, a uniformly accelerating observer sees the birth of particles with a constant temperature.

In 1980 Letaw and  Pfautsch \cite{LetawPfautsch} demonstrated  that particles  moving in uniform synchrotron motion will also be excited and radiate with a spectrum. However, in this case the spectrum will be distinct from a Planck spectrum because of angular velocity. Also, it is interesting that the spectrum is time independent in these two cases; but, this will not be true for all situations. For instance, if a particle will move rectilinearly with slowly increasing acceleration we will expect the Planckian spectrum, with a slowly growing temperature.

In 1981 Letaw \cite{Letaw:1980yv} found that the spectrum of a moving particle is time independent (include stationary spectrum) when the geodetic interval on the worldline between two points depends  only  on  the  proper  time  interval  between  them.   For  this  reason, the worldlines in our case are called stationary.  The geometric properties of these lines are important because they are constant proper accelerated trajectories. Letaw in his work  investigated the stationary world lines due to quantized field detectors in a vacuum which have time-independent excitation spectra.  To find  world lines Letaw  generalized the Frenet equations to Minkowski space and developed them. Also he showed that a proper acceleration and angular velocity of the world line can be found in the curvature invariants and the solutions are   the stationary world lines. Letaw was the first who classified stationary worldlines into the six types. Additionally he demonstrated the equivalence of  the stationary world lines and the timelike Killing vector field orbits. As a result of that, the classification scheme  extended the stationary coordinate systems  and Killing orbits  in flat spacetime.

This work is concerned with the properties of the radiation from a relativistically uniformly accelerated charge and based on a working paper \cite{OurWork}. The general formula for an arbitrary trajectory of the total radiated power and its angular distribution are specified to uniform accelerated motion with torsion and hypertorsion. In 1949  Schwinger \cite{Schwinger:1949ym} first calculated the angular distribution of power for point charges undergoing linear and circular acceleration for arbitrary dynamics and specification to uniformly accelerated circular motion.  Letaw \cite{Letaw:1980yv} found three previously unknown classes of uniformly accelerated world lines.  This work extends Schwinger's initial specification of circular motion to include the power distributions for these three fundamental cases of uniform acceleration.

\newpage
\section{Naming Convention} 

With an emphasis on torsion \cite{Letaw:1980yv}, the worldlines themselves can be called (e.g. a shortened use of the convention introduced by Rosu \cite{Rosu:1999ad}):
\begin{enumerate}
	\item Nulltor \hskip0.8cm	2. Ultrator  \hskip0.8cm 3. Parator \hskip0.8cm 4. Infrator  \hskip0.8cm 5. Hypertor
\end{enumerate}
 The worldline plots are in Appendix \ref{Appendix:Worldline} where we have also included their four-vector position parametrization $x^\mu(s)$ in proper time \cite{Letaw:1980yv}. 
 
Projected in space, the worldlines take the forms: 

\begin{enumerate}
	\item Line \hskip1.0cm	2. Circle  \hskip1.2cm 3. Cusp \hskip1.0cm 4. Catenary  \hskip1.0cm 5. Spiral
\end{enumerate}
where we have ignored the zero acceleration solution, $\kappa=0$, (the inertial path). These spatial projections are in Appendix \ref{Appendix:Spatial}, with the assigned axis convention.  

\section{Curvature Invariants}
The three curvature invariants are, $\kappa$, $\tau$, and $\nu$, (curvature, torsion, and hypertorsion, respectively). As mentioned above, the stationary world lines separate naturally into five classes according to the values of the curvature invariants, if we exclude case when particle  at the rest (calling the motions by their projections in space):
\begin{itemize}
  \item Line :   $\kappa \neq 0$, $\nu=\tau=0$;
  \item Circle : $|\kappa|<|\tau|$, $\nu=0$;
  \item Cusp :   $|\kappa|=|\tau|$, $\nu=0$;
  \item Catenary :  $|\kappa|>|\tau|$, $\nu=0$;
  \item Spiral : $\nu \neq 0$
\end{itemize}

\newpage

\section{Motivation}
Stationary background systems based on these world lines are of interest because the world lines are trajectories of time-like Killing vector fields \cite{Letaw:1980yv, Rosu:1999ad, LetawJ}. The less-known trajectories (3, 4, 5) are interesting in their own right because they are simple uniformly accelerated motions.  Their excitation spectra are connected to the question of coordinate dependence of thermodynamics and quantum field theory in flat spacetime (i.e. they may yield insights into the Unruh effect \cite{Unruh:1976db}, with crucial dependence on study of accelerating world lines).  

There is broad motivation for investigating the light and radiation from these fundamental motions: interesting avenues of research and connections between quantum theory and gravity have been made, like relativistic superfluidity \cite{Xiong:2014oga} or geometric creation of quantum vortexes \cite{Good:2014iua}, by pursuing the effects of the influence of quantum fields under external conditions. There are three important types of external effects: the Davies-Fulling effect\footnote{Consider the scale invariant accelerated solution in \cite{Good2013TimeMirrors} and the spin-statistics derivation from it \cite{Good:2012cp}.} \cite{Davies:1976hi}, the Hawking effect \cite{Hawking1975}, and the Parker effect \cite{Parker:1968mv}, i.e. respectively, moving mirrors, black holes, and expanding cosmologies. This program \cite{Parkerbook} continues to lead to progress that demonstrate strong correspondences \footnote{Example solutions are explored in the recent black hole-moving mirror correspondence contained in \cite{Good:2016oey}, temperature without a horizon in \cite{Good2017OnMirror}, and a black hole birth cry and death gasp in \cite{Good:2015nja}.} and simplifying physics \footnote{ The Kerr black hole has a temperature $2\pi T = g - k$ where $k = m\Omega^2$ is the spring constant and $g= (4m)^{-1}$ is the Schwarzschild surface gravity \cite{Good:2014uja}. }.  

Accelerated trajectories in flat spacetime and motion in curved spacetime are joined via the Equivalence Principle.  The thermodynamics `addendum' made to the Equivalence Principle, (e.g. Unruh effect), has been immensely helpful for further understanding the gravitational influence on quantum fields\footnote{Hawking's effect as a gravitational phenomena is further understood as a result of the quantum field phenomena of the Unruh effect.}. These motions are interesting and simple, and they deserve more attention particularly because of the fundamentally important and wide-reaching nature of uniform acceleration. 
\newcommand{\keyword}[1]{\textbf{#1}}
\newcommand{\tabhead}[1]{\textbf{#1}}
\newcommand{\code}[1]{\texttt{#1}}
\newcommand{\file}[1]{\texttt{\bfseries#1}}
\newcommand{\option}[1]{\texttt{\itshape#1}}
\let\cleardoublepage\clearpage
\chapter{Some Basics of Special Relativity } \label{sec:Worldline}
\section{Space and  Time}
The basic description of special relativity for this and next two section  was taken from  paper \cite{Andrey2015}. Space - time is the sum of all events. The event is described by what happened somewhere and someday. It is determined by four numbers - time and three space coordinates. Therefore, space-time is four-dimensional. The events that occurred with the object form its world line. The world line is the path traversed by the object in space-time. For simplicity, we will mainly consider the movement of an object in a straight line, then space-time in a two-dimension. The figure below shows the motion of  trains.
\begin{figure}[ht]
\begin{center}
{\rotatebox{0}{\includegraphics[width=3.0in]{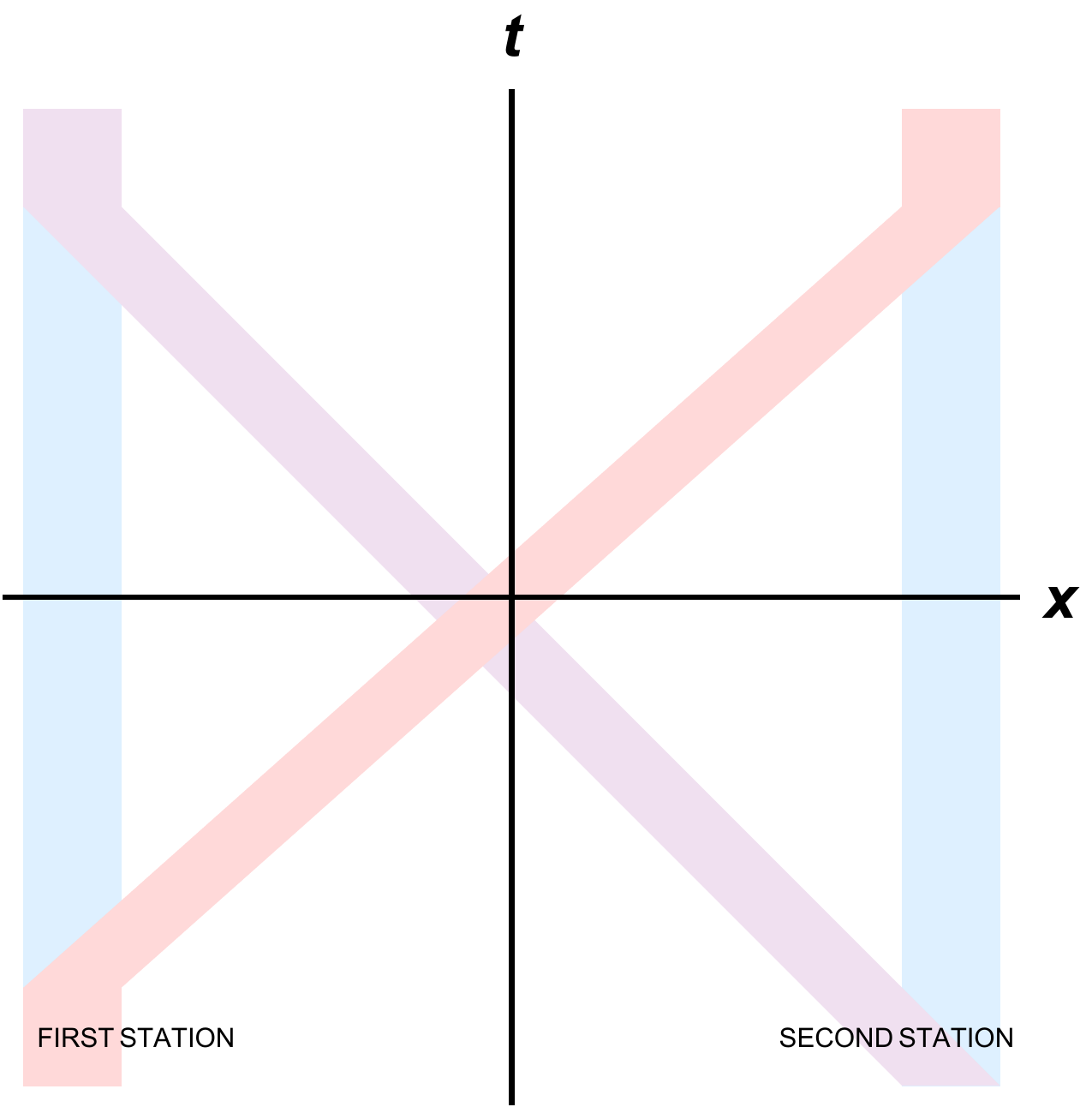}}} 
{\caption{ The motion is described by a two-dimensional region between the world lines.}} 
\end{center}
\end{figure}  
Each observer  can measure the time of events on his own world line.  The observer cannot directly measure the time of an event outside of its world line. Suppose observer $1$ sends a signal to observer $2$ at the time of $ t_1 $, then he  send a signal back Fig.~(\ref{ch2f2}). As result we can formulate three possibilities:
\begin{figure}
\begin{center}
{\rotatebox{0}{\includegraphics[width=3.0in]{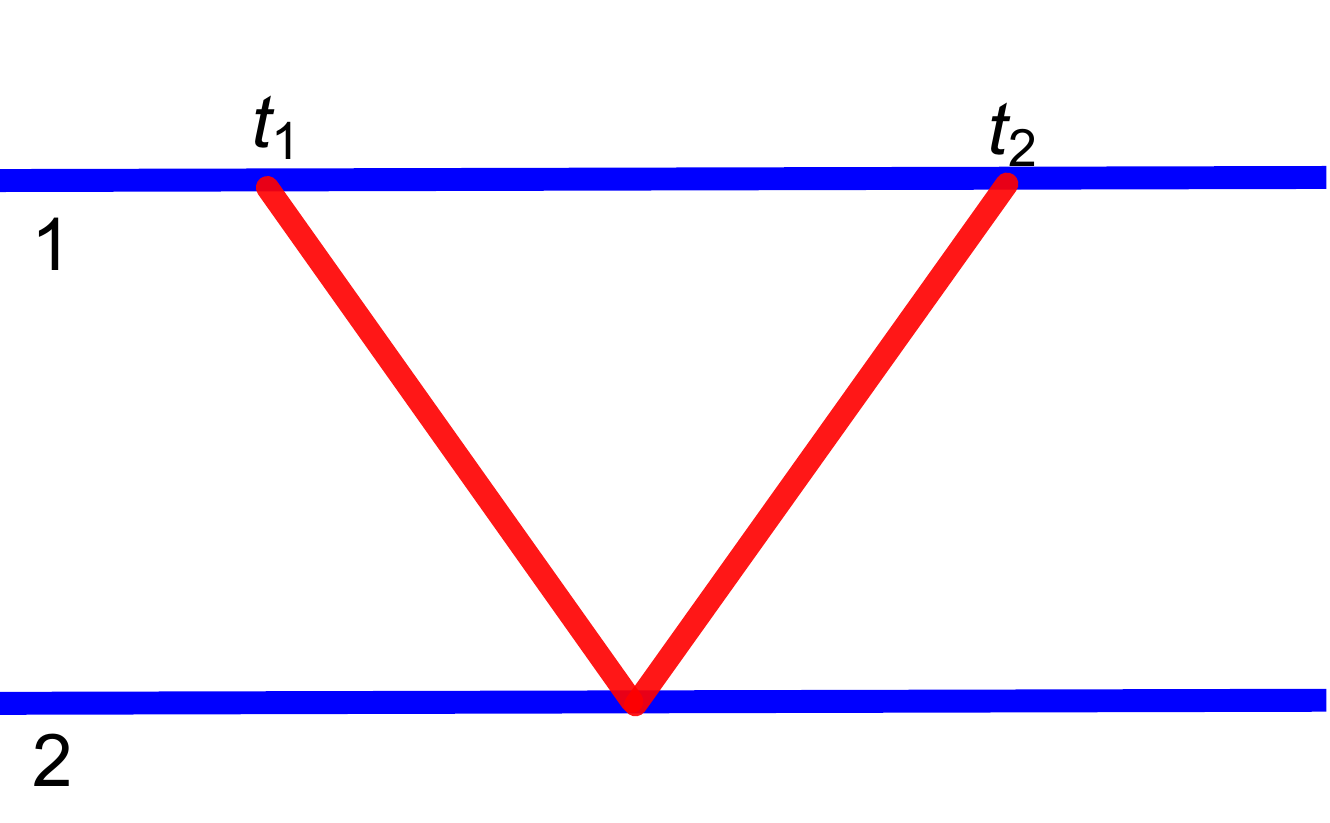}}} 
{\caption{\label{ch2f2} Signal exchange between observers.}} 
\end{center}
\end{figure}  
\begin{itemize}
  \item In case when $ t_2 <t_1 $, the response signal comes before sending the first signal. But, this opportunity is impossible.
  \item  When $ t_2 -  t_1 $ is positive, the difference can be very small. In this situation, we can introduce the absolute time, which the same for all observers. Before creating the theory of relativity,  everyone thought like this.
  \item In the last case when $ t_2 - t_1 $ can not be smaller than some positive value and it is the fastest signal. And this is true for the real world.
\end{itemize}
Such the fastest signal can be only light. Because photons are massless particle.  For example, we will image that event $ O $ be a explosion. The first thing that observer will see is the flash (light). On the world line light  emitted at $ O $ form the light half-cones of the future and past of this event. These area of space – time is called the future and past of $ O $.  Events inside and on the border of these half-cones can affect on the $ O $ event. The region outside the light cones is in the elsewhere of $ O $. Events in this area can not affect $ O $, and they can not influence to them.
\begin{figure}[ht]
\begin{center}
{\rotatebox{0}{\includegraphics[width=2.0in]{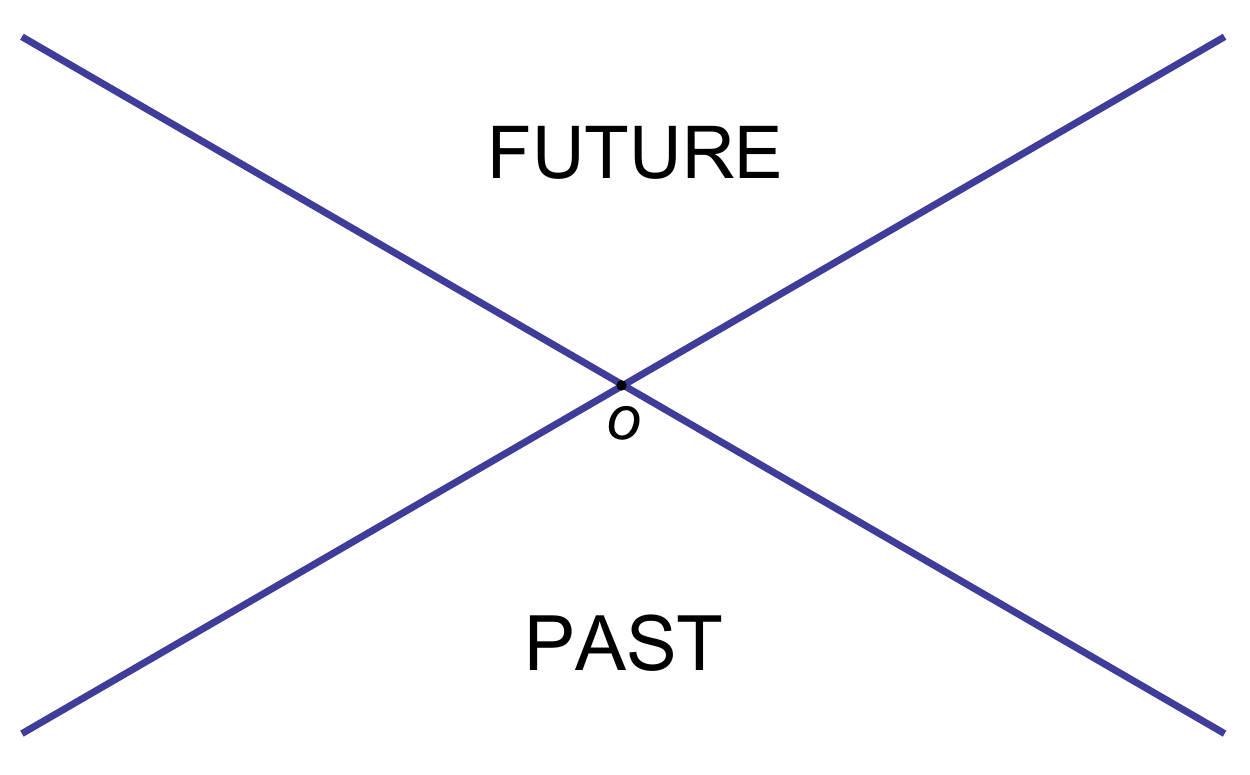}}} 
{\caption{\label{ch2f3} Past and Future  events of $ O $.}} 
\end{center}
\end{figure}  

\begin{figure}
\begin{center}
{\rotatebox{0}{\includegraphics[width=2.0in]{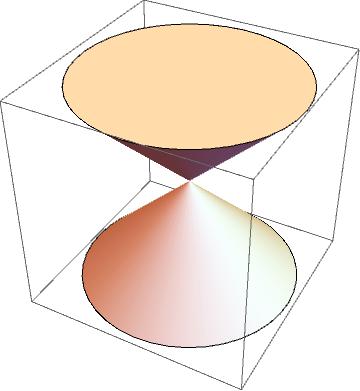}}}
{\caption{\label{ch2f4}Light cone in 3D space – time.}} 
\end{center}
\end{figure}

The light cone in three dimensional space – time is shown in the Fig.~(\ref{ch2f4}) with two space coordinates and one time coordinate. The light cone in two dimensional space – time is shown in the Fig.~(\ref{ch2f3}) with one space  and one time coordinate. The future events are inside the light of the semi-cone of the future, also the past events are inside the light half-cone of the past, while elsewhere events - outside the light cone.

\section{Lorentz transformation}
An object  moves by inertia when no forces act on it.  By  Galileo’s principle of relativity, all inertial observers are equal. If one inertial observer did some experiment and got some result, and the other inertial observer who did the same experiment  get the same result.
\begin{figure}[ht]
\begin{center}
{\rotatebox{0}{\includegraphics[width=2.5in]{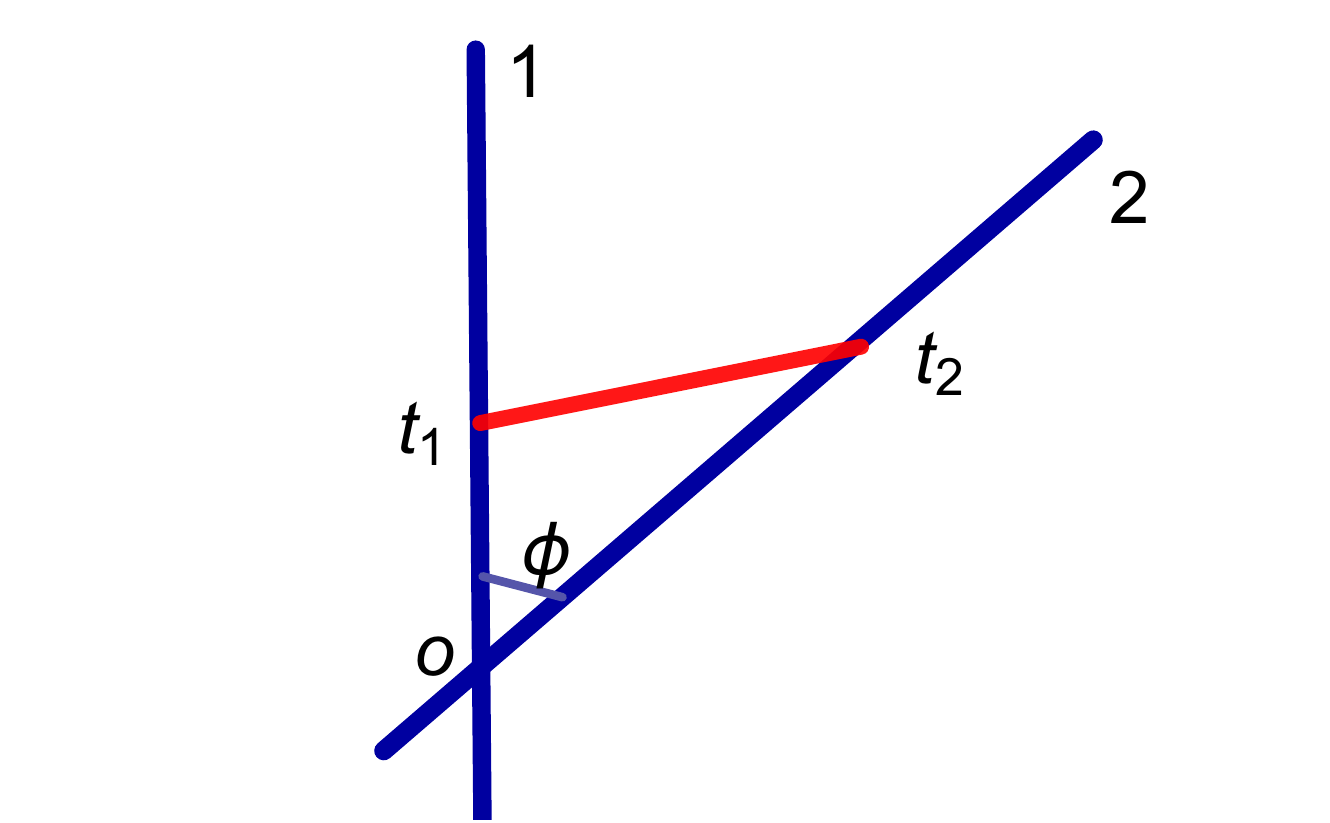}}} 
{\caption{ Two inertial observers.}} 
\end{center}
\end{figure} 
If  two inertial observers passed each other, and observer $ 1 $ sent a light signal at time $ t_1 $ and the observer $ 2 $ received it at the moment $ t_2 $  then we can write:
\begin{equation}
 t_2 = t_1 e^\phi, 
\end{equation}
where $\phi$ is the angle between the world lines of observers $1$ and $2$.
\begin{figure}[ht]
\begin{center}
{\rotatebox{0}{\includegraphics[width=2.5in]{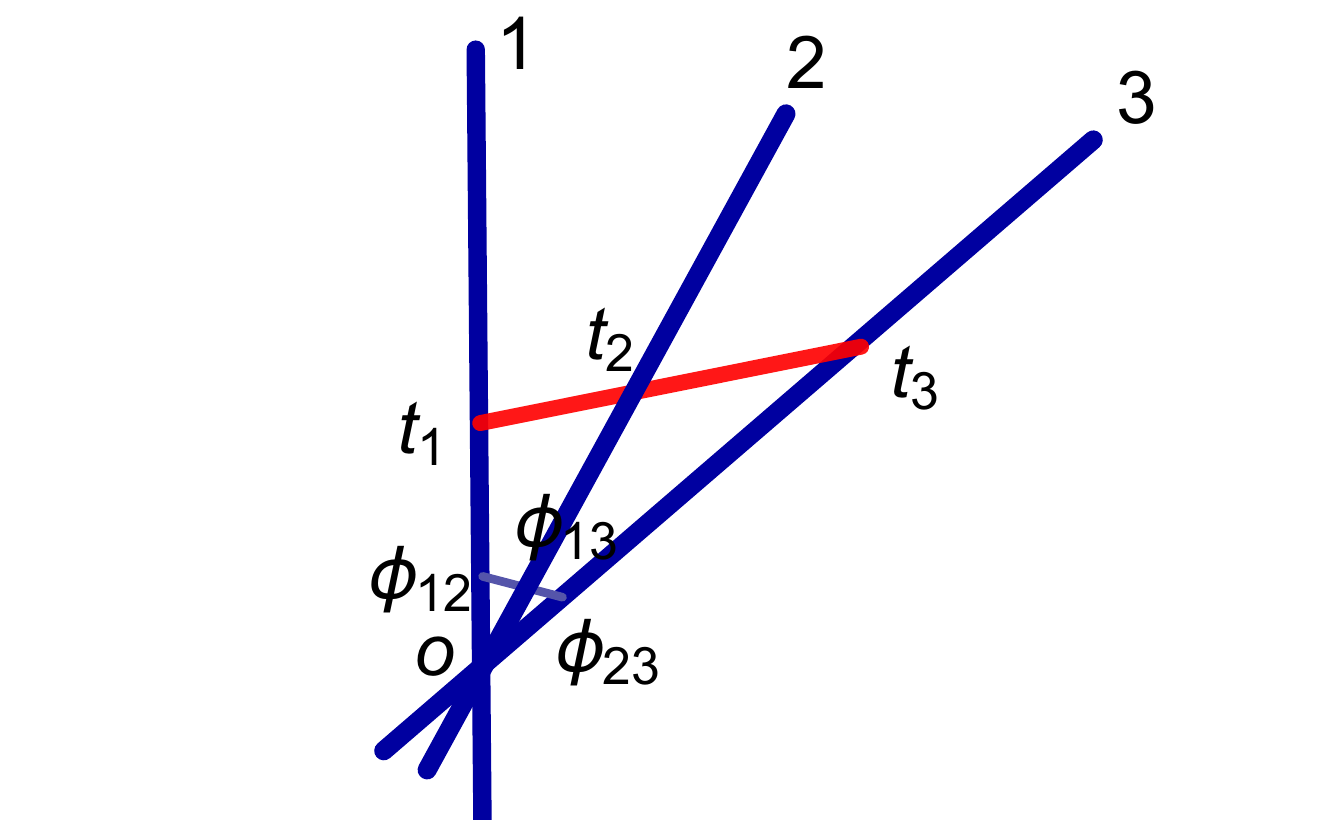}}} 
{\caption{ Three inertial observers in the same plane.}} 
\end{center}
\end{figure} 
Consider three inertial observers moving in the same plane. Let their world lines intersect at  point $ O $. Then:
\begin{equation}
 t_2 = t_1 e^{\phi_{12}}, \;\;\; t_3 = t_1 e^{\phi_{13}}=t_2e^{\phi_{23}},  
\end{equation}
\begin{equation}
\phi_{13}=\phi_{12}+\phi_{23},  
\end{equation}
as in Euclidean geometry. The addition of angles is true only for world lines in one plane.
\begin{figure}[ht]
\begin{center}
{\rotatebox{0}{\includegraphics[width=2.5in]{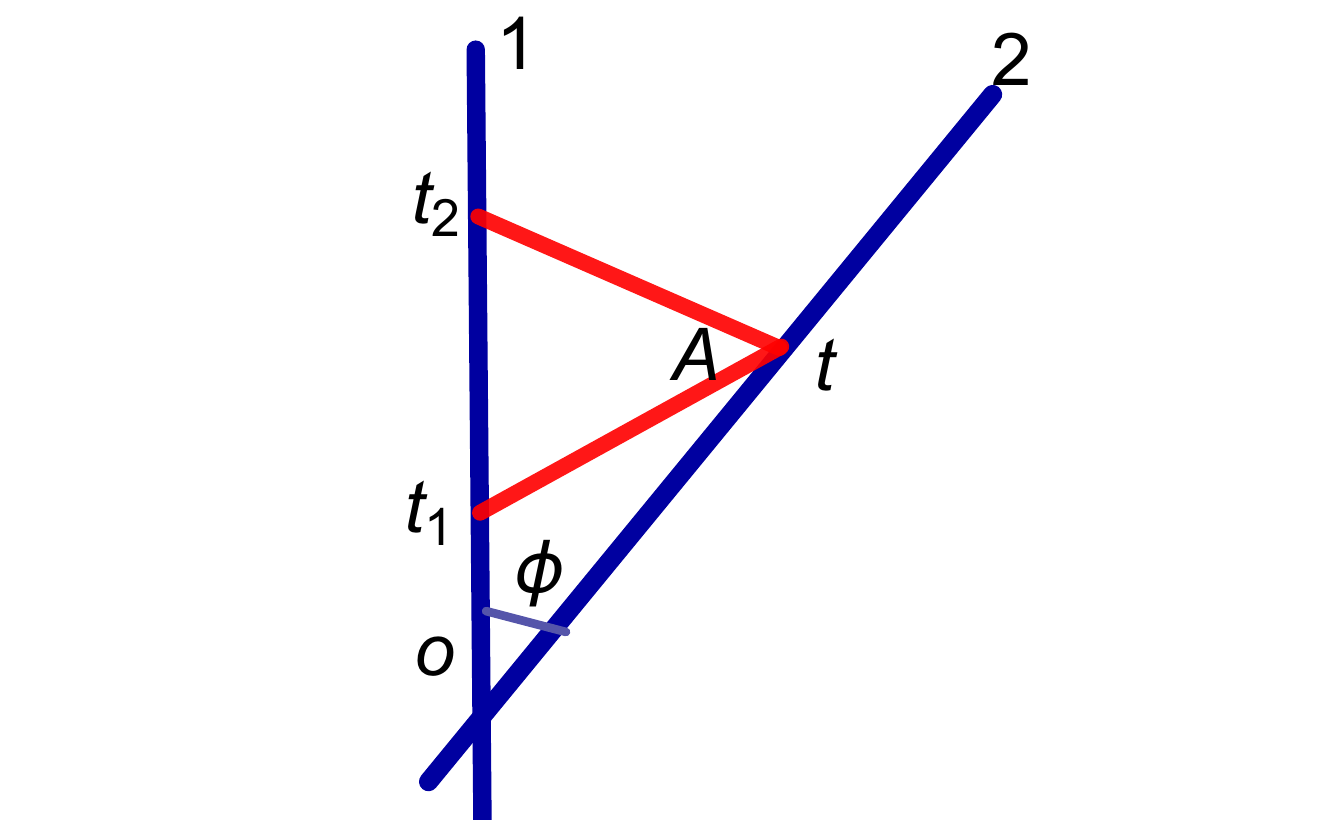}}} 
{\caption{\label{ch2f7} Signal exchange between two inertial observers.}} 
\end{center}
\end{figure} 
Now let the observer $ 1 $ send a light signal at the moment $ t_1 $ and the observer $ 2 $ receives it at $ t $ in event $A$ Fig.~(\ref{ch2f7}), and  send a response signal. The observer $ 1 $ receives it at the moment $ t_2 $. As we know, $ t = t_1 e ^ {\phi} $ ; besides, $ t_2 $ = $ t e ^ {\phi} $, then we get
\begin{equation}
 t_1 = t e^{-\phi}, \;\;\;\;\;\; t_2 = t e^{\phi}.  
\end{equation}
If event $A$ occurred outside the  world  line of observer $ 1 $, so observer cannot  measure the time . It is better to find the time of event $A$ from the point of view of observer $1$ in the interval $ [t_1, t_2] $, since the light signal propagates in both directions equally. Similarly we find the coordinate of the event $A$ from the  view of the observer $ 1 $ as half of this time interval. Thus,  the time and space coordinates of the event $A$ from the point of view of the observer $ 1 $ are
\begin{equation}
 x^0 = \frac{t_1+t_2}{2}, \;\;\;\;\;\; x^1 = \frac{t_2-t_1}{2}.  
\end{equation}
By substituting Eq. (2.4), we get
\begin{equation}\label{formul2.6}
 x^0 = t \cosh{\phi}, \;\;\;\;\;\; x^1 = t \sinh{\phi}.  
\end{equation}
We found  very important result for defining  the magnitude
\begin{equation}
 x^2 = (x^0)^2-(x^1)^2=t^2. 
\end{equation}
and it does not depend on how the observer $ 1 $  moves . The components of the $ x ^ 0 $, $ x ^ 1 $  from point $ O $ to point $ A $ are different for each observer.  The invariant quantity $ x ^ 2 $ is the square of the distance from $ O $ to $ A $ and it is equal to square of $ t $, according to  the observer $ 2 $ . Also compared to the usual Euclidean calculation there is difference in sign. Instead of the plus sign between two members there is a  minus sign. For this reason, the space – time geometry is called pseudo-Euclidean or Minkowski geometry. It can be seen from Eq.~(\ref{formul2.6}), where usual Newtonian velocity of the observer $ 2 $ with respect to the observer $ 1 $ is equal to
\begin{equation}
    u=\frac{x^1}{x^0}=\tanh{\phi}
\end{equation}
where $ u < 1 $, and tends to $ 1 $ when $ \phi \to \infty $. But  it is more convenient to use the angle $ \phi $ between the world lines.
\begin{figure}[ht]
\begin{center}
{\rotatebox{0}{\includegraphics[width=2.5in]{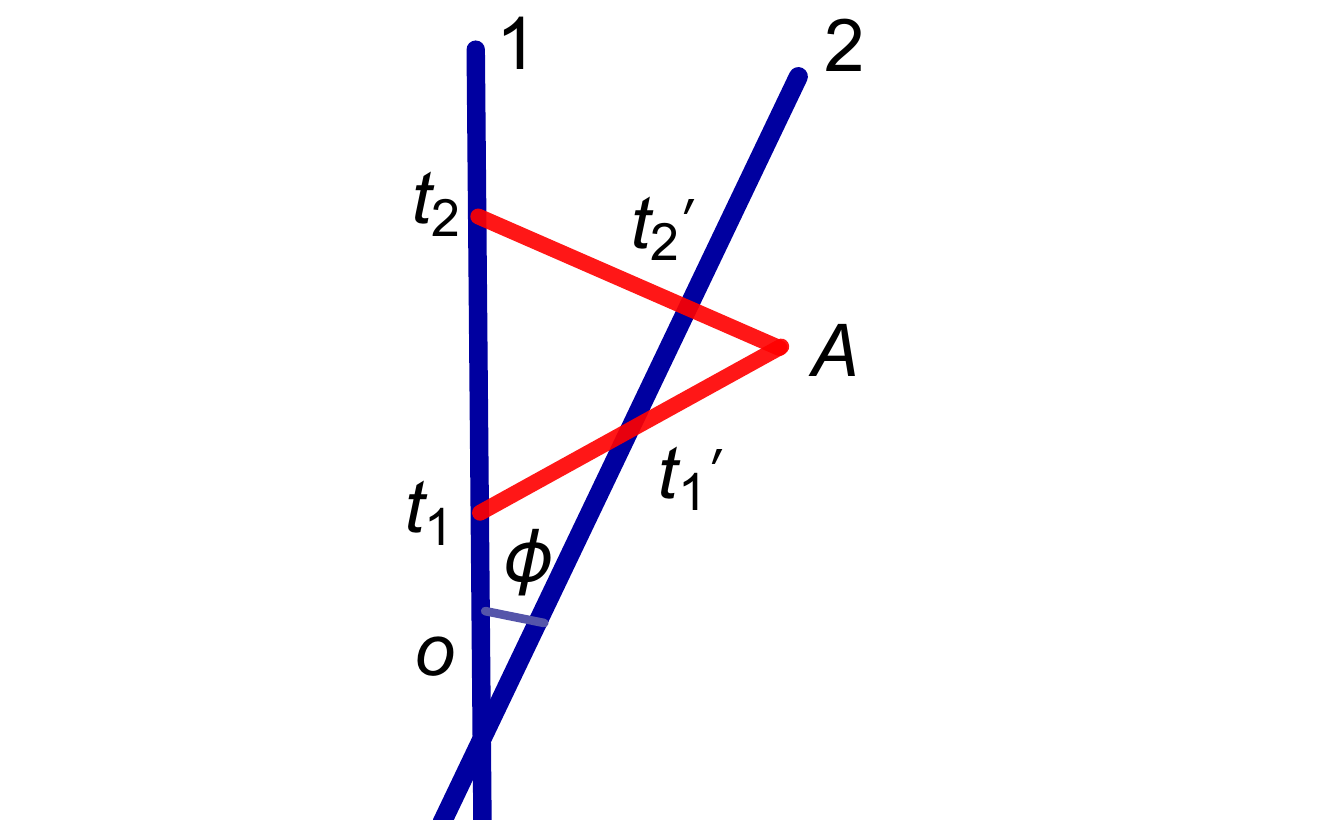}}} 
{\caption{\label{ch2f8} Lorentz transformation.}} 
\end{center}
\end{figure} 
In the  Fig.(\ref{ch2f8}) it is shown how are the coordinates of $x^0$, $x^1$  of events $A$ related to the observer $1$ and the coordinates $x^0$', $x^1$' of the same event from the view of the observer $2$. 
\begin{equation}
 x'^{0}=\frac{t_1'+t_2'}{2},\;x'^{1}=\frac{t_2'-t_1'}{2},  
\end{equation}
where
\begin{equation}
 t_1' = t_1 e^{\phi}, \;\;\;\;\;\; t_2' = t_2 e^{-\phi}.  
\end{equation}
The final step involves the substitution of:
\begin{equation}
 t_1 = x^0-x^1, \;\;\;\;\;\; t_2 = x^0+x^1.
\end{equation}
This will finally give us our answer, 
\begin{equation}
x'^{0} = x^0 \cosh \phi - x^1 \sinh \phi, \quad x'^{1} = -x^0 \sinh \phi + x^1 \cosh \phi,  
\end{equation}
which is the well-known \textbf{Lorentz transformation}.

\newpage
\section{Geometry of Minkowski}
Lets consider the vector $ x $ from points $ O $ to point $ A $ Fig.~(\ref{ch2f4}). Logically, we can conclude the 3 possibilities:
\begin{itemize}
  \item $ x ^ 2> 0 $ is a timelike vector and it can be directed to the future or to the past. By  Lorentz transformation, we can ensure that component $ x ^ 0 $ is  nonzero and others are zero .
  \item  $ x ^ 2 = 0 $ is a light-like vector, which  $A$ lies on the light cone of event $O$,  directed along the light cone. Two light-like vectors cannot be compared (which is longer and shorter), except when they are collinear.
  \item $ x ^ 2 <0 $ is a space-like vector. By Lorentz  transformation, we can find that the only nonzero component is $ x ^ 1 $ (when the  $ O $ and $ A $ events occur simultaneously).
\end{itemize}
Previously we described the vector $ x ^ 2 $. And now lets make the scalar product of two vectors ($ x \cdot y $), 
\be 
x \cdot y =\frac{(x+y)^2-x^2-y^2}{2}=x^0 y^0 - x^1 y^1.
\ee
In the four-dimensional space – time $x \cdot y = x^0 y^0 - x^1 y^1 - x^2 y^2 - x^3 y^3$.  Beside  the contravariant components $x^\mu$ of the vector $x(\mu = 0, 1, 2, 3)$, we introduce covariant components $x^\mu$:
\be x_0=x^0 , \;\;\;x_1=-x^1 , \;\;\;x_2=-x^2 , \;\;\;x_3=-x^3 . \ee
Then the dot product has a simple form:
\be x \cdot y = x^\mu y_\mu= x_\mu y^\mu, \ee
where the  index  on top and on the bottom means the summation from 0 to 3. The vectors $ x $ and $ y $ are orthogonal if there scalar product equal to zero  ($x\cdot y = 0$). 

\begin{figure}[ht]
\begin{center}
{\rotatebox{0}{\includegraphics[width=2.5in]{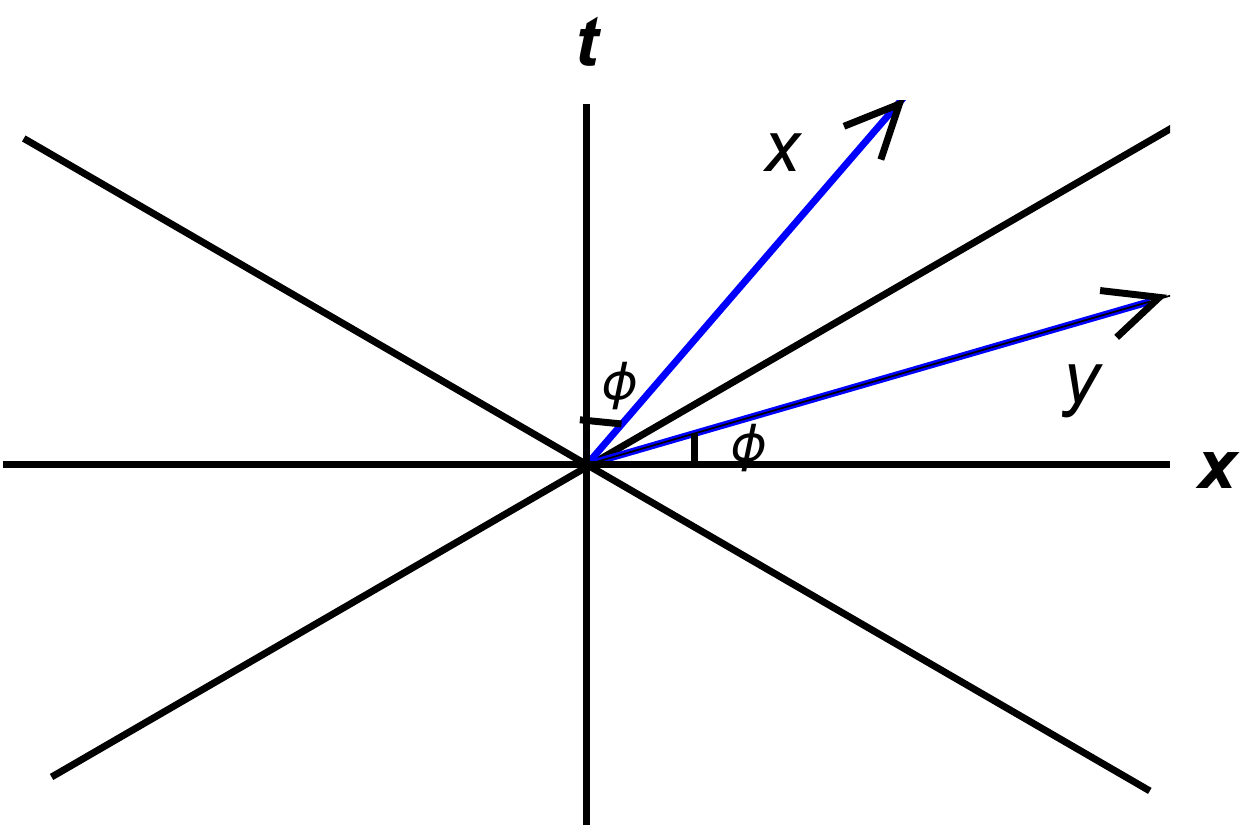}}} 
{\caption{\label{ch3f1} Orthogonal vectors.}} 
\end{center}
\end{figure}
\begin{figure}[ht]
\begin{center}
{\rotatebox{0}{\includegraphics[width=2.5in]{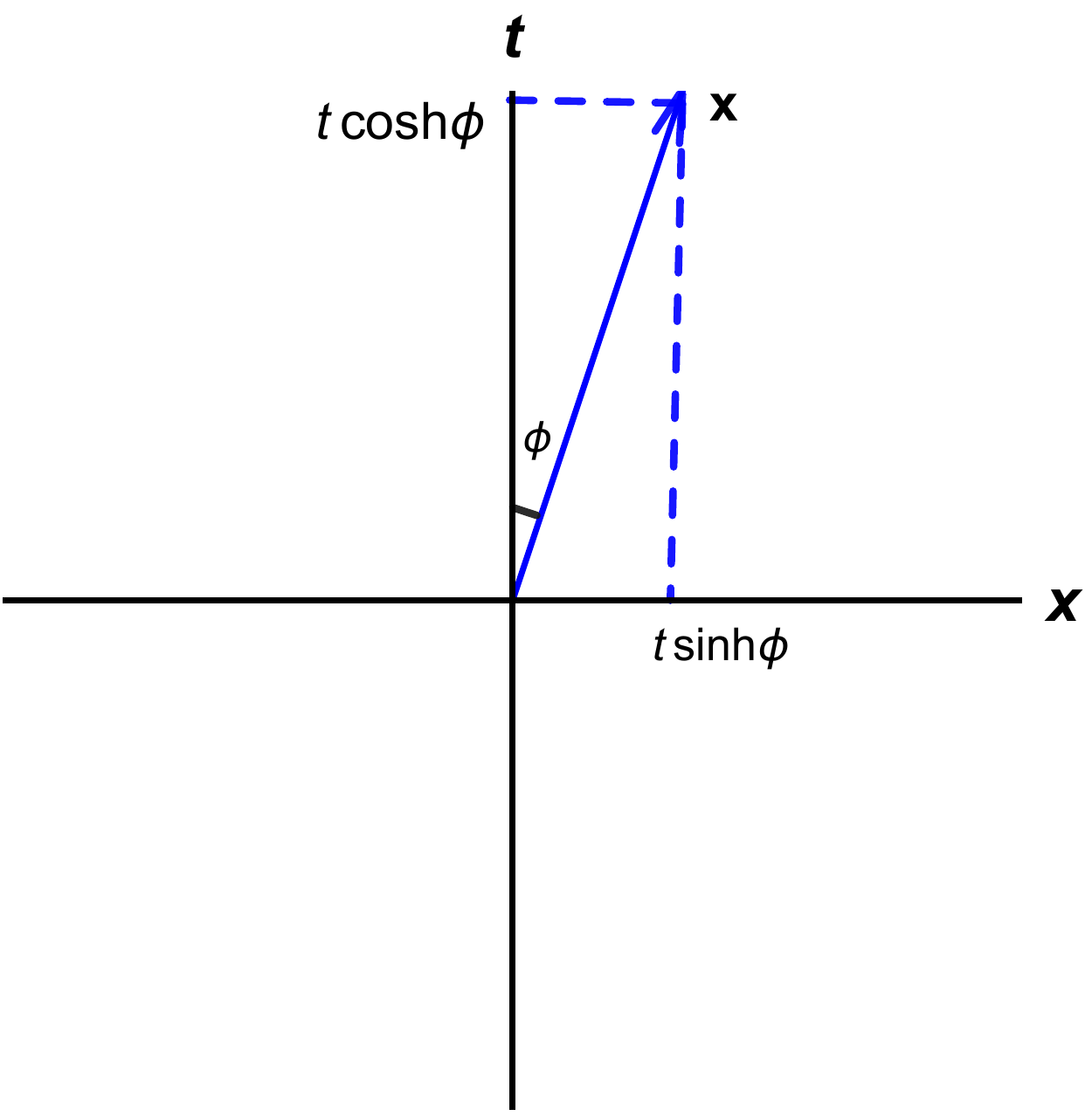}}} 
{\caption{\label{ch3f2} The projections of the timelike vector $ x $ on the coordinate axis.}} 
\end{center}
\end{figure}

 
Two timelike vectors can not be orthogonal to each other. If both have same direction to the future or  past their scalar product is greater than zero. If one have direction to the future and another to the  past their scalar product is less than zero.   The timelike vector $ x $ is orthogonal to the space vector $ y $, if their directions are symmetrical to each other with respect to the light-like straight line Fig.~(\ref{ch3f1}). Orthogonality is not violated if $y$ is multiplied by a scalar. Two space-like vectors can be orthogonal to each other, if the plane is space-like, then its geometry is Euclidean, and these two vectors can be orthogonal.
A timelike vector $ x $ of length $ t $ ($ x ^ 2 = t ^ 2 $), with angle $ \phi $ directed to the time axis, has the components $ x^\mu = t (\cosh \phi, \sinh \phi )$ Fig.~(\ref{ch3f2}).  Its projection in the direction $ e^\mu = (1, 0) $ equal to  $t\cosh\nu $. It is always greater or equal to $ t $, its equal when $ \phi = 0 $.

\section{Four-velocity and four-acceleration}

The main description for this section was taken from Rindler's book \cite{Rindler:1960zz}. When we consider motion along timelike directions we often  work with the invariant:

\be\label{2,4,1}
d \tau^2= \frac{ds^2}{c^2}=dt^2-\frac{dx^2+dy^2+dz^2}{c^2}
\ee
where $d\tau$ is a \textbf{interval of proper time}. Otherwise, we can write speed of the particle as $v$, then we have from Eq.(\ref{2,4,1}):
\be\label{2,4,2}
\frac{d \tau^2}{dt^2}=1-\frac{v^2}{c^2},\;\;\; \frac{dt}{d\tau}=\left(1-\frac{v^2}{c^2}  \right)^{-1/2} =\gamma.
\ee
Lets now consider four-vector motion with worldline $x^\mu=x^\mu(\tau)$. We denote vectors $v^\mu$ and $a^\mu$ as four-velocity and four-acceleration respectively:
\be
v^\mu=\frac{dx^\mu}{d\tau},\;\;\;\;\;\;a^\mu=\frac{d^2 x^\mu}{d\tau^2}=\frac{dv^\mu}{d\tau}.
\ee
From Eq.(\ref{2,4,2}) we find that:
\be 
v^\mu=\frac{dx^\mu}{d\tau}=\frac{dx^\mu}{dt}\frac{dt}{d\tau}=\gamma\frac{dx^\mu}{dt},
\ee
and 
\be 
a^\mu=\frac{dv^\mu}{d\tau}=\gamma\frac{dv^\mu}{dt}.
\ee
  The relation between the tree vector acceleration  $a$ and four acceleration $a^\mu$ can be written 
\be a^\mu=\frac{dv^\mu}{d\tau}=\gamma\frac{dv^\mu}{dt}= \gamma \frac{d}{dt}(\gamma c,\gamma v)= \gamma \left(\frac{d\gamma}{dt} c,\frac{d\gamma}{dt}v+\gamma a \right), \ee
and proper time is
\begin{equation}
a_{\mu}a^{\mu} = -\alpha^2. 
\end{equation}
From equation above we have:
\be a_{\mu}a^{\mu} = \gamma^2[\Dot{\gamma}^2 c^2-(\Dot{\gamma}v+\gamma a)^2], \ee
where $\gamma=\gamma(v)$. Using the relation $\Dot{\gamma}=\gamma^3 v\Dot{v}/c^2$ and three vector results like $(v^\mu)^2=v^2$ we find :
\be 
  \alpha^2=-a_{\mu}a^{\mu}= \gamma^2[\Dot{\gamma}^2 v^2+2\gamma \Dot{\gamma}v\Dot{v}+\gamma^2 a^2-\Dot{\gamma}^2 c^2]=\gamma^6 v^2 \Dot{v}^2/c^2 + \gamma^4 a^2.
 \ee
In an inertial frame in which the object is at the rest,  the 3-vector acceleration, combined with the zero time component, 4-object acceleration occurs, which makes it its own Lorentz-invariant acceleration. Thus, the concept is useful in the following cases: (i) with accelerated coordinates, (ii) at relativistic speeds and (iii) in curved spacetime.

In the unidirectional case where the acceleration of an object is parallel or antiparallel to its velocity to observer, the proper acceleration  $ \alpha $ and the acceleration of coordinates $ a $ are connected through the Lorentz factor $ \gamma $ for $ \alpha = \gamma ^ 3 a $. Therefore, the change in the proper velocity $ w = dx / d \tau $ is the integral of the proper acceleration with respect to the time of the stationary system $ t $, that is, $ \Delta w = \alpha \Delta t $ for the constant $ \alpha $. At low speeds, this comes down to  relationship between the coordinate velocity and the time of coordinate acceleration, that is, $ \Delta v = a \Delta t $.

For constant unidirectional proper acceleration, there are similar relations between the  rapidity $\eta$ and the elapsed proper time $\Delta \tau$, as well as between the Lorentz coefficient $\gamma$ and the distance $\Delta x$. Namely:
\be \alpha= \frac{\Delta w}{\Delta t}=c\frac{\Delta \eta}{\Delta \tau}=c^2 \frac{\Delta \gamma}{\Delta x}, \ee
where different velocity parameters are related by
\be \eta =\sinh^{-1}(\frac{w}{c})=\tanh^{-1}(\frac{v}{c})= \pm \cosh^{-1}(\gamma). \ee
These equations illustrate  some of the useful parameters of accelerated motion at high speed.

\section{Covariant Power}\label{sec:acc_basics}

Using the the relativistic covariant form of Larmor's formula, an accelerated point charge has power, 
\begin{equation}
\label{Power1} P = \frac{2}{3} q^2 \alpha^2, 
\end{equation}
where $\alpha$  is the proper acceleration (as we will see in Eq.~(\ref{properacc})) and $q$ is the charge.  A constant proper acceleration yields constant power. The proper acceleration is the scalar invariant magnitude 
\begin{equation}
\alpha^2 \equiv -a_{\mu}a^{\mu}, 
\end{equation}
of the four-acceleration $a^\mu$,
\begin{equation}
     a^{\mu} = \frac{d^2 x^{\mu}}{d\tau^2},
\end{equation}
most easily expressed (see e.g. \cite{ Rindler:1960zz, Proceedings}), in two nice forms:
\bea\label{properacc}
     \alpha^2
	&=& \gamma^4a^2 + \gamma^6(\boldsymbol{v}\cdot \boldsymbol{a})^2,\\
  &=& \gamma^6a^2 - \gamma^6(\boldsymbol{v} \times \boldsymbol{a})^2 \;.
  \eea
In the case of straight-line motion, parallel vectors $\boldsymbol{v}\times \boldsymbol{a} = \boldsymbol{0}$ yield $\alpha^2 = \gamma^6 a^2$. In the case of circular motion, the three-acceleration vector is perpendicular to the three-velocity vector, $\boldsymbol{v}\cdot \boldsymbol{a} = 0$, and one obtains $\alpha^2 = \gamma^4 a^2$. The acceleration ratio, $A_\textrm{WL} \equiv \alpha^2/a^2$, is given for each worldline in Appendix \ref{Appendix:Acceleration}.

In the instantaneous rest frame of the accelerating particle, the proper acceleration, $\alpha$, is the acceleration as measured by a hand-held accelerometer.  It is what is `felt' and is the property of the particle, a Lorentz scalar invariant in all frames. For a stationary worldline the Lorentz scalar remains constant and the proper acceleration, $\alpha = \kappa$ is the curvature, the simplest of the curvature invariants for stationary worldlines (the others being torsion, $\tau$, and hypertorsion, $\nu$ \cite{Letaw:1980yv}). The power is consequently,
\begin{equation}
 P = \frac{2}{3} q^2 \kappa^2.    
\end{equation}

\chapter{Stationary Worldlines } 
 
\section{Nulltor}
The Nulltor is a linear motion at the world line in 4-dimensional space-time.  Parameters of the particle which moves along this trajectory can be described through hyperbolic functions :
\begin{equation}
    x^\mu (s) = \kappa^{-1} \left(\sinh(\kappa s), 0, 0, \cosh(\kappa s)\right).
\end{equation} 
where $s$ is a proper time. The four-dimensional velocity vector is a time like  vector and lies inside the light cone. Due to derivation along time we can find three components of velocities on the $x,y$ and $z$ axis:
\bea v_x &=& \frac{ds}{dt} \frac{d x}{ds} = 0,\\v_y &=& 0, \\ 
v_z &=& \frac{ds}{dt} \frac{d z}{ds} = \tanh{( \kappa s )},\eea
and total velocity is:
\be\label{speedNull} v^2 = v_x^2 +v_y^2+ v_z^2 = \tanh ^2{( \kappa s )}.\ee
Three dimensional acceleration components could be found also through derivation of velocity :
\bea a_x &=& \frac{ds}{dt} \frac{d v_x}{ds} = 0\,,\\  a_y &=&0 ,\\
a_z &=& \frac{ds}{dt} \frac{d v_z}{ds} = \kappa \sech^3(\kappa s) .\eea
and total acceleration is:
\be a^2 = a_x^2+a_y^2 + a_z^2 = \kappa^2 \sech^6(\kappa s).\ee
Proper time $s$ can  described by $\gamma$ through equation :
\bea
1-\gamma^{-2} &=&   \tanh ^2{( \kappa s )} ,
\eea
here we substitute total velocity Eq. (\ref{speedNull}) with $\beta$ function which can be described through $\gamma$. Finally we will get:
\bea
s &=& \sqrt{\frac{1}{\kappa} \arctan(1-\gamma^{-2})} . 
\eea
Due to expression above we can find more shorter description of velocity and acceleration, like in Appendix \ref{Appendix:Derivatives}.

\newpage

\section{Ultrator}
Synchrotron  radiation , also can be called Ultrator motion, is    electromagnetic radiation and evaporated by charged particle. Synchrotron  radiation moving with relativistic speed along circular trajectory and  bent by a magnetic field. Parametric representation of synchrotron  radiation :
\begin{equation}
    x^\mu (s) = \rho^{-2} \left(\tau \,\rho \,s, \kappa \cos\rho s, 0, \kappa \sin \rho s\right).
\end{equation}
where $\rho^{2}=\tau^2-\kappa^2$ . Velocity components in three dimensional space:
\bea v_x &=& \frac{ds}{dt} \frac{d x}{ds} = -\frac{\kappa  \sin (\rho  s)}{\tau },\\v_y &=& 0 ,\\ 
v_z &=& \frac{ds}{dt} \frac{d z}{ds} = \frac{\kappa  \cos (\rho  s)}{\tau },\eea
\be v^2 = v_x^2 +v_y^2+ v_z^2 = \frac{\kappa ^2}{\tau ^2} .\ee
 Components of the acceleration:
\bea a_x &=& \frac{ds}{dt} \frac{d v_x}{ds} = -\frac{\kappa  \rho ^2 \sin (\rho  s)}{\tau ^2},\\  a_y &=&0 ,\\
a_z &=& \frac{ds}{dt} \frac{d v_z}{ds} = -\frac{\kappa  \rho ^2 \cos (\rho  s)}{\tau ^2}.\eea
%
Total acceleration  we find us: 
\be a^2 = a_x^2+a_y^2 + a_z^2 =  \frac{\kappa ^2 \rho ^4}{\tau ^4}.\ee
Also we can find expression of $\gamma$ in terms of $\kappa$ and $\tau$: 
 \be
 \gamma=\frac{1}{\sqrt{1-\frac{\kappa ^2}{\tau ^2}}}.
 \ee
In Appendix  \ref{Appendix:Derivatives} it is shown components of velocity and acceleration , in case when proper time is zero.
 
 \newpage
\section{Parator}
Here we consider a parator motion which moves along the cusp trajectory. Cusp - a singular point at which the curved line is divided into two  branches that have the same direction vector at this point. That is, the branches at a given point have a common tangent and the movement along them initially occurs in the same direction.
Parametric representation of cusp motion in flat space:
\be x^\mu (s) = \left(s + \frac{1}{6} \kappa^2 s^3, \frac{1}{2} \kappa s^2, 0, \frac{1}{6} \kappa^2 s^3\right). \ee
Derivation of the velocity components in three dimensional space: 
\bea v_x &=& \frac{ds}{dt} \frac{d x}{ds} = \frac{2 \kappa  s }{\kappa ^2 s ^2+2},\\v_y &=& 0 ,\\ 
v_z &=& \frac{ds}{dt} \frac{d z}{ds} = \frac{\kappa ^2 s ^2}{\kappa ^2 s ^2+2}.\eea
This gives total velocity:
\be\label{speedSq} v^2 = v_x^2 +v_y^2+ v_z^2 = \frac{\kappa ^2 \tau ^2 \left(\kappa ^2 \tau ^2+4\right)}{\left(\kappa ^2 \tau ^2+2\right)^2}\;.\ee
Further we examine the components of the acceleration at the same way,
%
\bea a_x &=& \frac{ds}{dt} \frac{d v_x}{ds} = \frac{8 \kappa -4 \kappa ^3 \tau ^2}{\left(\kappa ^2 \tau ^2+2\right)^3}\,,\\  a_y &=&0 ,\\
a_z &=& \frac{ds}{dt} \frac{d v_z}{ds} = \frac{8 \kappa ^2 \tau }{\left(\kappa ^2 \tau ^2+2\right)^3}.\eea
%
The vector summation gives us: 
\be\label{acc1} a^2 = a_x^2+a_y^2 + a_z^2 = \frac{16 \kappa ^2 \left(\kappa ^4 \tau ^4+4\right)}{\left(\kappa ^2 \tau ^2+2\right)^6}.\ee
Through Eq. (\ref{acc1}) we can find relation of the proper time as $s=s(\gamma)$:   
\be \gamma = 1+\frac{\kappa^2\tau^2}{2}\quad\Leftrightarrow\quad \tau = \kappa^{-1}\sqrt{2\gamma-2} , \ee
obtained  by adding Eq. (\ref{speedSq}) into the $\gamma$ definition. 

Finally, we can describe velocity and acceleration components in more convenient way, through relation between $\gamma$ and curvature invariants $\kappa$ and $\tau$, as it is shown in Appendix \ref{Appendix:Derivatives}.

\newpage
\section{Infrator}
In this section we consider Infrator motion , which has a trajectory of curved chain and also can be called catenary motion. Catenary  is a line, the shape of which takes a flexible homogeneous  chain (hence the name) with fixed ends in a uniform gravitational field. In geometry, the catenary is the graph of the hyperbolic cosine function. Parametric representation of catenary or infrator motion :
\begin{equation}
    x^\mu (s) = \sigma^{-2} \left(\kappa \sinh (s \sigma ), \kappa \cosh (s \sigma ), 0, s \tau \sigma\right).
\end{equation}
where $\sigma^2=\kappa^2-\tau^2$.Derivation of the velocity in 3-dimensional space:
%
%
\bea v_x &=& \frac{ds}{dt} \frac{d x}{ds} = \tanh (s \sigma ),\\v_y &=& 0 \; ,\\ 
v_z &=& \frac{ds}{dt} \frac{d z}{ds} = v_R\; \text{sech}(s \sigma ) ,\eea
where $v_R=\frac{\tau}{\kappa}$ is minimum velocity. Then total velocity :
\be v^2 = v_x^2 +v_y^2+ v_z^2 =\frac{\tau ^2 \text{sech}^2(s \sigma )}{\kappa^2}+\tanh ^2(s \sigma ) .\ee
Further we derive the components of the acceleration :
\bea
 a_x &=& \frac{ds}{dt} \frac{d v_x}{ds} =\frac{\sigma ^2 }{\kappa}\text{sech}^3(s \sigma ),
\eea
\bea 
 a_y &=&0 ,
\eea
\bea
a_z &=& \frac{ds}{dt} \frac{d v_z}{ds} =  - \frac{\sigma ^2 \tau  }{\kappa^2}\tanh (s \sigma ) \text{sech}^2(s \sigma ).
\eea

 Sum of the vectors give us: 
\be a^2 = a_x^2+a_y^2 + a_z^2 = \frac{\sigma ^4 \tau ^2 \tanh ^2(s \sigma ) \text{sech}^4(s \sigma )}{\kappa^4}+\frac{\sigma ^4 \text{sech}^6(s \sigma )}{\kappa^2}.\ee
If substitute total velocity with $\beta$ function we can find expression of proper time: 
\bea
1-\gamma^{-2} &=&  \frac{\tau ^2 \text{sech}^2(s \sigma )}{\kappa^2}+\tanh ^2(s \sigma ) ,
\eea
where proper time $s$ :
\bea
s&=&\frac{1}{2 \sigma }\textrm{ln} \left(2 \frac{\gamma ^2}{\gamma_R ^2} -2\frac{\gamma }{\gamma_R } \sqrt{\frac{\gamma ^2}{\gamma_R ^2}-1}-1\right),
\eea
where $\gamma_R^{-1}=\sqrt{1-v_R^2}$.  Finally, we can rewrite equation of total acceleration in more convenient way :
\bea a&=&\frac{\sqrt{\gamma ^2 \tau ^2+\kappa^2}}{\gamma ^3}. \eea

\newpage

\section{Hypertor}

The hypertor motion is a superposition of a uniform synchrotron motion and a constant linear motion, ultimately forming a spiral trajectory. The path of particle which moving along this world line has a helix form with slope that decrease to zero when its proper time equal to zero and increase when proper time grows. The hypertor motion is specified in world line coordinates by parametric equations of the form:
\begin{multline}
    x^\mu (s) =  \left(\frac{\Delta}{RR_{+}} \sinh (R_{+}s ),\, \frac{\Delta}{RR_{+}} \cosh (R_{+}s ),\right.\\ \left.\frac{\kappa \tau}{R\Delta R_{-}}\cos(R_{-}s),\, \frac{\kappa \tau}{R\Delta R_{-}}\sin(R_{-}s) \right),
\end{multline}
where all constants are the combination of three invariant variables:
\begin{equation*}  \Delta^2 \equiv \frac{1}{2}( R^2 +\kappa^2 + \tau^2 + \nu^2 ) \,\,,
\end{equation*}
 \begin{equation*} R^2 \equiv R_+^2 + R_-^2 \,\,,\end{equation*}
 \begin{equation*}
 R_{\pm}^2 \equiv {\sqrt{a^2+b^2} \pm a}\,\,,
 \end{equation*}
 \begin{equation*}
 a=\frac{1}{2} \left(\kappa ^2-\nu ^2-\tau ^2\right) \,\,,\qquad b= \kappa  \nu\,.
 \end{equation*}
Here we find the velocity of Hypertor in three coordinates as well as others:
\bea v_x &=& \frac{ds}{dt} \frac{d x}{ds} = \tanh \left(R_+ s\right),\\v_y  &=& \frac{ds}{dt} \frac{d x}{ds}= v_{\textrm{min}} \sin \left(R_- s\right) \text{sech} \left(R_+ s\right) \; ,\\ 
v_z &=& \frac{ds}{dt} \frac{d z}{ds} =v_{\textrm{min}} \cos \left(R_- s\right) \text{sech} \left(R_+ s\right) .\eea
where $v_{\textrm{min}}=\frac{\kappa \tau}{\Delta^2}$ is a minimum velocity,
\be v^2 = v_x^2 +v_y^2+ v_z^2 = 1-\frac{\text{sech}^2(R_{+}s)}{\gamma_{\textrm{min}}^2},\ee
where $\gamma_{\textrm{min}}^{-1}=\sqrt{1-v_{\textrm{min}}^2}$.  Further we take derivation again and find acceleration :
\be
a_x=\frac{R R_{+}}{\Delta}\text{sech}^3(R_{+} s),
\ee
\be 
a_y=\frac{R  v_{\textrm{min}}}{\Delta }\text{sech}^2(R_{+} s)[R_{+}\sin (R_{-} s)\tanh (R_{+} s)-R_{-}\cos (R_{-} s)],
\ee
\be
a_z = - \frac{R v_{\textrm{min}}  }{\Delta }\text{sech}^2(R_{+} s)[R_{+} \cos (R_{-} s) \tanh (R_{+} s)+R_{-} \sin (R_{-} s)],
\ee

\be a^2 = a_x^2+a_y^2 + a_z^2 = \frac{R^2 \text{sech}^4(R_{+} s) }{\Delta ^2} \left( R^2 v_{\textrm{min}}^2+R_{+}^2 \frac{\text{sech}^2(R_{+}s)}{ \gamma_{\textrm{min}}^2}\right).\ee

Then substitute total velocity with $\beta$ and find proper time : 
\be
1-\gamma^{-2} =  1-\frac{\text{sech}^2(R_{+}s)}{\gamma_{\textrm{min}}^2} ,
\ee
where proper time $s$ :
\be
s=-\frac{1}{R_+}\cosh ^{-1}\left(\frac{\gamma}{\gamma_{\textrm{min}}}\right),
\ee
and we can also rewrite acceleration components:
\be a^2=\frac{\gamma _{\min }^4 }{\gamma
   ^4 \Delta ^2}\left(R^4 v_{\min }^2+\frac{R^2 R_{+}^2}{\gamma ^2}\right). \ee

\newpage

\section*{Vacuum Spectra}
Stationary world lines are special in part because each have a unique spectrum, e.g. Parator is exactly calculable and distinctly non-Planckian.  Each unique motion is a stationary world line solution of the Frenet equations when the curvature invariants of proper acceleration and proper angular velocity are constant.  There is a dearth of papers that have treated these worldlines over the last four decades\footnote{A more recent treatment discussing the difficulties in finding a characterization of the notion of higher order time derivatives of constant proper $n-$acceleration itself as a Lorentz invariant statement and its relationship to the Frenet-Serret formalism is given by Pons and Palol \cite{Josep}.}, for instance the exactly solvable motion has been treated in only a handful of papers: 
\cite{Rosu:1999ad, Padmanabhan:1983ub,Takagi:1986kn,Audretsch:1995iw,Sriramkumar:1999nw,Leinaas:2000mh,Rosu:2005iu,Louko:2006zv,Obadia:2007qf,Russo:2009yd}. 

As an illustration of the importance of investigating these worldlines, it is good to remark that the vacuum spectra differ among the stationary worldlines. For instance, two can be calculated exactly: the Nulltor and Parator motions which are both one parameter motions ($\kappa$ only).  The vacuuum fluctuation spectrum for the zero torsion case \cite{Letaw:1980yv}, is the same Planckian distribution as found by Unruh \cite{Unruh:1976db} for uniform acceleration radiation, and scales (with $\omega/\kappa \gg 1$): 
\begin{equation}
\frac{1}{e^{2\pi \omega/\kappa}-1} \approx e^{-2\pi \omega/\kappa}, \quad \textrm{with} \quad T = \frac{\kappa}{2\pi} = 0.159\kappa.
\end{equation}
The spectrum for the paratorsional worldline is Letawian, with a hotter temperature (by a factor of $\pi/\sqrt{3} = 1.81$), scaling:
\begin{equation}
    e^{-\sqrt{12} \omega/\kappa}, \quad \textrm{with} \quad T = \frac{\kappa}{\sqrt{12}} = 0.289\kappa. 
\end{equation} 
The Nulltor and Parator cases both have uniform acceleration of $\kappa$, with an exactly analytic spectrum, recently confirmed in \cite{Proceedings}.

\chapter{Calculating the Power Distribution} \label{sec:re_stress}

\section{Lienard-Wiechert Potentials}
In order to correctly calculate the power distribution, we must take into account the relativistic effects of a point charge in a field. In this section, we will find the potentials for moving a point charge, and then in the following sections we will find the electric and magnetic fields, where we will use them to find the angular distribution and finally the power distribution.Also we will do all our calculation in natural units, then we must remember that $c=1,\e=1/4\pi,\mu_0=4\pi$, where $c$ is the speed of light, $\e$ and $\mu_0$ is a electric and magnetic constants respectively. Further we will consider moving particle in the radiated sphere . Lets firstly specified trajectory,
\be  \bd{w}(t) = \textrm{position of } q \textrm{ at time } t .\ee
The retarded time (time that take a particle to move from the center) can be found by the equation 
\be |\bd{r}-\bd{w}(t)|=t-t_r ,\ee
where the left side is the distance that radiation travel and right side is the time it takes to it. We call $\bd{w}(t)$ as the \textbf{retarded position} of the particle, then $\brcurs$ is the vector from \textbf{retarded position} to the field point $\bd{r}$:
\be \brcurs = \bd{r} - \bd{w}(t). \ee
Now, we can write basic potential equation like:
\be V(\bd{r},t)= \int \frac{\rho(\bd{r'},t_r)}{\rcurs} d\tau' \ee
where $\rho$  and $\tau'$  is a density and volume respectively. We can suggest that retarded potential of point charge is 
\begin{equation*}
 V(\bd{r},t)=\frac{q}{\rcurs}.
\end{equation*}
However it is not true, because it did not account relativistic effect. It is true that $\rcurs$ comes outside the integral\footnote{There is $\rcurs$ have  change in its functional dependence: $\rcurs=|\bd{r}-\bd{r'}|$ before integration, $\bd{r'}=\bd{w}(t_r)$ and $\rcurs=|\bd{r}-\bd{w}(t_r)|$ after integration is a function of $\bd{r}$ and t.}, but charge not equal to:
\begin{equation}
q=\int \rho(\bd{r'},t_r) d\tau'.
\end{equation}
Here we must remember about retarded time ($t_r=t-\rcurs$), and for extended particle the retardation will be written with factor $(1-\hat{\brcurs} \cdot \bd{v})^{-1}$, where $\bd{v}$ is the velocity of the charge:
\be 
\int \rho(\bd{r'},t_r) d\tau' =\frac{q}{1-\hat{\brcurs} \cdot \bd{v}} , \ee
where by extended we mean  relativistic growth. For example when linear object become longer when in move faster or like in our case, 
\be 
\tau'=\frac{\tau}{1-\hat{\brcurs} \cdot \bd{v}},
\ee
where $\tau$ is actual volume and $\tau'$ is a apparent volume, and $\hat{\brcurs}$ is a unit of vector from object to observer. Finally our potential will be :
\be V(\bd{r},t)= \frac{q}{\rcurs-\brcurs \cdot \bd{v}}. \ee
where $\brcurs=\hat{\brcurs}\rcurs$ is the vector from retarded position to the field point $\bd{r}$ and $\bd{v}$ is the velocity of charge at the retarded time. Also vector potential will be:
\be  A(\bd{r},t)= \int \frac{\rho(\bd{r'},t_r) \bd{v}(t_r)}{\rcurs} d\tau'=\frac{\bd{v}}{\rcurs} \int \rho(\bd{r'},t_r) d\tau', \ee 
or
\be  A(\bd{r},t)= \frac{q \bd{v}}{\rcurs-\brcurs \cdot \bd{v}}= \bd{v} V(\bd{r},t).\ee
These equations are \textbf{ Lienard-Wiechert potentials} for a moving point charge.

\newpage

\section{The field of a moving charge}

To calculate the electric and magnetic fields of charge in natural units ($c=1,\e=1/4\pi,\mu_0=4\pi$), we are use the Lienard-Wiechert potentials:
\be 
V(\bd{r},t)= \frac{q }{(\rcurs  - \brcurs \cdot \bd{v})} , \;\;\;\;\; \bd{A}(\bd{r},t)= v\; V(\bd{r},t),
\ee
and substitute them in Maxwell equations for electric and magnetic fields:
\be
\bd{E}= -\nabla V - \frac{\partial \bd{A}}{\partial t}, \;\;\;\;\; \bd{B}= \nabla \cross \bd{A}.
\ee
Further we will use the equations like:
\be   \brcurs=\bd{r}-\bd{w}(t_r),\;\;\textrm{and} \;\;\; \bd{v}=\bd{w}(t_r), \ee
where $t_r$ is retarted time. We can write fuction of $\bd{r}$ and $t$ like:
\be  |\bd{r}-\bd{w}(t_r)|=t-t_r. \ee
Firstly let's start from gradient of potential $V$:
\be\label{nablaV} 
\nabla V = - \frac{q}{(\rcurs  - \brcurs \cdot \bd{v})^2} \nabla(\rcurs  - \brcurs \cdot \bd{v}).   \ee
By taking derivation of $\rcurs=t-t_r$,
\be\label{5.6} \nabla \rcurs = - \nabla t_r. \ee 
Second term of product gives,
\be\label{5.1} \nabla(\brcurs \cdot \bd{v}) = (\brcurs \cdot \nabla)\bd{v}+(\bd{v}\cdot \nabla)\brcurs+\brcurs \cross(\nabla \cross \bd{v}) +\bd{v}\cross(\nabla \cross \brcurs). \ee
 We will derive  each term of Eq.~(\ref{5.1}) separately and add it together at the end. First term of  Eq.~(\ref{5.1}) gives
 
 \begin{equation}
\begin{aligned}
 (\brcurs \cdot \nabla)\bd{v} = \left( \rcurs_x \frac{\partial}{\partial x}+\rcurs_y \frac{\partial}{\partial y}+\rcurs_z \frac{\partial}{\partial z} \right) \bd{v}(t_r)=\\  \rcurs_x \frac{d \bd{v}}{d t_r} \frac{\partial t_r}{\partial x}+\rcurs_y \frac{d \bd{v}}{d t_r} \frac{\partial t_r}{\partial y}+\rcurs_z \frac{d \bd{v}}{d t_r} \frac{\partial t_r}{\partial z} = \bd{a}(\rcurs \cdot \nabla t_r),
\end{aligned}
\end{equation}
 
where $a=\bd{\Dot{v}}$ is the acceleration at the retarded time of the particle. Second term of Eq.~(\ref{5.1}) gives
\be 
(\bd{v}\cdot \nabla)\brcurs = (\bd{v} \cdot \nabla)\bd{r} - (\bd{v} \cdot \nabla)\bd{w}, \ee
and
\begin{equation}
\begin{aligned}
(\bd{v} \cdot \nabla)\bd{r} = \left( v_x \frac{\partial}{\partial x}+v_y \frac{\partial}{\partial y}+v_z \frac{\partial}{\partial z} \right)(x \hat{\bd{x}}+y \hat{\bd{y}}+z \hat{\bd{z}})=\\v_x \hat{\bd{x}}+v_y \hat{\bd{y}}+v_z \hat{\bd{z}}=\bd{v}
\end{aligned}
\end{equation}
while
\be
(\bd{v} \cdot \nabla)\bd{w}=\bd{v}(\bd{v}\cdot \nabla t_r).\ee
Third term of  Eq.~(\ref{5.1}) gives
\begin{equation}
\begin{aligned}
\nabla \bd{v} =\left(\frac{\partial v_z}{\partial y} - \frac{\partial v_y}{\partial z} \right) \hat{\bd{x}}+\left(\frac{\partial v_x}{\partial z} - \frac{\partial v_z}{\partial x} \right) \hat{\bd{y}}+\left(\frac{\partial v_y}{\partial x} - \frac{\partial v_x}{\partial y} \right) \hat{\bd{z}}= \\ \left(\frac{d v_z}{dt_r}\frac{\partial t_r}{\partial y}-\frac{d v_y}{dt_r}\frac{\partial t_r}{\partial z} \right) \hat{\bd{x}}+\left(\frac{d v_x}{dt_r}\frac{\partial t_r}{\partial z}-\frac{d v_z}{dt_r}\frac{\partial t_r}{\partial x} \right) \hat{\bd{y}}+\left(\frac{d v_y}{dt_r}\frac{\partial t_r}{\partial x}-\frac{d v_x}{dt_r}\frac{\partial t_r}{\partial y} \right) \hat{\bd{z}} = \\ - \bd{a} \cross \nabla t_r .  
\end{aligned}
\end{equation}
Finally forth term gives
\be \nabla \cross \brcurs = \nabla \cross \bd{r} - \nabla \cross \bd{w}, \ee
where $\nabla \cross \bd{r}=0$, while 
\be \nabla \cross \bd{w} = - \bd{v} \cross \nabla t_r. \ee
If we put all terms together and reduce them by using "BAC-CAB" rule, we get
\begin{equation}
\begin{aligned}
\nabla(\brcurs \cdot \bd{v}) = \bd{a}(\brcurs \cdot \nabla t_r)+ \bd{v}-\bd{v}(\bd{v} \cdot \nabla t_r)-\brcurs \cross(\bd{a} \cross \nabla t_r)+ \bd{v}\cross(\bd{v} \cross \nabla t_r)= \\  \bd{v}+(\brcurs \cdot \bd{a}-v^2)\nabla t_r.
\end{aligned}
\end{equation}
Collecting together  Eq.~(\ref{5.1}) and Eq.~(\ref{nablaV}), we find
\be
   \nabla V =  \frac{q}{(\rcurs  - \brcurs \cdot \bd{v})^2}\left[\bd{v}+(1-v^2+\brcurs \cdot \bd{a})\nabla t_r \right].  
\ee
Now we must find $\nabla t_r$. This can be done by expanding Eq.~(\ref{5.6}):
\begin{equation}
\begin{aligned}
 - \nabla t_r=\nabla \rcurs=\nabla \sqrt{\brcurs \cdot \brcurs}=\frac{1}{2\sqrt{\brcurs \cdot \brcurs}}\nabla(\brcurs \cdot \brcurs)=\\ \frac{1}{\rcurs}\left[(\brcurs\cdot \nabla)\brcurs + \brcurs \cross(\nabla \cross \brcurs) \right].
\end{aligned}
\end{equation}
and 
\be (\brcurs\cdot \nabla)\brcurs= \brcurs-\bd{v}(\brcurs \cdot \nabla t_r), \ee
while
\be \nabla \cross \brcurs =(\bd{v} \cross \nabla t_r). \ee
Thus
\be
-\nabla t_r =  \frac{1}{\rcurs}\left[\brcurs - \bd{v}(\brcurs \cdot \nabla t_r)+\brcurs \cross (\bd{v} \cross \nabla t_r)\right]=\frac{1}{\rcurs}\left[\brcurs-(\brcurs \cdot \bd{v})\nabla t_r \right],
\ee
and hence 
\be 
\nabla t_r=-\frac{\brcurs}{\rcurs- \brcurs \cdot \bd{v}}. 
\ee
Finally gradient of potential $V$ will be :
\be
   \nabla V = \frac{q }{(\rcurs  - \brcurs \cdot \bd{v})^3}\left[(\rcurs  - \brcurs \cdot \bd{v})\bd{v}-(1-v^2+\brcurs \cdot \bd{a})\brcurs \right].  
\ee
By using the same way of calculation we can find derivation of vector potential $A$ by time $t$:
\be
\frac{\partial \bd{A}}{\partial t}= \frac{q }{(\rcurs  - \brcurs \cdot \bd{v})^3}\left[(\rcurs  - \brcurs \cdot \bd{v})(-\bd{v}+\rcurs\bd{a})+\rcurs(1-v^2+\brcurs \cdot \bd{a})\brcurs \right].
\ee
By combing our results we get electric field,
\begin{equation}
    \bd{E}(r,t)= \frac{ q \rcurs }{(\brcurs \cdot \bd{u})^3}[(1-v^2)\bd{u}+\brcurs \times (\bd{u} \times \bd{a})],
\end{equation}
where vector $u=\hat{\brcurs}-\bd{v}$. Also we can find magnetic field of particle:
\be
B=\nabla \cross \bd{A}=\nabla \cross(V \bd{v})=V(\nabla \cross \bd{v})-\bd{v}\cross(\nabla V), \ee
here we know the answers of $\nabla \cross \bd{v}$ and $\nabla V$ from previous calculation.
\be
B=\nabla \cross \bd{A}=\frac{q}{(\bd{u} \cdot \brcurs)^3}\brcurs \cross [(1-v^2)\bd{v}+(\brcurs \cdot \bd{a})\bd{v}+(\brcurs \cdot \bd{u})\bd{a}], \ee
further we use "BAC-CAB" rule to combine and reduce description of magnetic field. Since $\brcurs$ crossed to all terms, we can change $\bd{v}$ into $-\bd{u}$ and get,
\be \bd{B}(\bd{r},t)=\hat{\brcurs}\cross \bd{E}(\bd{r},t).
\ee
Here we can see that $B$ is always perpendicular to $E$ and to the vector from retarded point.

\newpage

\section{The angular distribution of radiated power}
In this section we  derived angular distribution of radiated power in natural  units ($c=1,\e=1/4\pi,\mu_0=4\pi$).  Griffiths \cite{Griffiths}  demonstrated   derivation of the electric field of a point  charge in arbitrary motion, by using the Lienard-Wiechert potentials, where he found:
\begin{equation}
    \bd{E}(r,t)= \frac{ q \rcurs }{(\brcurs \cdot \bd{u})^3}[(1-v^2)\bd{u}+\brcurs \times (\bd{u} \times \bd{a})],
\end{equation}
where $\brcurs$ is a retarded vector (from retarded position to the field point $\bd{r}$), and $u=\hrcurs-\bd{v}$. The first term called \textbf{velocity field}, and the second is \textbf{acceleration field}
\be
\bd{B}(r,t)=\hrcurs \times \bd{E}(r,t). \ee
By adding them in Poynting vector :
\be
 \bd{S}= \frac{1}{4\pi} ( \bd{E} \times  \bd{B})= \frac{1}{4\pi} [ \bd{E} \times (\hrcurs \times \bd{E})]=\frac{1}{4\pi} [ E^2 \hrcurs -(\hrcurs \cdot \bd{E})\bd{E} ]
\ee
The surface of radiated sphere is proportional to $\rcurs^2$, so any results of Poynting vector that goes like $1/\rcurs^2$ will give us finite answer, however other terms like $1/\rcurs^3$, $1/\rcurs^4$ and so on will not give us nothing in the case $\rcurs$ goes to infinity. For this reason only second term called \textbf{acceleration field}(other name  \textbf{radiation field}) give us real solution of field:
\be
\bd{E}_{rad}=   \frac{q  \rcurs }{(\brcurs \cdot \bd{u})^3} [\brcurs \times (\bd{u} \times \bd{a})] \ee
As the radiation field is perpendicular to $\hrcurs$, second terms of Poynting vector will be zero:
\be
\bd{S}_{rad}=\frac{1}{4\pi} E^2_{rad} \hrcurs. \ee
Now we must count that the rate of energy which pass through the sphere is different from energy rate which left the charge. Situation is the same as in the Doppler effect. Because of moving sphere  we will add in our calculation extra term like:
\be
\frac{\brcurs \cdot \bd{u}}{\rcurs}=1- \hrcurs \cdot \bd{v}
\ee
The power emitted by particle in area $\rcurs^2 \sin{\theta} d\theta d\phi=\rcurs^2 d\Omega$ on the sphere finally is given by:
\be
\frac{dP}{d\Omega}= \left(\frac{\brcurs \cdot \bd{u}}{\rcurs} \right) \frac{1}{4\pi} E^2_{rad} \rcurs^2=\frac{q^2}{4 \pi} \frac{|\hrcurs \times (\bd{u} \times \bd{a})|^2}{(\hrcurs \cdot \bd{u})^5},
\ee
where $\Omega$ is the solid angle in which direction the power is emitted.

\newpage

\section{The radiated power }
For a charge in relativistic motion the power distribution has a well-known (e.g. Jackson \cite{DavidJackson}), result (using the notation of Griffiths \cite{Griffiths}) with the intricate vector algebra details in Appendix \ref{Appendix:Angular},
\begin{equation}
\label{AngDistr}  \frac{dP}{d\Omega} = \frac{q^2}{4\pi} \frac{|\hrcurs \times (\bd{u} \times \bd{a})|^2}{(\hrcurs \cdot \bd{u})^5}. 
\end{equation}
Two types of relativistic effects are present: (1) the spatial relationship between the velocity and acceleration and (2) the transformation between the instantaneous rest frame of the particle to the observer.\footnote{The denominator terms constitutes this dominant effect for ultra-relativistic motion.} The relevant derivatives for each worldline are given in Appendix \ref{Appendix:Derivatives}.
We define the power distribution function $I_{\textrm{WL}}(\theta,\phi)$,
\begin{equation}
\frac{dP}{d\Omega} \equiv  \frac{2}{3} q^2 \kappa^2 I_{\textrm{WL}}(\theta,\phi), 
\end{equation}  
so that for any particular worldline, the function $I_\mathrm{WL}$ satisfies the relation,
\begin{equation}
    \label{unitone} 
\int_0^{2\pi} \int_0^\pi I_{\textrm{WL}} \; \textrm{sin}\theta d\theta d\phi =1.
\end{equation}
The total radiating power is therefore given by a constant,
\begin{equation}
P = \frac{2}{3} q^2 \kappa^2. 
\end{equation}
Computing the power distribution, Eq.~(\ref{AngDistr}), requires straightforward but involved vector algebra.  We have done this work and the results for each class of motion are in Appendix \ref{Appendix:Power}, where we include the simplified exact algebraic answers, the main results of this paper.  Furthermore, we resolve two dimensional polar plots $\phi = 0$, and 3D plots of the shape of the light for various speeds in Appendix \ref{Appendix:Polar} and Appendix \ref{Appendix:3D}, respectively.

Recently, a claim of experimental evidence of thermalized radiation reaction  at the Fulling - Davies - Unruh temperature $2\pi T = \kappa$, has been announced \cite{Morgan2019} .  The researchers discovered a vacuum Larmor formula and power spectrum that are thermalized by the acceleration.  Their results underscore the importance of investigating Larmor radiation with uniform acceleration (i.e. the rectilinear case of Nulltor), in the context of the Unruh effect.

\chapter{Some Further Results}
\section{Scaling at High Speeds}
The intensity of maximum radiation in the ultra-relativistic case, ($\gamma \to \infty$), 
\be f \equiv \frac{(\left. d P/d\Omega \right|_{\theta=\theta_{\textrm{max}}})_{\textrm{ultra-rel}}}{(\left.d P/d\Omega\right|_{\theta=\theta_{\textrm{max}}})_{\textrm{rest}}}, \ee
is known for rectilinear motion, see e.g. Griffiths \cite{Griffiths}.  The angle $\theta_{\textrm{max}}$ at which maximum radiation is emitted occurs when
\be \frac{d}{d\theta} \Big[\frac{dP}{d\Omega}\Big] = 0. \ee
Calculation of derivative for all world lines was done with respect to the angle $\theta$ in the plane $\phi = 0$, except hypertor, since its  $\phi$ range is between $0$ and $2\pi$. After numerical and graphical analysis we find the solution angles and expand them for the high speed limit, 
\begin{eqnarray}
\theta_{\textrm{max}}^{\textrm{nulltor}} = \frac{1}{2\gamma}+ \mathcal{O}\left(\frac{1}{\gamma^2}\right)^{3/2}\,,\quad
\end{eqnarray}
\be \label{maxangle}\theta^{\textrm{parator}}_{\textrm{max}} = 
\sqrt{\frac{2}{\gamma}}  + \mathcal{O}\left(\frac{1}{\gamma}\right)^{3/2}, \ee
For the case of Nulltor and Parator motions, the answer does not depend on curvature invariants, and we will always get the same maximum angle. For the case of Infrator and Hypertor motions we removed the constant variables $\kappa, \tau$, and $\nu$, because they do not affect the final scaling results with respect to $\gamma$.  If we substitute values for the curvature invariants the same series by order  gamma with different numerical coefficients result. For this reason, the maximum angle for  Infrator and Hypertor motions below are shown to leading order in gamma,

\be \label{maxangle1}\theta^{\textrm{infrator}}_{\textrm{max}} \sim 
\frac{1 }{\gamma}+ \mathcal{O}\left(\frac{1}{\gamma^2}\right)^{3/2}, \ee

\be \label{maxangle11}\theta^{\textrm{hypertor}}_{\textrm{max}} \sim 
\frac{1}{\gamma}+ \mathcal{O}\left(\frac{1}{\gamma^2}\right)^{3/2}. \ee
The case of Ultrator is trivial as the radiation is always maximally beamed along $\theta_{\textrm{max}}^{\textrm{ultrator}}=0$ for a stationary worldline with circular spatial projection, as $\phi$ ranges from 0 to $2\pi$. These expressions demonstrate the maximum angle scaling for the aforementioned trajectories, placing emphasis on the $\gamma^{-1/2}$ scaling of Parator.  The slow scaling (relative to rectilinear motion) here explains the physics behind the delayed collimation of radiation at high speeds that can be seen in the polar plot, $\phi = 0$, in Appendix \ref{Appendix:Polar}.  The outlier case of Parator is also corroborated by the different scaling in the simple calculation of Thomas precession for the worldlines given in Appendix \ref{Appendix:Thomas}.  The $\gamma^{-1}$ result for Infrator and Hypertor corresponds to the usual beaming that occurs in rectilinear motion at high speeds.

\newpage
\section{Thomas precession}
Tomas precession is a kinematic effect of the special  relativity in the flat spacetime, which demonstrates  the changes in the orientation of vectors in the non-inertial frame. Used by  Llewellyn Thomas in 1926 to explain the spin-orbit interaction of an electron in the atom. If a force acting on a rotating gyroscope and  its speed changes , but there is no moment of force, then in classical mechanics such moving gyroscope will maintain the orientation of its own angular momentum (spin). In the theory of relativity, this is no longer the case, and as the gyroscope changes its speed, its spin vector will also change. Mathematically, this effect is associated with the group properties of the Lorentz transformations.

In 1926 in "Nature" journal,  Thomas published a notes \cite{ThomasNature}, where he explained the deviation of the measurement data by a factor of $ 1/2 $ , from the predictions of the structure of the hydrogen atom linked by spin-orbit splitting with Larmor precession. The work attracted great attention and the effect of the precession of the coordinate system during accelerated motion became known as the “Thomas precession”. The only source that was known to Thomas was De Sitter’s work on the precession of the moon, published by Arthur Eddington \cite{Eddington1924}.

Lets consider particle in the lab frame where observer can measure its relativistic motion. At each moment of time our particle has an  inertial frame where it will be at the rest. In this lab frame velocity of the particle will be bounded by the speed of light $(0 \leq v(t) < c )$ , where $t$ is a coordinate time. Speed of the particle can have any magnitude , except of the upper limit $c$,  and not necessarily  constant. Precession of the particle with angular velocity will give:

\be 
{\displaystyle {\boldsymbol {\omega }}_{\text{T}}={\frac {1}{c^{2}}}\left({\frac {\gamma ^{2}}{\gamma +1}}\right)\mathbf {a} \times \mathbf {v} }
\ee
where sign between $a$ and $v$ is the cross product and $\gamma$ is a Lorentz factor, a function of the instantaneous velocity.

\be
{\displaystyle \gamma ={\dfrac {1}{\sqrt {1-{\dfrac {|\mathbf {v} (t)|^{2}}{c^{2}}}}}}},
\ee
here angular velocity $(\omega_T)$ is a pseudovector and its magnitude is radians per second with direction along the rotation axis. In  cross product we use the right-hand rule. In Appendix \ref{Appendix:Thomas} we find six type of Thomas precession in all case of motions with approximations.

\let\cleardoublepage\clearpage

\chapter{Conclusions}

We have found the power distributions for light emitted by a point charge undergoing uniform acceleration for all five classes of stationary worldlines, including those with torsion and hypertorsion. After computing their proper accelerations and confirming that all of stationary motions have values equal to $\kappa$ (as they should be!), we have calculated acceleration ratios, minimum velocities and confirmed the results with constant power emission.  

We have graphically illustrated the emitted radiation from the uniformly accelerated trajectories.  Using the widest range of speeds, ranging from rest to near the speed of light, the illustrations are made in two and three dimensional space. After numerical and graphical analysis we found the maximum angle at which radiation is emitted at high speeds. 

Additionally, in the Appendices of this work we have included worldline plots of all systems with four-vector functions  $x^{\mu}(s)$, specifying the coordinates of each proper time point $s$ on the curve and special projections of their general trajectory equations including curvature, torsion and hypertorsion. 

Finally, we have derived Thomas precession for all  classes of stationary worldliness, as it is important as a relativistic correction and a kinematic effect in the flat spacetime of special relativity.  Its importance is underscored by its algebraic origin as a result of the non-commutativity of Lorentz transformations, and its accounting of relativistic time dilation between the electron and nucleus of an atom, giving the right correction to the spin-orbit interaction in quantum mechanics.

Uniform accelerated motion was crucial for construction of the equivalence principle in general relativity.  Motivated by the fact that investigations into uniform acceleration has yielded insights into gravity, thermodynamics and quantum field theory, we hope that this study may be an incipient direction that can lead toward further insights into the basic radiation emitted from fundamental uniformly accelerated motion (in electrodynamics\footnote{For an interesting and recent paper on paradoxes involving the electric field of a uniformly accelerating charge see \cite{DavidG} and references therein.} or otherwise). 

These solutions are exact results in classical electrodynamics and may be a first step toward a better understanding (or disentangling) related radiation effects like Letaw's vacuum spectra, Unruh's uniform acceleration radiation with torsion, or Davies-Fulling moving mirror radiation.  

Further work along this direction includes the computation of radiation reaction for each of the five classes of stationary worldline motions. This can be interesting to address because results may be used to contrast with the consensus example that Nulltor radiates without radiation reaction (a nice book on uniformly accelerated point charges including much discussion of this example was written by Lyle \cite{Lyle}).  Other extensions include investigation into the intensity spectrum distribution, and maximum intensity measure.

\let\cleardoublepage\clearpage

\appendix

\chapter{Worldline Plots}\label{Appendix:Worldline}

\subsection*{Nulltor}
\begin{equation}
    x^\mu (s) = \kappa^{-1} \left(\sinh(\kappa s), 0, 0, \cosh(\kappa s)\right).
\end{equation} 
\begin{figure}[ht]
\begin{center}
{\rotatebox{0}{\includegraphics[width=1.0in]{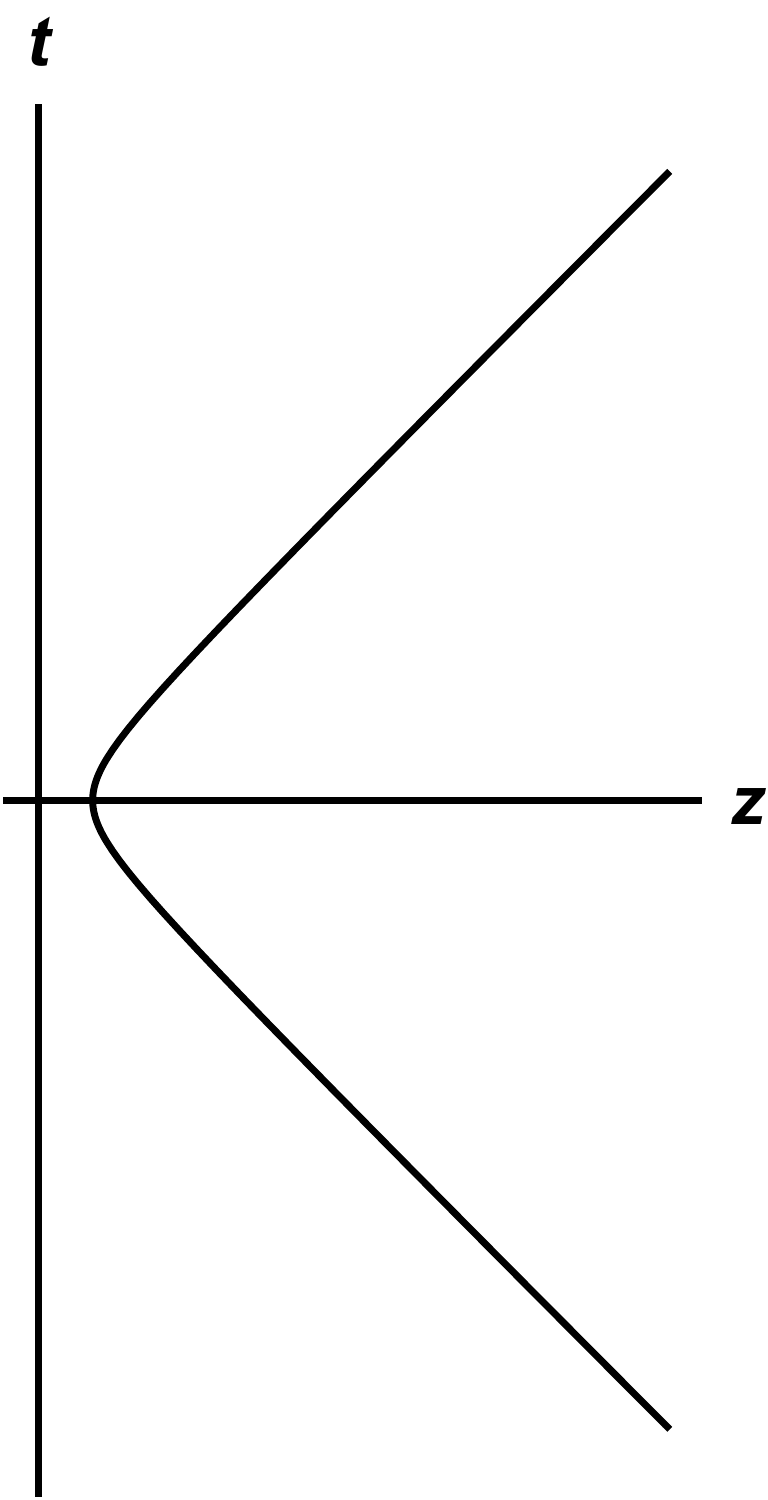}}} 
{\caption{ The hyperbola,  of the uniformly accelerated spatially rectilinear worldline.}} 
\end{center}
\end{figure}  
\subsection*{Ultrator}
\begin{equation}
    x^\mu (s) = \rho^{-2} \left(\tau \,\rho \,s, \kappa \cos\rho s, 0, \kappa \sin \rho s\right).
\end{equation} 
\begin{figure}[ht]
\begin{center}
\includegraphics[width=1.5in]{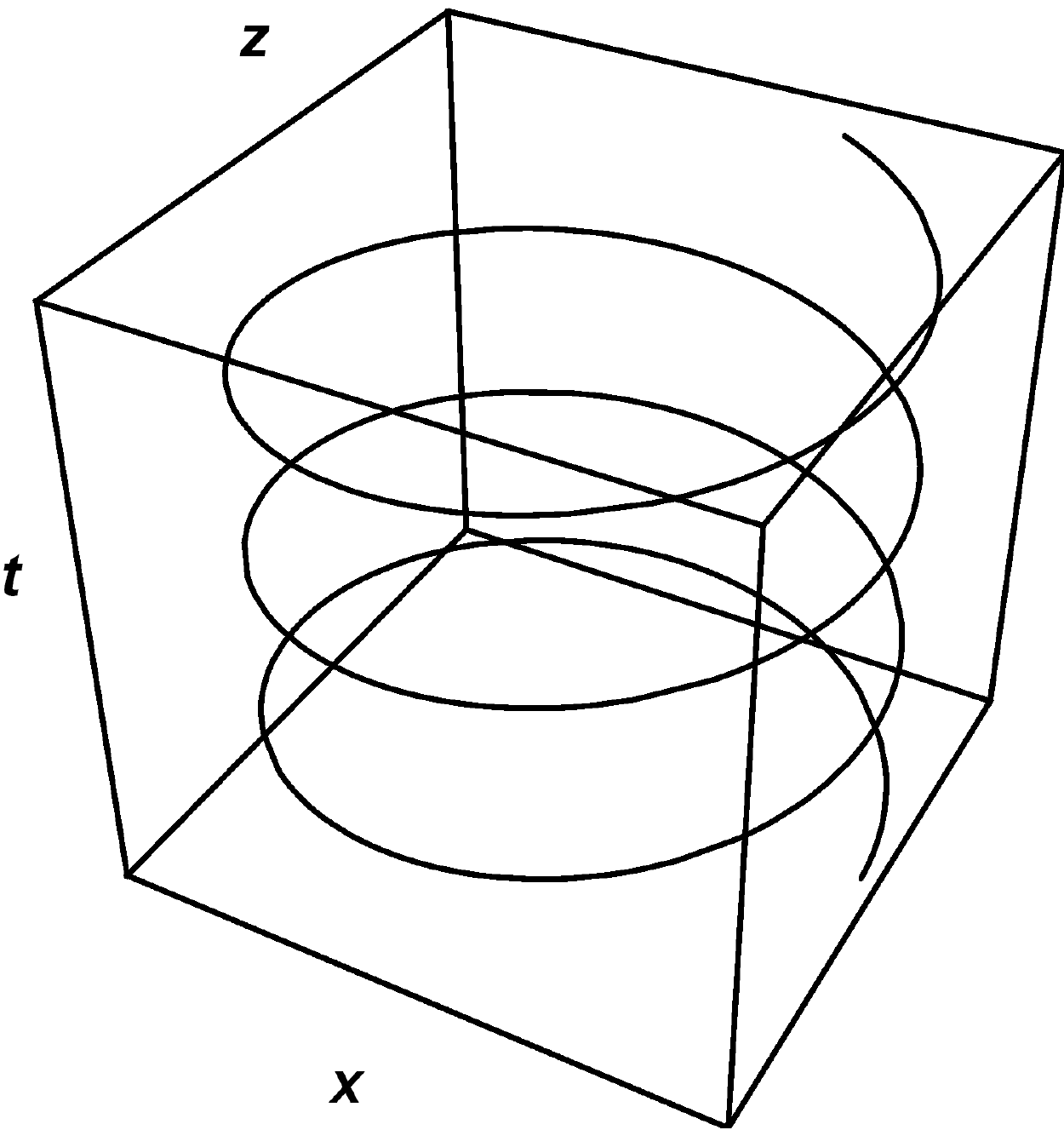}
%
{\caption{ The helix,  3D parametric plot of the uniformly accelerated spatially circular worldline.}} 
\end{center}
\end{figure}

\newpage

\subsection*{Parator}
\begin{equation}
    x^\mu (s) = \left(s + \frac{1}{6} \kappa^2 s^3, \frac{1}{2} \kappa s^2, 0, \frac{1}{6} \kappa^2 s^3\right).
\end{equation} 
\begin{figure}[ht]
\begin{center}
\includegraphics[width=2.0in]{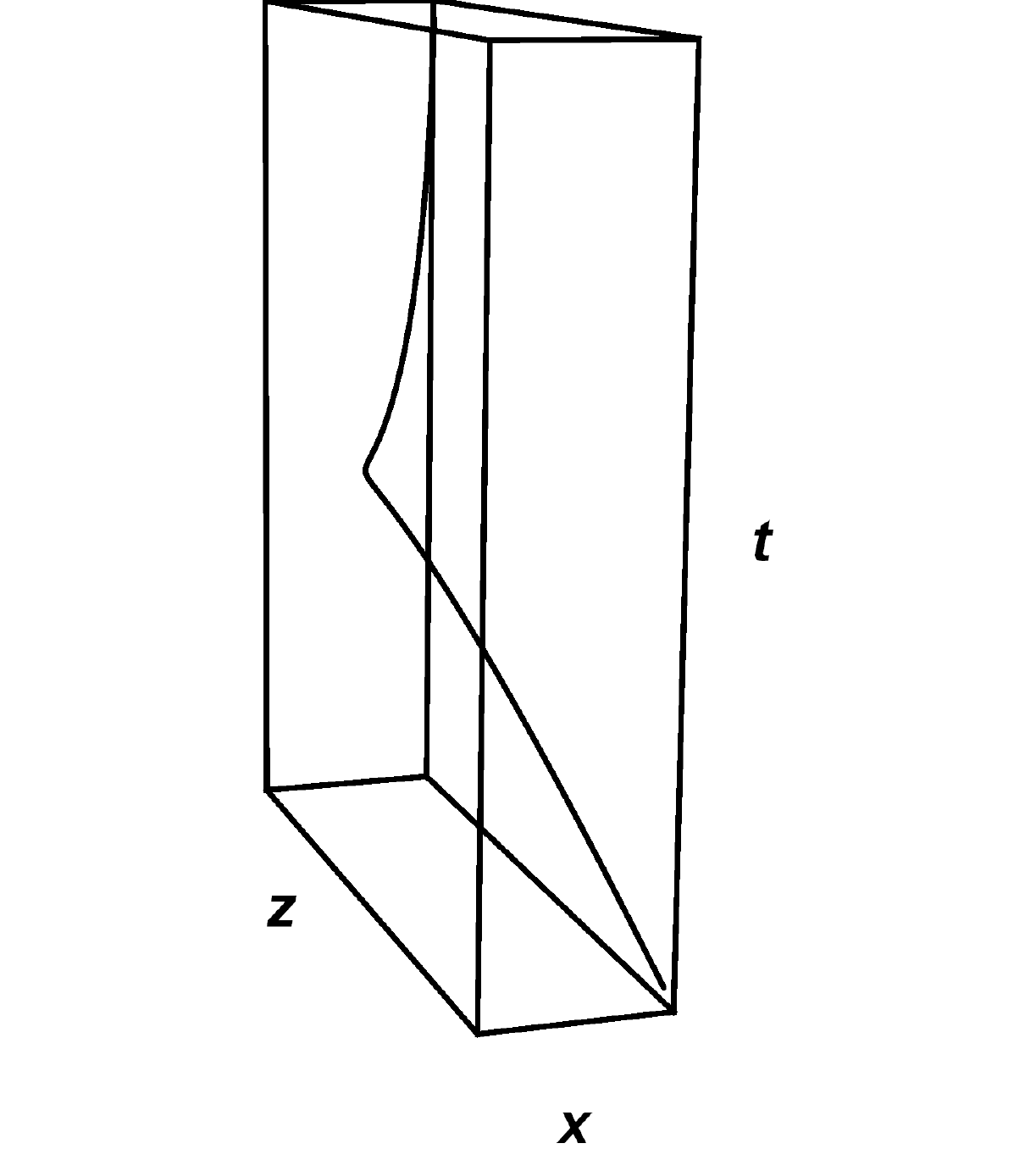}
{\caption{ The cusp world line which is spatially a semi-cubic parabola plotted in a 3D parametric plot.
}} 
\end{center}
\end{figure}

\subsection*{Infrator}
\begin{equation}
    x^\mu (s) = \sigma^{-2} \left(\kappa \sinh (s \sigma ), \kappa \cosh (s \sigma ), 0, s \tau \sigma\right).
\end{equation} 
\begin{figure}[ht]
\begin{center}
\includegraphics[width=2.0in]{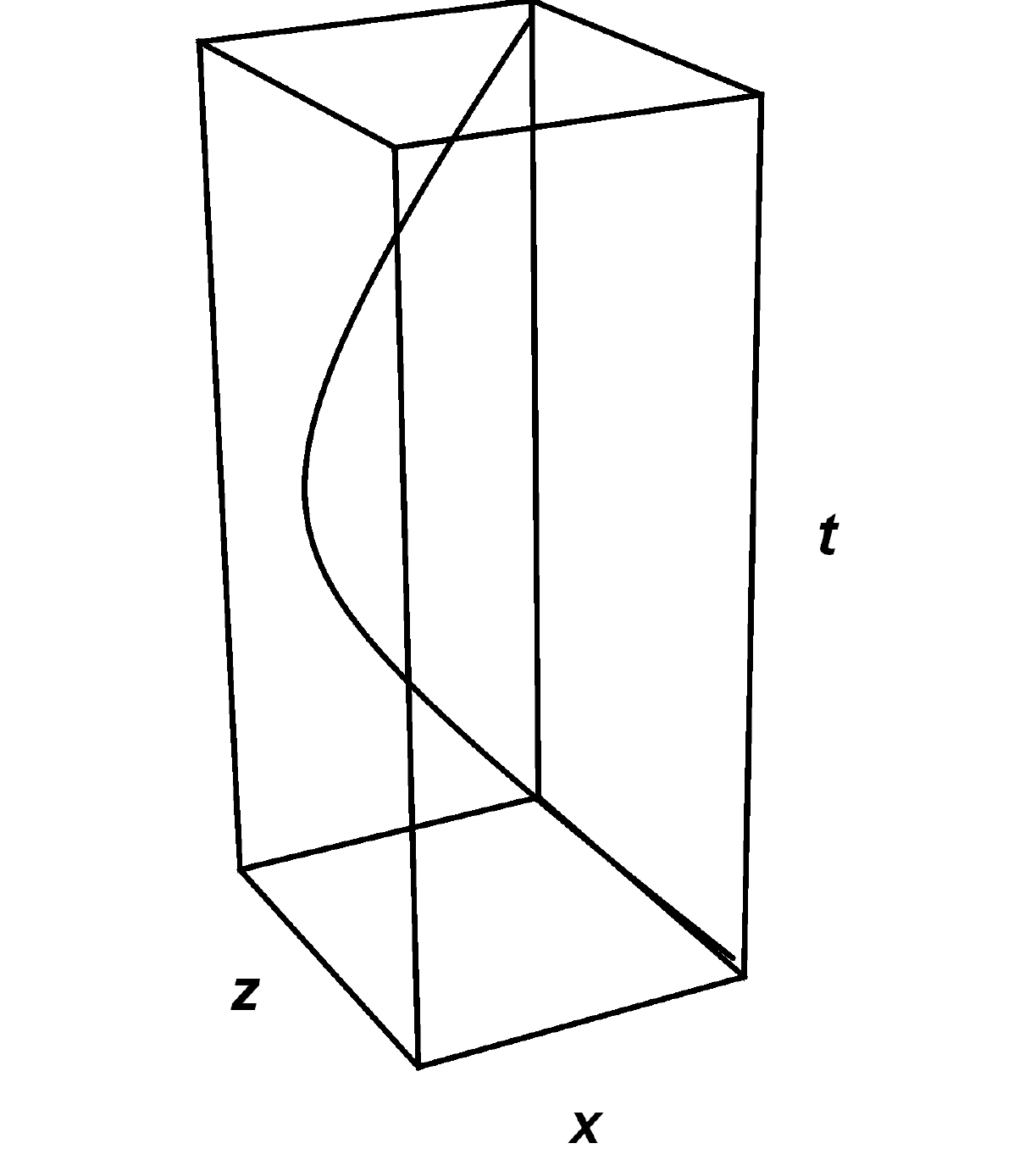}
{\caption{The infrator worldline (spatial catenary) plotted in a 3D parametric plot.} } 
\end{center}
\end{figure}  

\subsection*{Hypertor}
The Hypertor worldline is 3+1 dimensional, and so cannot be plotted on a 3D plot.  Its worldline function is written down explicitly in Letaw (1980): \cite{Letaw:1980yv}. 
\newpage

\chapter{Spatial Projections}\label{Appendix:Spatial}
\section*{Nulltor: Line}
For the zero torsional worldline whose spatial projection is just a line, its direction has been assigned to the x-axis in this work. 

\section*{Ultrator: Circle}
\begin{equation}
    z=\pm\sqrt{\frac{\kappa^2}{(\tau^2-\kappa^2)^2}-x^2}
\end{equation} 
\begin{figure}[ht]
\begin{center}
\includegraphics[width=2.5in]{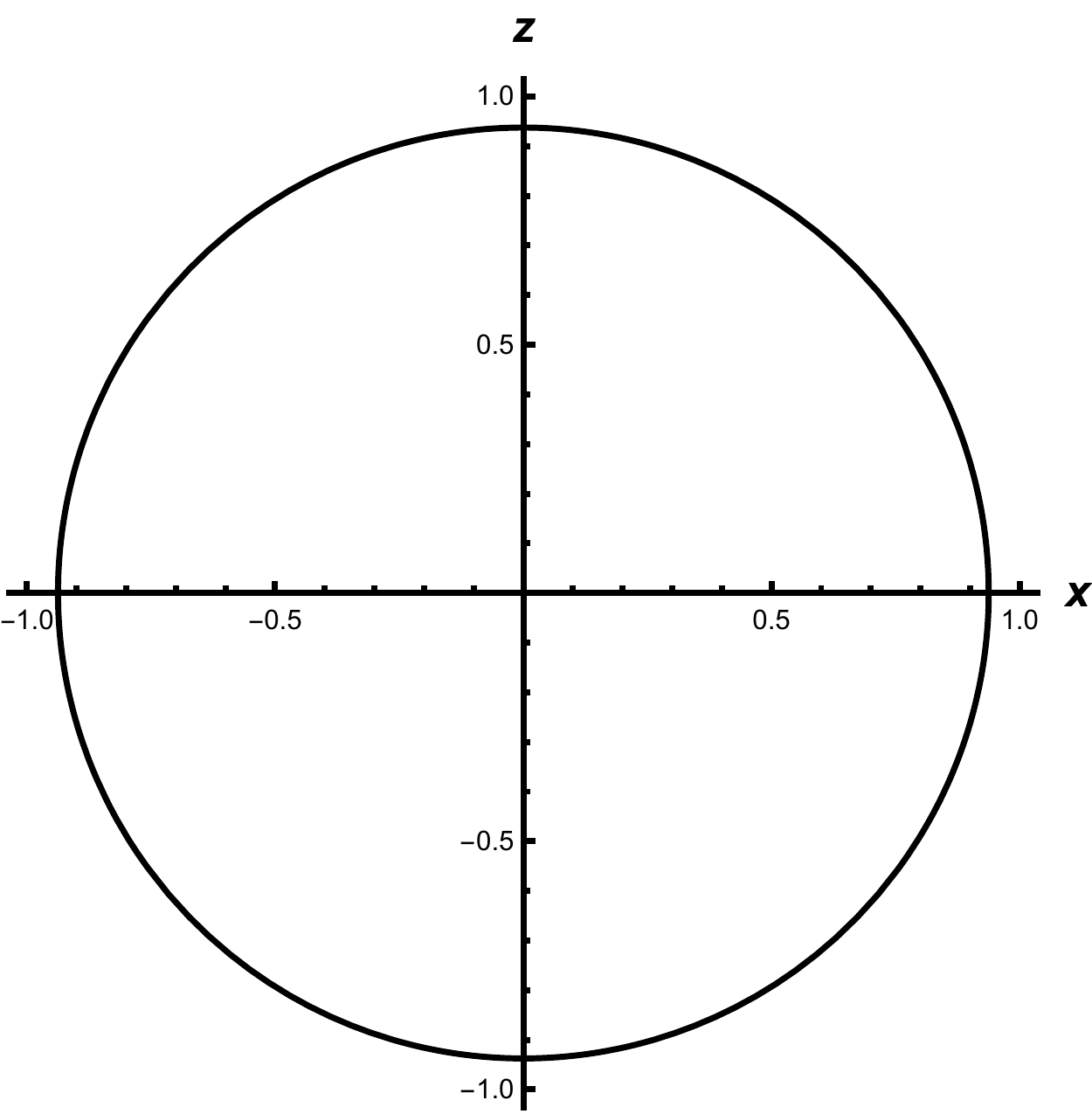}
{\caption{ The spatial Ultrator plot with invariants $\tau > \kappa$ for $\tau = 1$ and  $\kappa = 3/5$, with radius $R=\frac{\kappa}{\rho^2}=15/16 $.}} 
\end{center}
\end{figure}  
\newpage
\section*{Parator: Cusp}
\begin{equation}
z = \frac{\sqrt{2 \kappa}}{3} x^{3/2}
\end{equation}
\begin{figure}[ht]
\begin{center}
\includegraphics[width=2.5in]{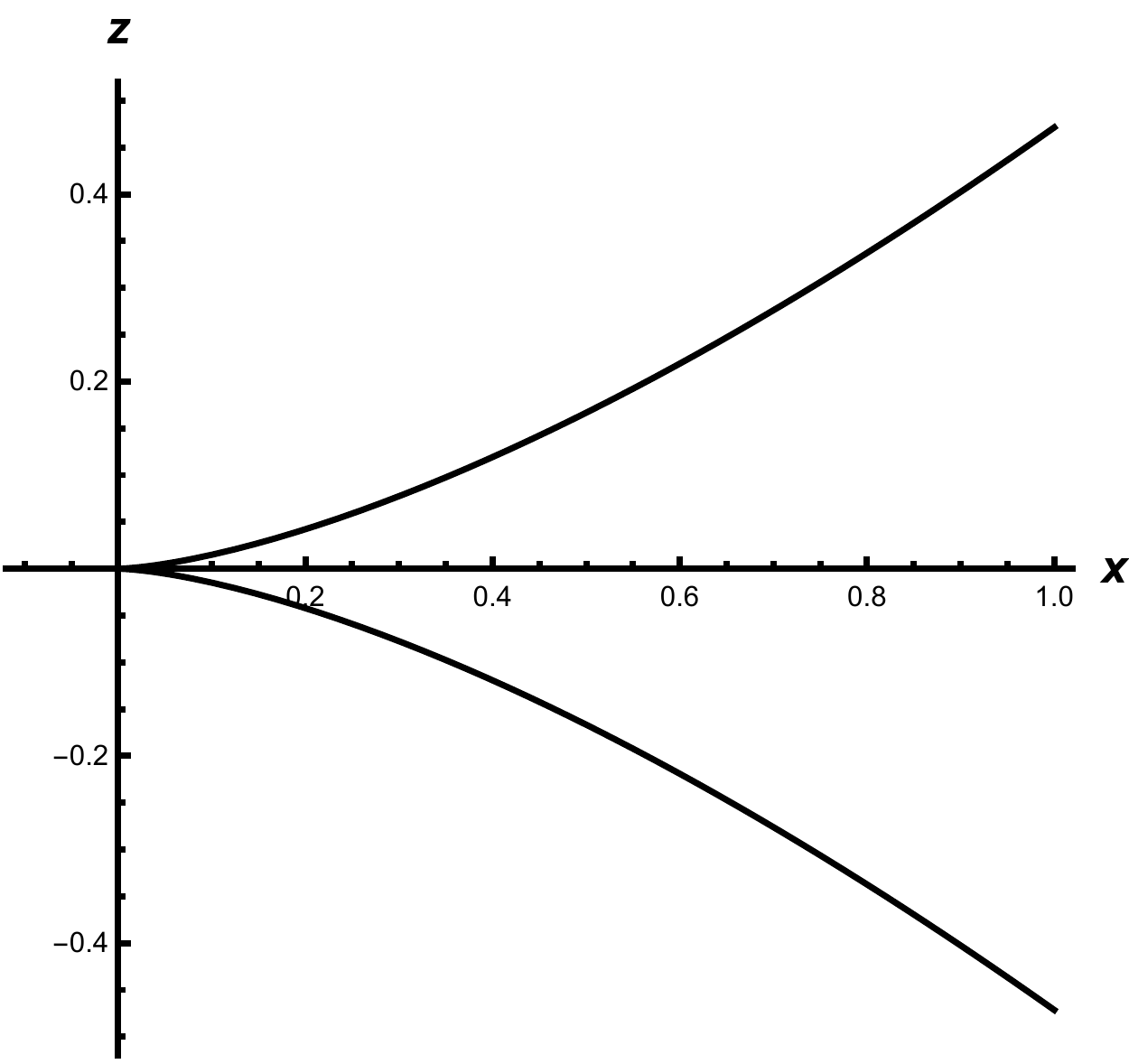}
{\caption{ The spatial Parator plot with invariants $\kappa =\tau= 1$.}} 
\end{center}
\end{figure}

\section*{Infrator: Catenary}
\begin{equation}
z= \frac{\tau}{\kappa^2-\tau^2}\cosh^{-1}\left[ \frac{x(\kappa^2-\tau^2)}{\kappa}\right]  \end{equation}
\begin{figure}[ht]
\begin{center}
\includegraphics[width=3.0in]{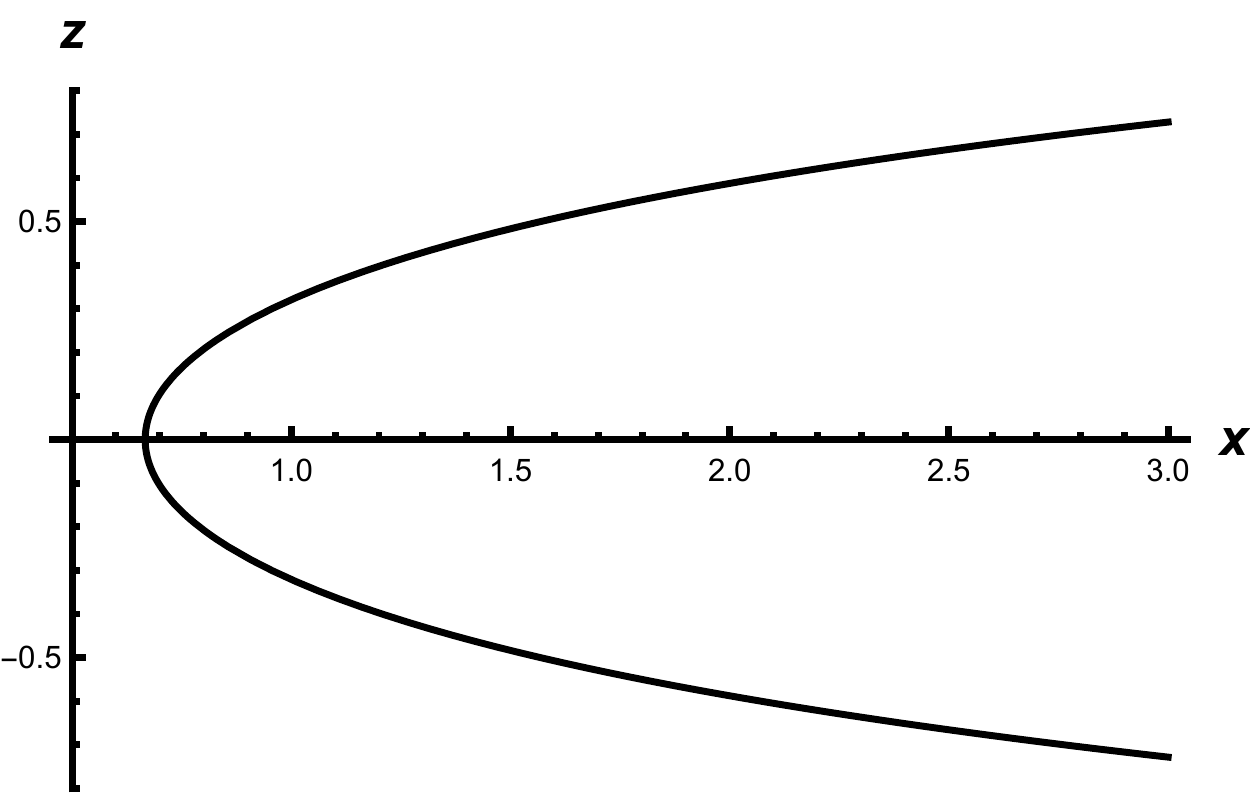}
{\caption{ Two dimensional catenary  plot with invariants $\tau < \kappa$ for $\tau = 1$,  $\kappa = 2$ of the Infrator motion.}} 
\end{center}
\end{figure}  

\newpage
\section*{Hypertor: Helix with Variable Pitch}
\begin{figure}[ht]
Parametric equations:
\begin{center}
\be x = \frac{\Delta}{RR_{+}} \cosh (R_{+}s ),\ee
\be y = \frac{\kappa \tau}{R\Delta R_{-}}\cos(R_{-}s),\ee
\be z = \frac{\kappa \tau}{R\Delta R_{-}}\sin(R_{-}s). \ee
where variable $\Delta$ and others are shown in Appendix \ref{Appendix:Acceleration}.

\includegraphics[width=3.0in]{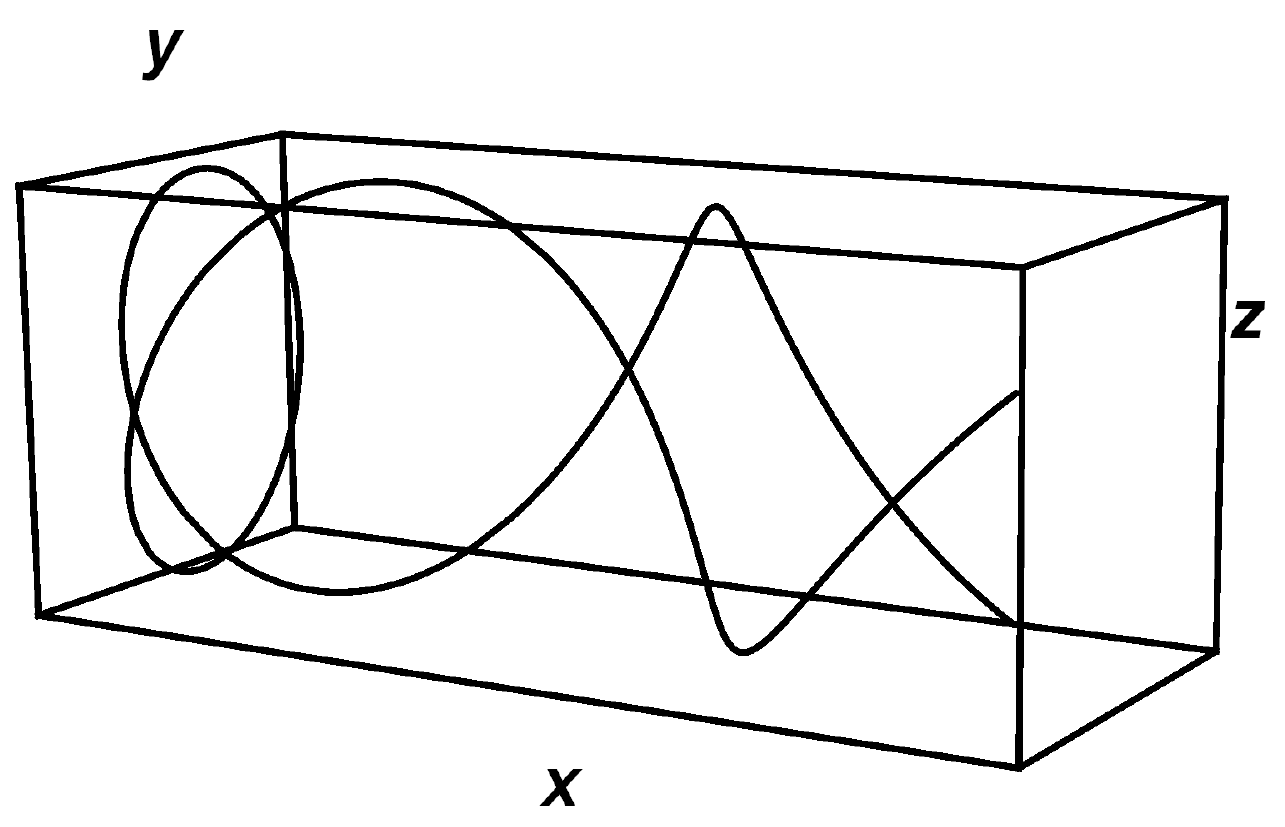}
{\caption{ Spatial Projection of the Hypertorsional worldline with $\tau = 10$, $\nu = \kappa = 1$.  Proper time ranges from $-1$ to $1$.}} 
\end{center}
\end{figure}  

\newpage

\chapter{Acceleration Ratios}\label{Appendix:Acceleration}
The acceleration ratio is defined simply as $A^2 \equiv \alpha^2/a^2$. 
\subsection*{Nulltor}
\begin{equation}
A^2_{\textrm{line}} = \gamma^6. 
\end{equation}
\subsection*{Ultrator}
\begin{equation} 
A^2_{\textrm{circle}} = \gamma^4, \quad v \equiv \kappa/\tau. 
\end{equation}
\subsection*{Parator}
\begin{equation} 
A^2_{\textrm{cusp}} = \frac{\gamma^6}{\gamma^2 - 2\gamma + 2}, \quad \kappa = \tau. 
\end{equation}
\subsection*{Infrator}
\begin{equation}
A^2_{\textrm{cat}} = \frac{\gamma^6}{1+ \gamma^2v_R^2}, \quad v_R \equiv \tau/\kappa.
\end{equation}

\subsection*{Hypertor}
\begin{equation}
A^2_{\textrm{hyper}} = \frac{\gamma ^4  \kappa ^3 \tau}{R^2 \gamma _{\textrm{min} }^4 v_{\textrm{min}} \left(R^2 v_{\textrm{min} }^2+\frac{R_+^2}{\gamma ^2}\right)}\,\,,  
\end{equation}
\begin{equation} \gamma_{{\textrm{min}}} \equiv (1-v_{\textrm{min}}^2)^{-1/2} , \qquad v_{\textrm{min}} \equiv \frac{\kappa \tau}{\Delta^2} \,\,,
\end{equation}
\begin{equation}  \Delta^2 \equiv \frac{1}{2}( R^2 +\kappa^2 + \tau^2 + \nu^2 ) \,\,,
\end{equation}
 \begin{equation} R^2 \equiv R_+^2 + R_-^2 \,\,,\end{equation}
 \begin{equation}
 R_{\pm}^2 \equiv {\sqrt{a^2+b^2} \pm a}\,\,,
 \end{equation}
 \begin{equation}
 a=\frac{1}{2} \left(\kappa ^2-\nu ^2-\tau ^2\right) \,\,,\qquad b= \kappa  \nu\,.
 \end{equation}


\let\cleardoublepage\clearpage

\chapter{Details to find Angular Distribution}\label{Appendix:Angular}

The necessary tools for determining the stationary worldline power distributions, 
\begin{equation}
    \frac{dP}{d\Omega} = \frac{q^2}{4\pi} \frac{ |\hrcurs \times (\bd{u} \times \bd{a})|^2}{(\hrcurs \cdot \bd{u})^5}, 
\end{equation} 
are shown in the following.  Here 
\begin{equation}
    \hrcurs \times(\bd{u} \times \bd{a}) = (\hrcurs \cdot \bd{a})\bd{u} - (\hrcurs \cdot \bd{u})\bd{a},
\end{equation} 
so that
\begin{equation}
    |\hrcurs \times (\bd{u} \times \bd{a})|^2 = (\hrcurs \cdot \bd{a})^2 u^2 - 2 (\bd{u}\cdot \bd{a})(\hrcurs \cdot \bd{a})(\hrcurs \cdot \bd{u}) + (\hrcurs \cdot \bd{u})^2 a^2.
\end{equation}  
Using the definitions, including $\brcurs \equiv \bd{r} - \bd{r}'$, (the vector from a source point $\bd{r}'$ to a field point $\bd{r}$), where the source point is our origin, $\bd{r}'=0$,
\be
    \hrcurs \equiv \frac{\bd{r} - \bd{r}'}{|\bd{r} - \bd{r}'|} = \bd{\hat{r}} \equiv \sin\theta\cos\phi \bd{\hat{x}} + \sin\theta\sin\phi \bd{\hat{y}} + \cos\theta \bd{\hat{z}},
\ee
\begin{equation}
    \bd{u} \equiv \hrcurs - \bd{v},
\end{equation} 
one finds the terms in the order that they appear,
\begin{equation}
    \hrcurs \cdot \bd{a} = a_x \sin\theta\cos\phi + a_y\sin\theta\sin\phi + a_z \cos\theta,
\end{equation} 
\begin{equation}
    u^2 = (\hrcurs - \bd{v})^2 = 1 - 2( v_x \sin\theta\cos\phi + v_y\sin\theta\sin\phi + v_z \cos\theta) + v^2,
\end{equation} 

\bea
    \bd{u}\cdot \bd{a} 
    &=& (\hrcurs - \bd{v})\cdot \bd{a},\\ 
    &=& a_x \sin\theta \cos\phi + a_y\sin\theta\sin\phi + a_z \cos\theta\\ 
    &&-v_x a_x -v_y a_y - v_z a_z,
\eea 

\begin{equation}
    \hrcurs \cdot \bd{u} = \hrcurs^2 - \hrcurs\cdot \bd{v} = 1- v_x \sin\theta\cos\phi -v_y\sin\theta\sin\phi - v_z \cos\theta,
\end{equation} 
\begin{equation}
    a^2 = a_x^2 + a_y^2 + a_z^2, \quad v^2 = v_x^2 + v_y^2 + v_z^2.
\end{equation} 

\let\cleardoublepage\clearpage

\chapter{Derivatives of the Stationary Worldlines}
\label{Appendix:Derivatives}
\subsection*{Nulltor}
The rectilinear values for the component velocities and accelerations are,
\be a_x = 0, \quad a_y = 0, \quad a_z = -\frac{\kappa }{\gamma ^3},\ee
\be v_x = 0, \quad v_y = 0, \quad v_z = \beta.\ee
\subsection*{Ultrator}

The values for the component velocities and accelerations are (at proper time moment $s = 0$, axis chosen such that $v_z = v \hat{z}$ and $a_x=a \hat{x}$), using $\beta = \kappa/\tau$,
\be a_x = -\frac{\kappa }{\gamma ^2}, \quad a_y = 0, \quad a_z = 0,\ee
\be v_x = 0, \quad v_y = 0, \quad v_z = \beta.\ee

\subsection*{Parator}
The values for the component velocities and accelerations are,
\be a_x = -\frac{(\gamma -2) \kappa }{\gamma ^3}, \quad a_y = 0, \quad a_z = \frac{\sqrt{2\gamma -2} \kappa }{\gamma ^3},\ee
\be v_x = \gamma^{-1}\sqrt{2(\gamma-1)}, \quad v_y = 0, \quad v_z = 1-\gamma^{-1}.\ee

\newpage
\subsection*{Infrator}
The values for the component velocities and accelerations are,
\be a_x =\frac{\sigma ^2 }{\kappa}\text{sech}^3(s \sigma ) , \quad a_y = 0, \quad a_z = - \frac{\sigma ^2 \tau  }{\kappa^2}\tanh (s \sigma ) \text{sech}^2(s \sigma ),\ee
\be v_x = \tanh (s \sigma ), \quad v_y =0 , \quad v_z = v_R \text{sech}(s \sigma ),\ee
\\
where $s=\frac{1}{2 \sigma }\textrm{ln} \left(2 \frac{\gamma ^2}{\gamma_R ^2} -2\frac{\gamma }{\gamma_R } \sqrt{\frac{\gamma ^2}{\gamma_R ^2}-1}-1\right)$.

\subsection*{Hypertor}
The hypertor values with proper time 
 for the component velocities and accelerations are,
\begin{multline}
\qquad a_x = \frac{R R_{+} }{\Delta }\text{sech}^3(R_{+} s)\,\,, \qquad \\ a_y = \frac{R  v_{\textrm{min}  } }{\Delta } \text{sech}^2(R_{+} s) [ R_{+} \sin (R_{-} s) \tanh (R_{+} s)-R_{-}\cos (R_{-} s)]\,\,,\\  a_z =- \frac{R v_{\textrm{min} }  }{\Delta }\text{sech}^2(R_{+} s)[R_{+} \cos (R_{-} s) \tanh (R_{+} s)+R_{-} \sin (R_{-} s)]\,\,,
 \end{multline}
 
\begin{multline}
    v_x = \tanh \left(R_+ s\right), \quad v_y = v_{\textrm{min}} \sin \left(R_- s\right) \text{sech} \left(R_+ s\right), \quad \\ v_z = v_{\textrm{min}} \cos \left(R_- s\right) \text{sech} \left(R_+ s\right),
\end{multline} 

where  $s=-\frac{1}{R_+}\cosh ^{-1}\left(\frac{\gamma}{\gamma_{\textrm{min}}}\right)$ .

\let\cleardoublepage\clearpage

\chapter{Compilation of Power Distributions}\label{Appendix:Power}
\subsection*{Nulltor}
\be \label{LineHS} I_{\textrm{null}} = \frac{3}{8\pi \gamma^6} \frac{\textrm{sin}^2\theta}{(1-\beta\textrm{cos}\theta)^5}. \ee
\subsection*{Ultrator}
\be I_{\textrm{ultra}} = \frac{3}{8\pi} \frac{(1-\beta\textrm{cos}\theta)^2-(1-\beta^2)\textrm{sin}^2\theta\textrm{cos}^2\phi}{\gamma^4(1-\beta\textrm{cos}\theta)^5}. \ee
\subsection*{Parator}
\be\label{even_func} I_{\textrm{para}} \equiv 
\frac{\lambda_1
+\lambda_2 \textrm{cos}{\phi} 
+\lambda_3\textrm{cos}^2\phi }{ 
\left( \textrm{cos}\phi + \lambda_4 \right){}^5},\ee
%
%
\begin{subequations}
\begin{eqnarray}
\lambda_0&=&-\frac{3}{8\pi \gamma\sqrt{2\gamma _1}^5 \textrm{sin}^5\theta},\\
\lambda_1&=&\lambda_0
\left[2 + \gamma  \gamma _2 + \gamma _1 \textrm{cos}\theta 
\left(\gamma_3 \textrm{cos}\theta-2 \gamma_2\right)\right],\\
\lambda_2&=& -2\lambda_0 \sqrt{2\gamma _1}\textrm{sin}\theta \left(\gamma _1-\gamma _2 \textrm{cos}\theta\right),\\
\lambda_3&=& 2\lambda_0 \gamma_{3/2}\,\textrm{sin}^2\theta,\\
\lambda_4&=& \frac{\gamma _1 \textrm{cos}\theta - \gamma}{\sqrt{2\gamma _1}\textrm{sin}\theta},
\end{eqnarray}
\end{subequations}
\be \gamma_n \equiv \gamma-n \ee
\be n=1,2,3, 3/2 \ee
\newpage
\subsection*{Infrator}
\begin{eqnarray}
I_{\textrm{infra}}=\frac{3}{8 \pi R(\theta,\phi)^5} \left[C_2 \, F(\theta,\phi)^2+2\, C_1 F(\theta,\phi)G(\theta,\phi)-\frac{1}{\gamma ^2} G(\theta,\phi)^2\right]\,\,, 
\end{eqnarray}

\begin{subequations}
\begin{eqnarray}
 G(\theta,\phi)&=&\frac{\sin (\theta ) \, F(\theta,\phi)}{R(\theta,\phi)}\,\left( \cos (\phi )-v_R \, \sin (\phi )\, \sinh (\alpha)\right)\,\,,\\
F(\theta,\phi)&=&(1-v_R^2)\,\text{sech}^3(\alpha)\, R(\theta,\phi)\,\,,\\
R(\theta,\phi)&=&1-v_R \, \sin (\theta ) \, \sin (\phi )\, \text{sech}(\alpha)+\,\sin (\theta )\, \cos (\phi ) \, \tanh (\alpha ) \,\,, 
\end{eqnarray}
\end{subequations}

\begin{subequations}
\begin{eqnarray}
C_1&=&(1-v_R^2)\tanh (\alpha)\,\,, \\
C_2&=& v_R^2 \sinh ^2(\alpha)+1 \,\,,
\end{eqnarray}
\end{subequations}
\begin{subequations}
\begin{eqnarray}
 \alpha&=& \frac{1}{2}\, \textrm{ln}(2\,\xi \,\eta-1)\,\,, \\
 \eta &=&\sqrt{\xi^2-1}+\xi\,\,, \\
 \xi &=&\gamma \sqrt{1-v_R^2} \,\,.
\end{eqnarray}
\end{subequations}

\subsection*{Hypertor}

\begin{eqnarray}
I_\mathrm{hyper}=\frac{3}{8\pi \kappa^2 Q^5(\theta,\phi)} \left[C_1 Q^2(\theta,\phi)+2\, C_2 Q(\theta,\phi)P(\theta,\phi) -\frac{1}{\gamma^2}P^2(\theta,\phi)\right]\,\,,
\end{eqnarray}

\begin{subequations}
\begin{eqnarray}
Q(\theta,\phi)&=&1 - A_1\,\cos(\theta) - A_2\, \sin(\theta)\cos(\phi) +  A_3\,\sin(\theta) \sin(\phi)\,\,,\\
P(\theta,\phi)&=& F_2\,\cos(\theta) + B_2\, \sin(\theta)\cos(\phi) + F_1\,\sin(\theta) \sin(\phi)\,\,,
\end{eqnarray}
\end{subequations}

\begin{subequations}
\begin{eqnarray}
A_1&=& v_{\textrm{min}} b\cos(\Omega_\gamma)\,\,,\\
A_2&=&\sqrt{1-b^2}\,\,,\\
A_3&=& v_{\textrm{min}}b\sin(\Omega_\gamma)\,\,,
\end{eqnarray}
\end{subequations}

\begin{subequations}
\begin{eqnarray}
B_1&=& b(A_2A_3D_1-A_1D_2)\,\,,\\
B_2 &=& D_1 b^3 \,\,,\\
B_3 &=& b(A_3D_2+A_2A_1D_1) \,\,,
\end{eqnarray}
\end{subequations}

\begin{subequations}
\begin{eqnarray}
C_1&=& B_2^2+B_1^2+B_3^2\,\,,\\
C_2&=& -A_2 B_2+A_3B_1+A_1B_3\,\,,
\end{eqnarray}
\end{subequations}

\begin{subequations}
\begin{eqnarray}
D_1&=& \frac{R\,R_{+}}{\Delta}\,\,, \\
D_2 &=& \frac{R\,R_{-}}{\Delta}\,\,,
\end{eqnarray}
\end{subequations}

\begin{subequations}
\begin{eqnarray}
\Omega_\gamma&=& \frac{R_{-}}{R_{+}}\cosh^{-1}(\frac{1}{b}) \,\,,\,\,\,\,\,\,\, 
b=\frac{\gamma_{\textrm{min}}}{\gamma}\,\,,
\end{eqnarray}
\end{subequations}

\begin{subequations}
\begin{eqnarray}
v_{\textrm{min}}=\frac{\kappa\,\tau}{\Delta^2} \,\,,\,\,\,\, \gamma_{\textrm{min}}= \frac{1}{\sqrt{1-v_\textrm{min}^2}} \,\,,
\end{eqnarray}
\end{subequations}

\begin{subequations}
\begin{eqnarray}
R=\sqrt{R_{+}^2 + R_{-}^2}\,\,,
\end{eqnarray}
\end{subequations}

\begin{subequations}
\begin{eqnarray}
R_{+}^2&=& \frac{1}{2}(\kappa^2-\nu^2-\tau^2) +\sqrt{\kappa^2\,\tau^2+\frac{1}{4}(\kappa^2-\nu^2-\tau^2)^2}  \,\,,\\
R_{-}^2&=& -\frac{1}{2}(\kappa^2-\nu^2-\tau^2) +\sqrt{\kappa^2\,\tau^2+\frac{1}{4}(\kappa^2-\nu^2-\tau^2)^2} \,\,.
\end{eqnarray}
\end{subequations}

\let\cleardoublepage\clearpage

\chapter{Thomas Precession}\label{Appendix:Thomas}
Using the crossproduct definition of proper acceleration, $\alpha^2 = \gamma^6 a^2 - \gamma^6(\bd{v}\times\bd{a})^2$, the definition of Thomas precession, $(\gamma+1)^2\omega_T^2 = \gamma^4(\bd{v}\times\bd{a})^2$, and the fact that stationary worldlines have constant acceleration, $\alpha = \kappa$, one finds:
\be \gamma^2(\gamma+1)^2 \omega_T^2  = \gamma^6 a^2 - \kappa^2. \ee
Applying this to the worldlines gives the Thomas precession and the lowest order term in $\beta$: 

\subsection*{Nulltor}
\be \omega_T = 0. \ee

\subsection*{Ultrator}
\be \omega_T = \frac{\kappa \beta}{\gamma+1} \approx \frac{\kappa \beta}{2}. \ee

\subsection*{Parator}
\be \omega_T =  \frac{\kappa}{\gamma}\left(\frac{\gamma-1}{\gamma+1}\right)\approx \frac{\kappa \beta^2}{4}.\ee

\subsection*{Infrator}
\be \omega_T = \frac{\tau}{\gamma+1} \approx \frac{\tau}{2} = \frac{\kappa \beta_R}{2}. \ee

\subsection*{Hypertor}
\be \omega_T=\frac{\sqrt{\gamma^6 a_H^2-\kappa^2}}{\gamma(\gamma+1)}. \ee
The quantity $a_H$ is the magnitude of the 3-acceleration of Hypertor \footnote{The variable  $\gamma_{min}$ and others are described in Appendix \ref{Appendix:Power}.}:
\begin{equation*}
a_H \equiv \frac{\gamma_{min}^2}{\gamma^2} \frac{R}{\Delta}\sqrt{\frac{R_{+}^2}{\gamma^2 }+R^2 v_{min}^2}.
\end{equation*}

\chapter{Polar Distribution Plots}\label{Appendix:Polar}
\section*{Nulltor}
\begin{figure}[ht]
\begin{center}
{\rotatebox{0}{\includegraphics[width=1.2in]{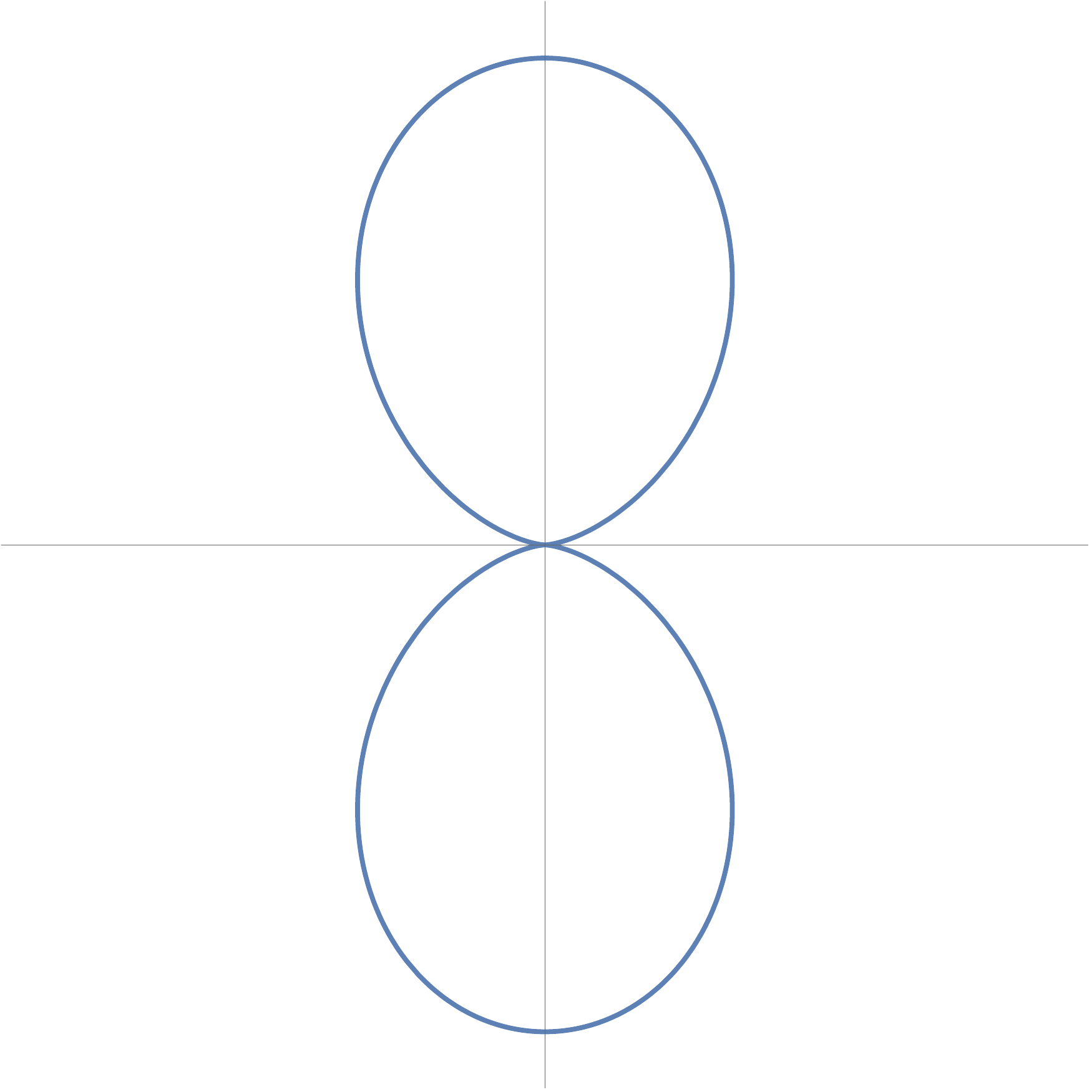}}\quad  \rotatebox{0}{\includegraphics[width=1.2in]{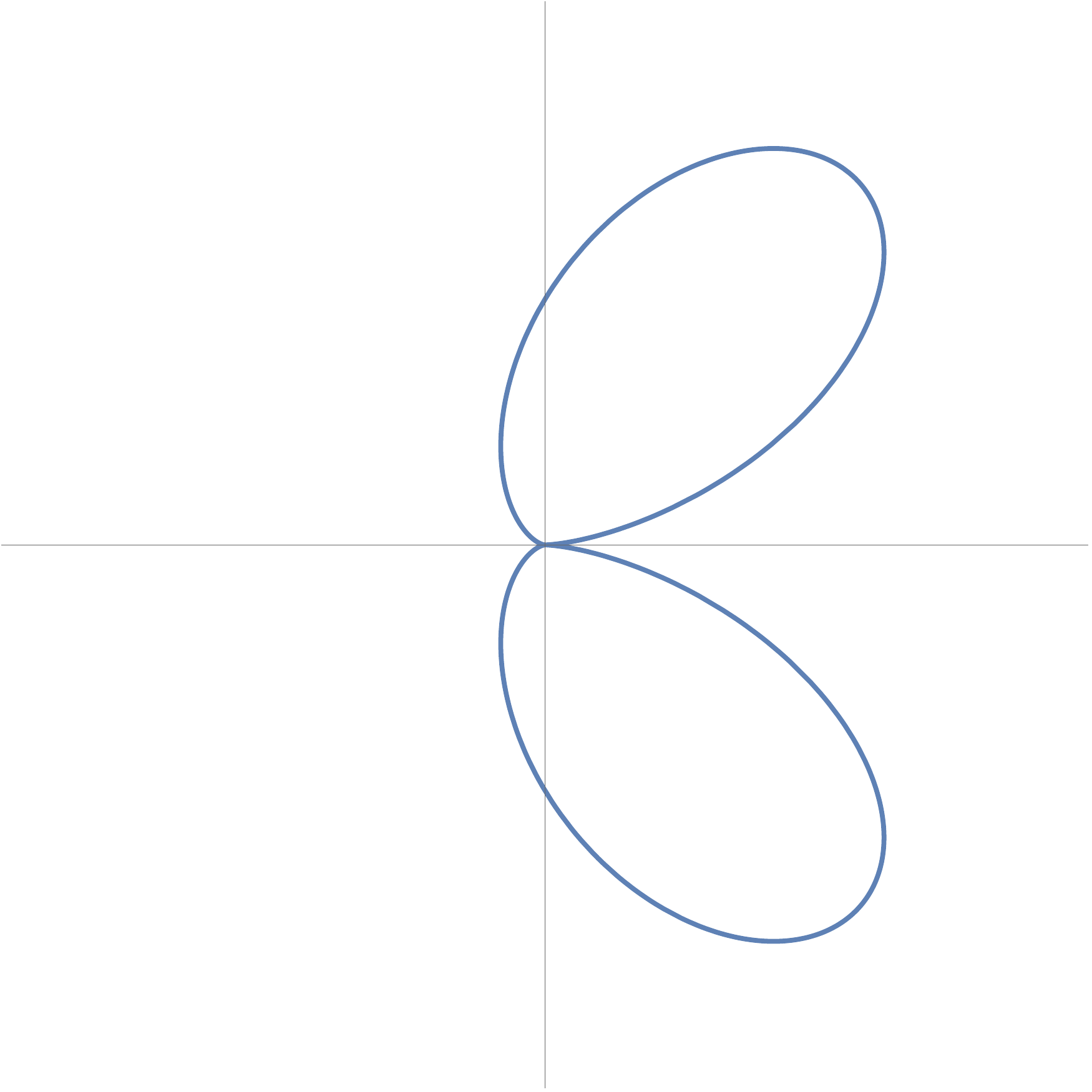}}\quad  \rotatebox{0}{\includegraphics[width=1.2in]{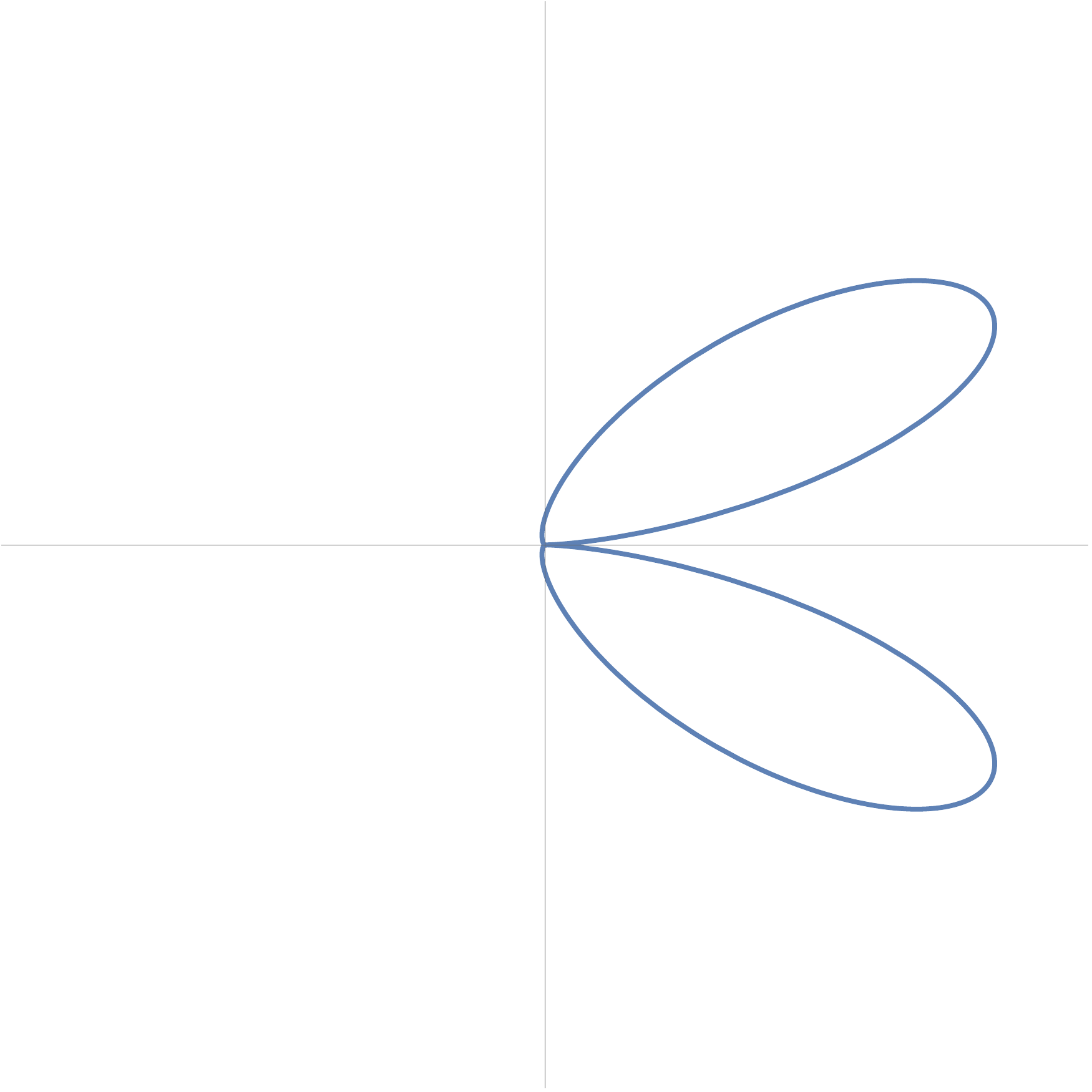}}\quad  \rotatebox{0}{\includegraphics[width=1.2in]{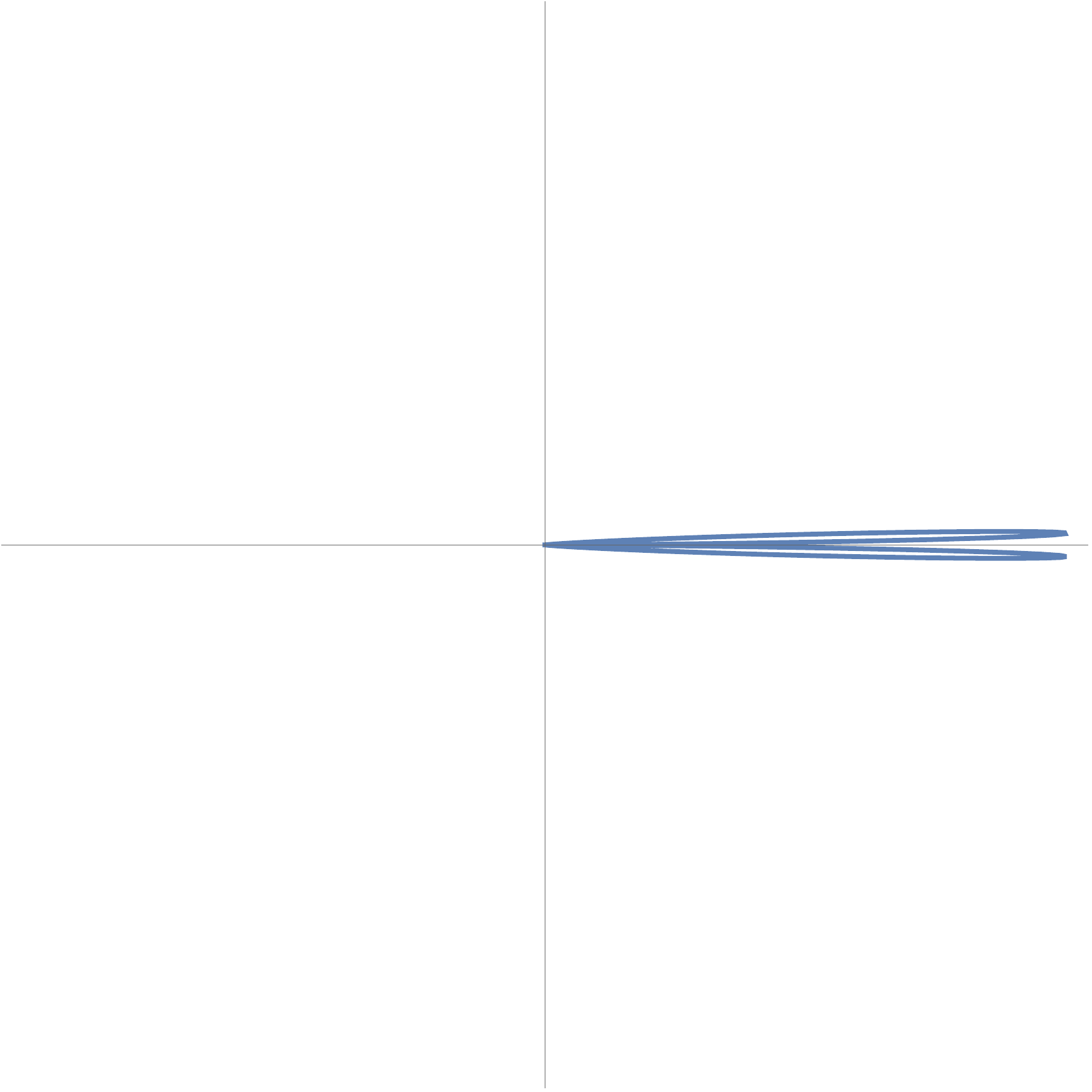}}} 
\caption{Rectilinear Angular Distribution; A polar plot with  $\beta=0.0, 0.33,$ $0.66, 0.99$. The vertical axes is $x$ and the horizontal axis is $z$. The electron moves forward along the $z$ straight line direction.} 
\end{center}
\end{figure} 

\section*{Ultrator}
\begin{figure}[ht]
\begin{center}
{\rotatebox{0}{\includegraphics[width=1.2in]{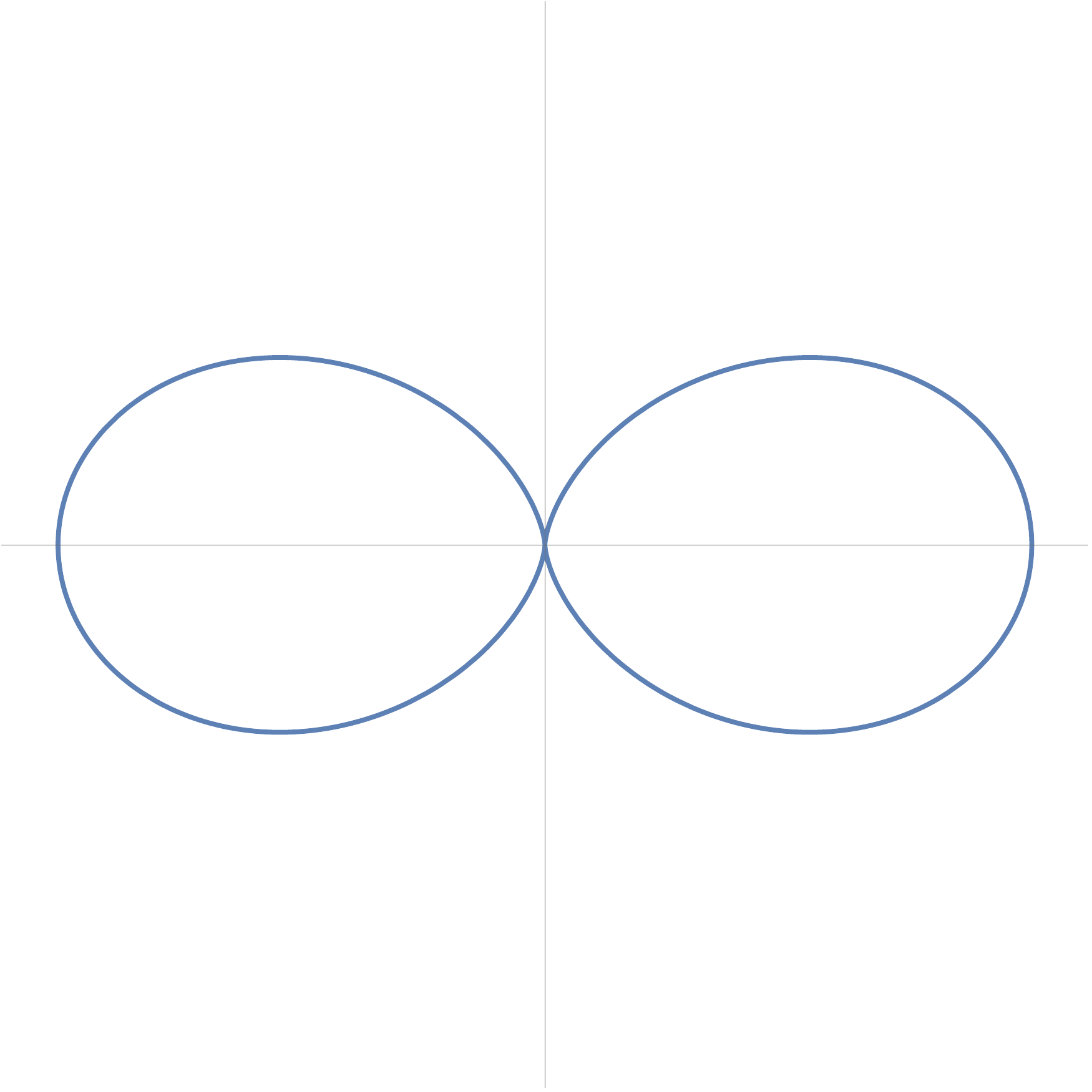}}\quad  \rotatebox{0}{\includegraphics[width=1.2in]{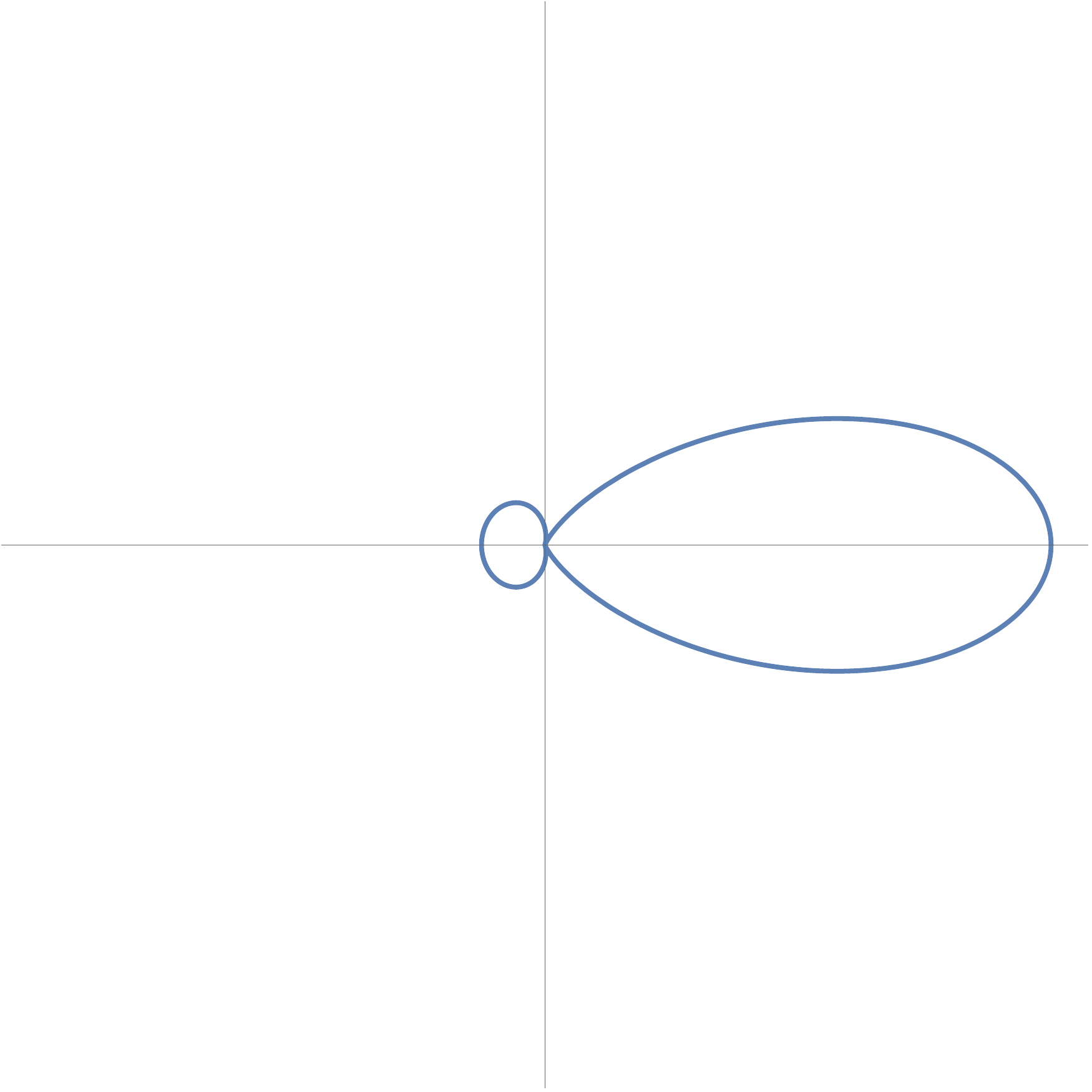}}\quad  \rotatebox{0}{\includegraphics[width=1.2in]{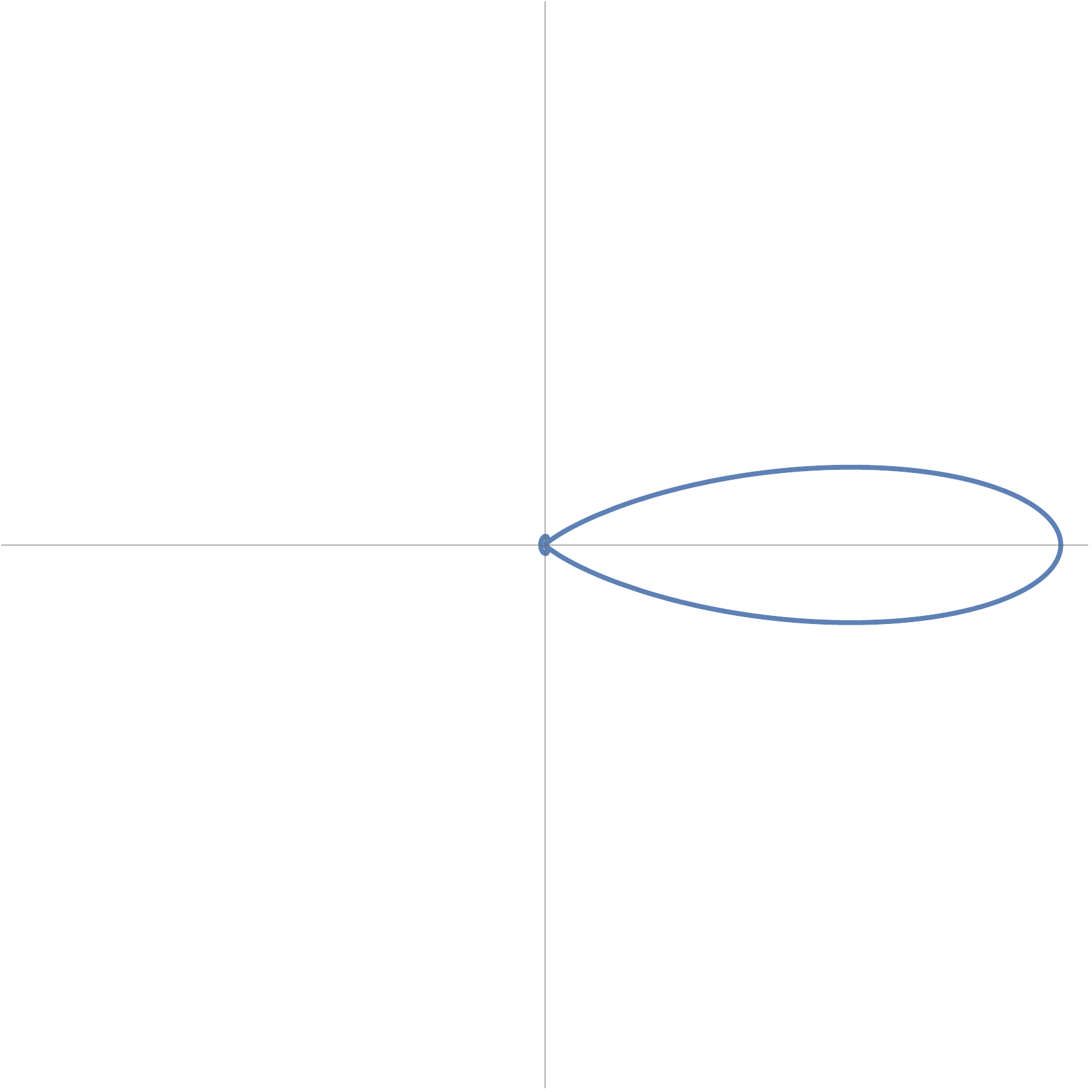}}\quad  \rotatebox{0}{\includegraphics[width=1.2in]{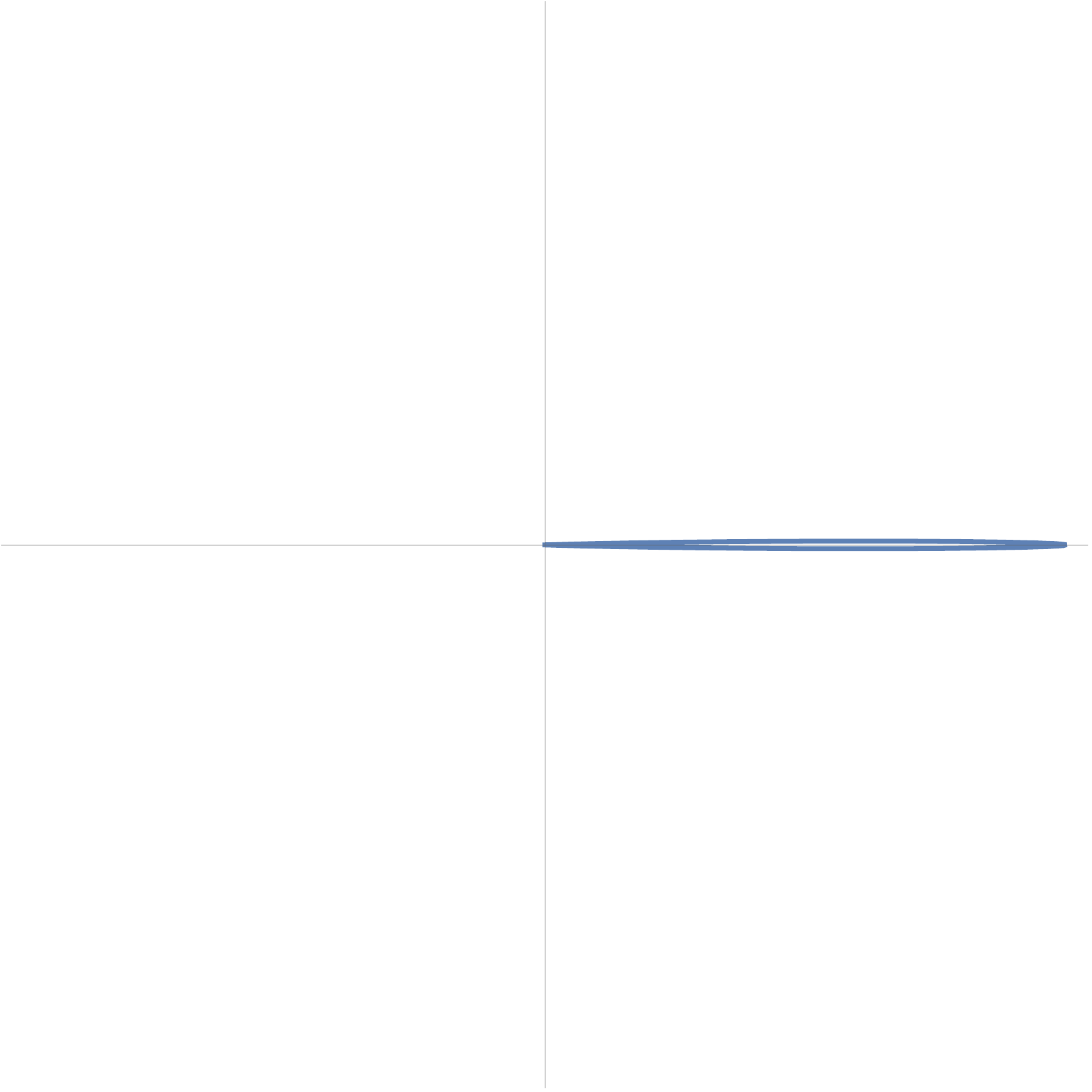}}} 

\caption{ Synchrotron Angular Distribution; A polar plot with 
$\beta=0,0.33,$ $0.66,0.99$. The vertical axes is $x$ and the horizontal axis is $z$. The electron moves forward in a circle toward the horizontal $z$ axis. Notice each graph is a different worldline as the shape remains the same along a constant $v$ stationary worldline.  } 

\end{center}
\end{figure} 

\newpage
\section*{Parator}
\begin{figure}[ht]
\begin{center}
{\rotatebox{0}{\includegraphics[width=1.2in]{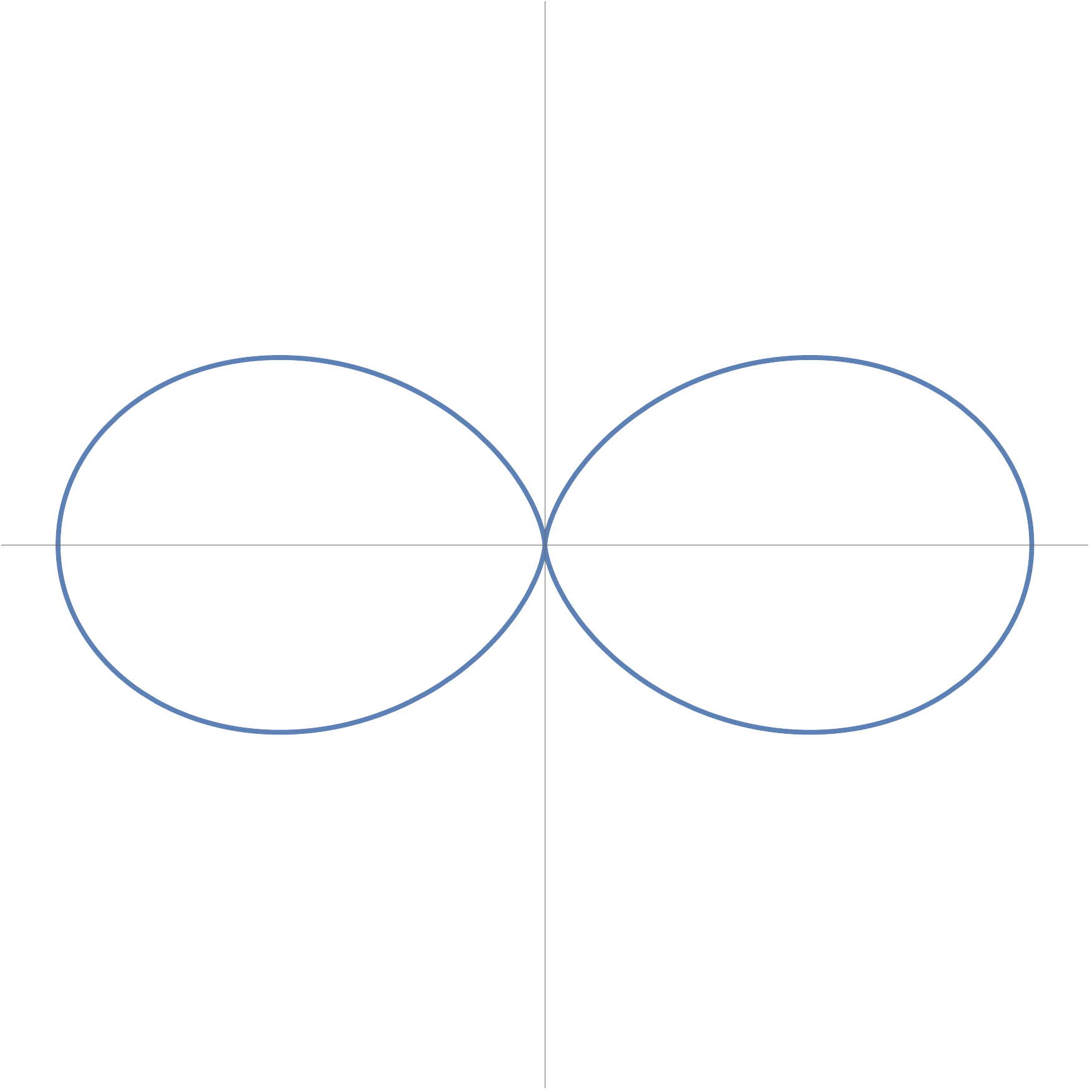}}\quad  \rotatebox{0}{\includegraphics[width=1.2in]{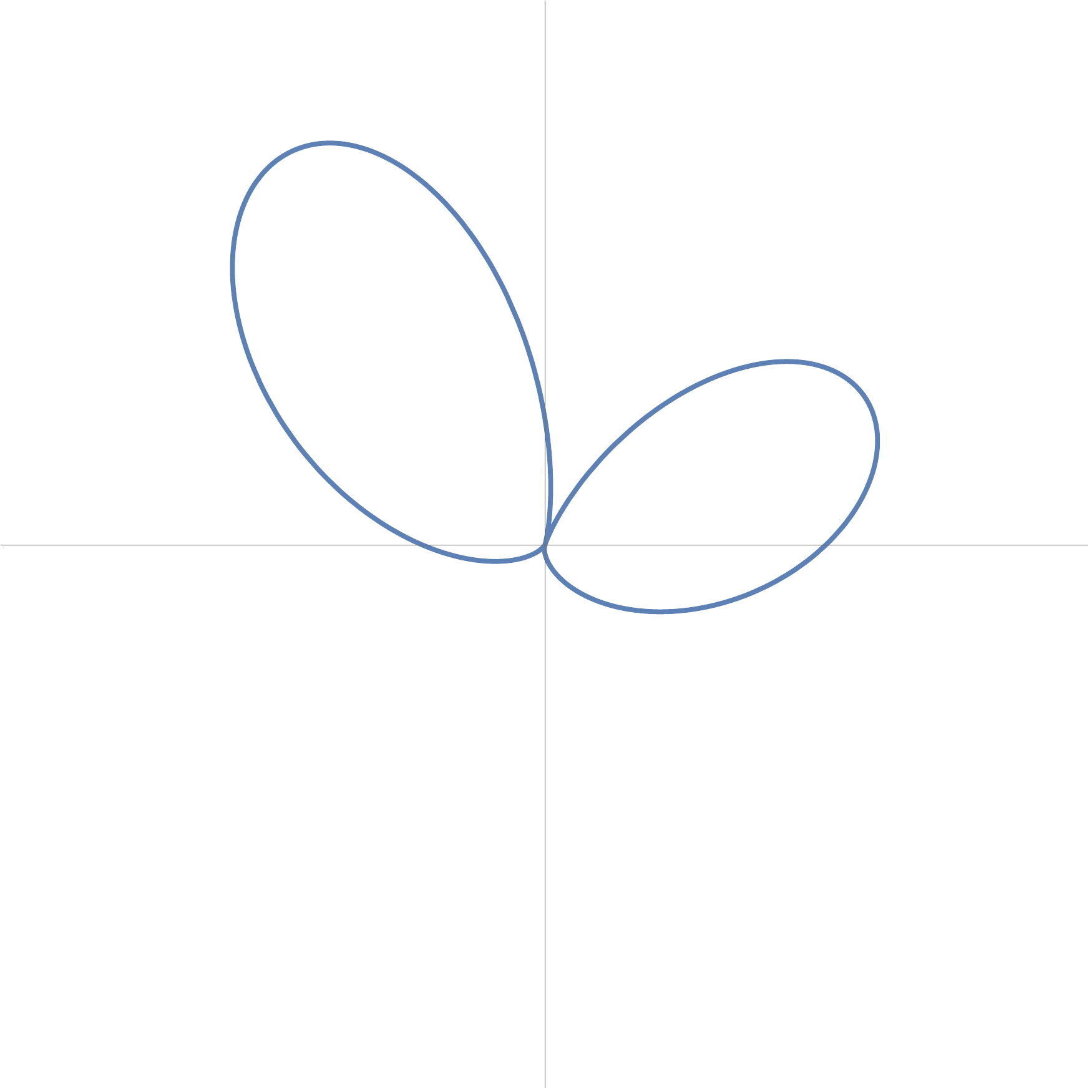}}\quad  \rotatebox{0}{\includegraphics[width=1.2in]{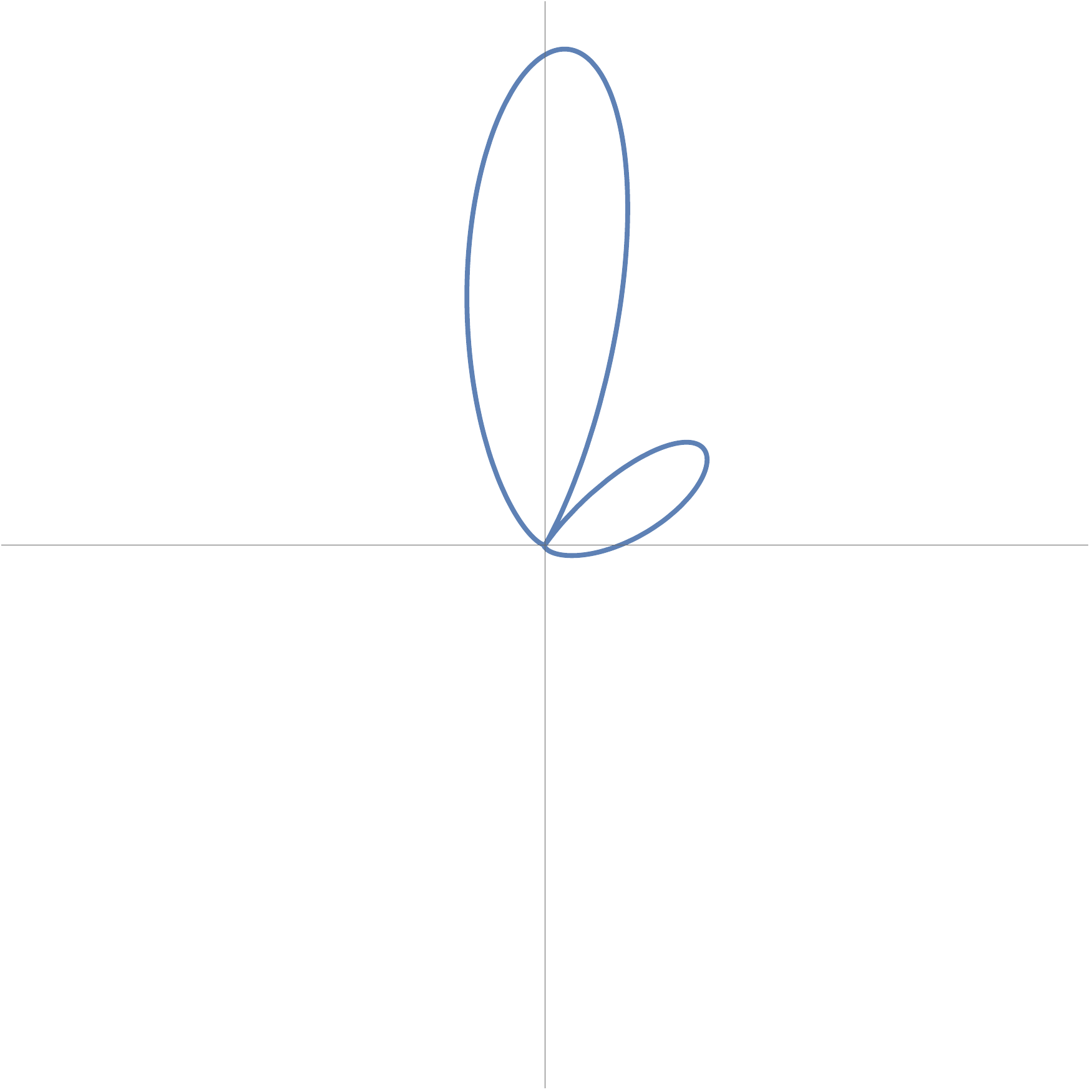}}\quad  \rotatebox{0}{\includegraphics[width=1.2in]{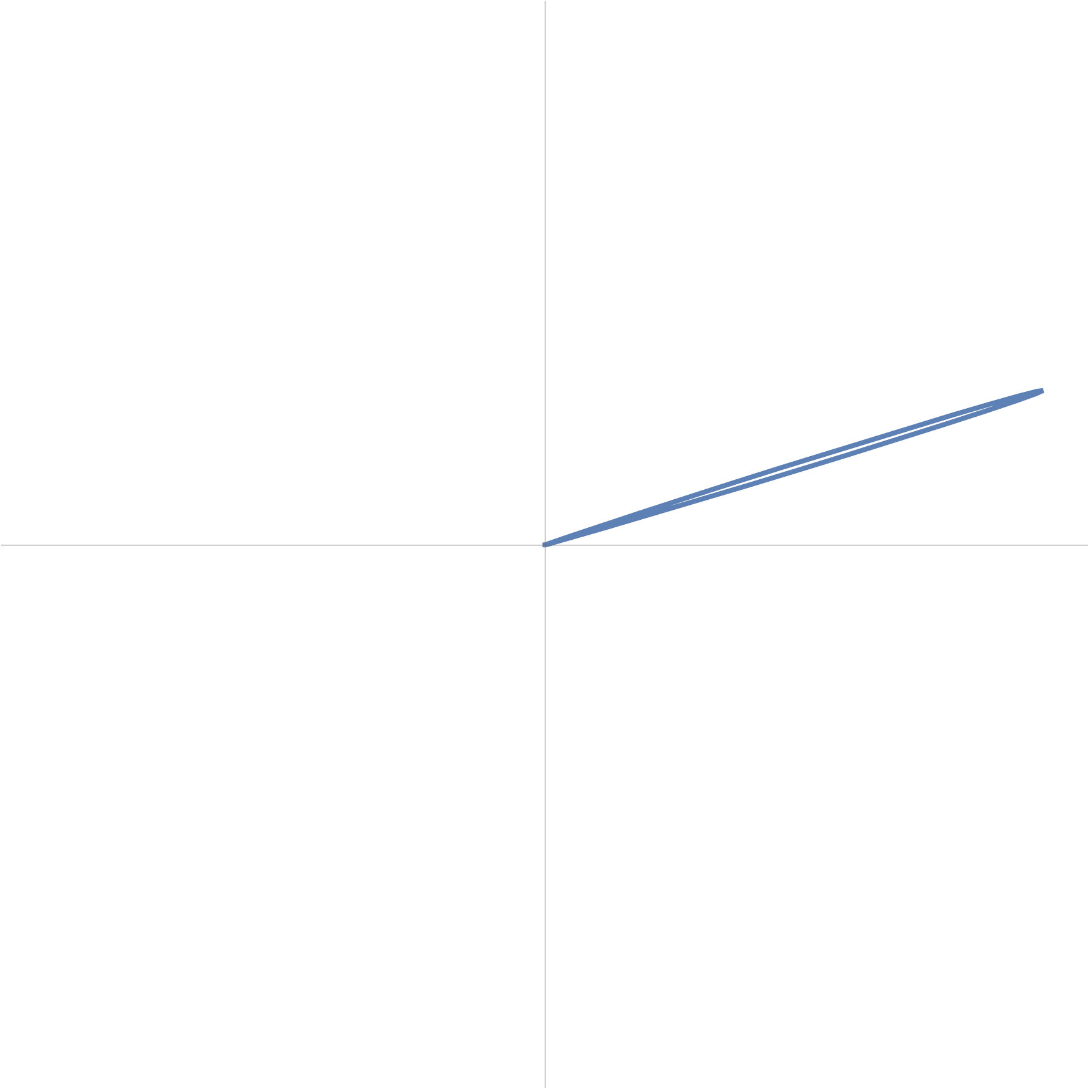}}} 
\caption{ Cusp Angular Distribution; A polar plot with $\beta=0.0,0.333,$ $0.666,0.999$. The vertical axes is $x$ and the horizontal axis is $z$. The electron moves in both dimensions.  } 
\end{center}
\end{figure}

\section*{Infrator}
\begin{figure}[ht]
\begin{center}
{\rotatebox{0}{\includegraphics[width=1.2in]{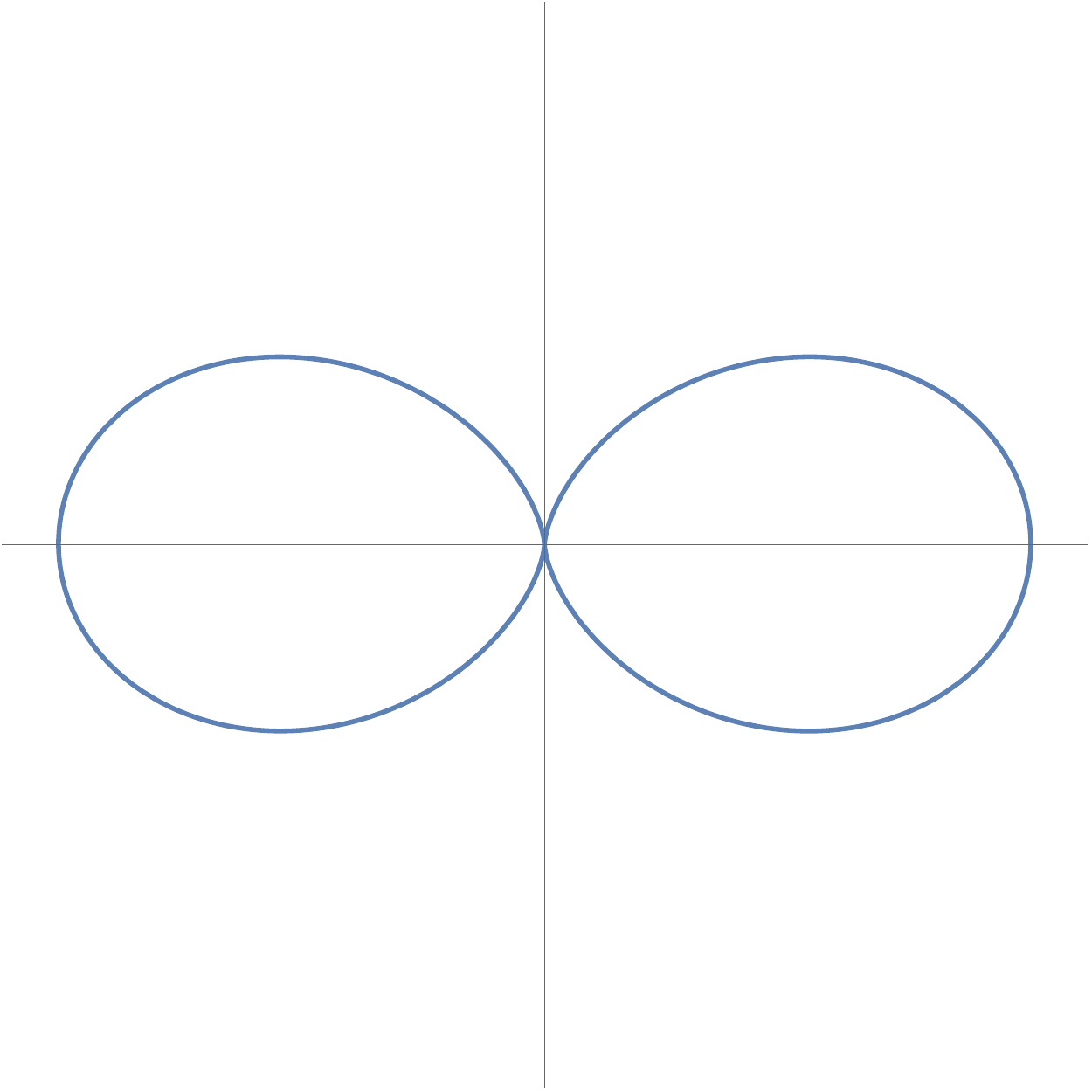}}\quad  \rotatebox{0}{\includegraphics[width=1.2in]{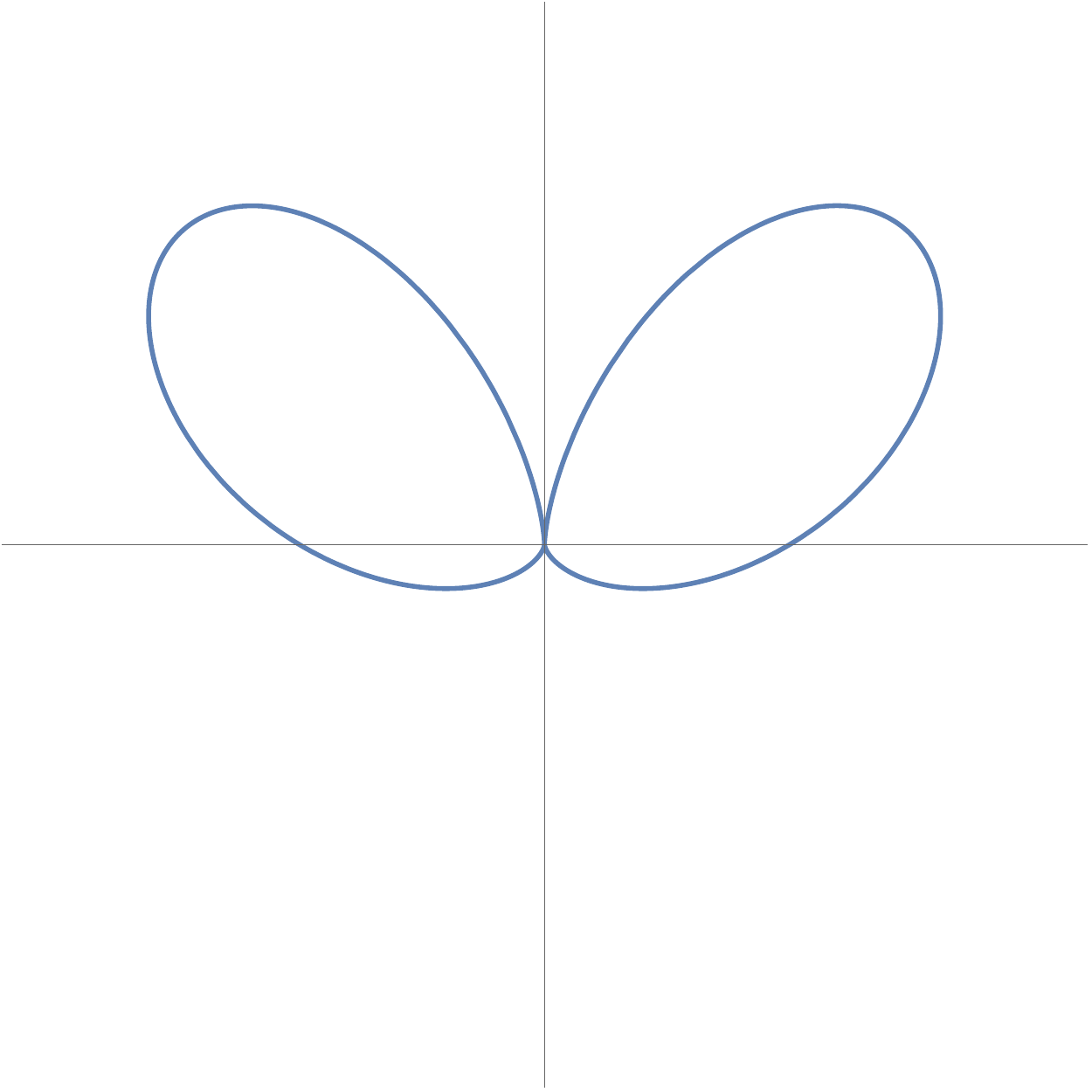}}\quad  \rotatebox{0}{\includegraphics[width=1.2in]{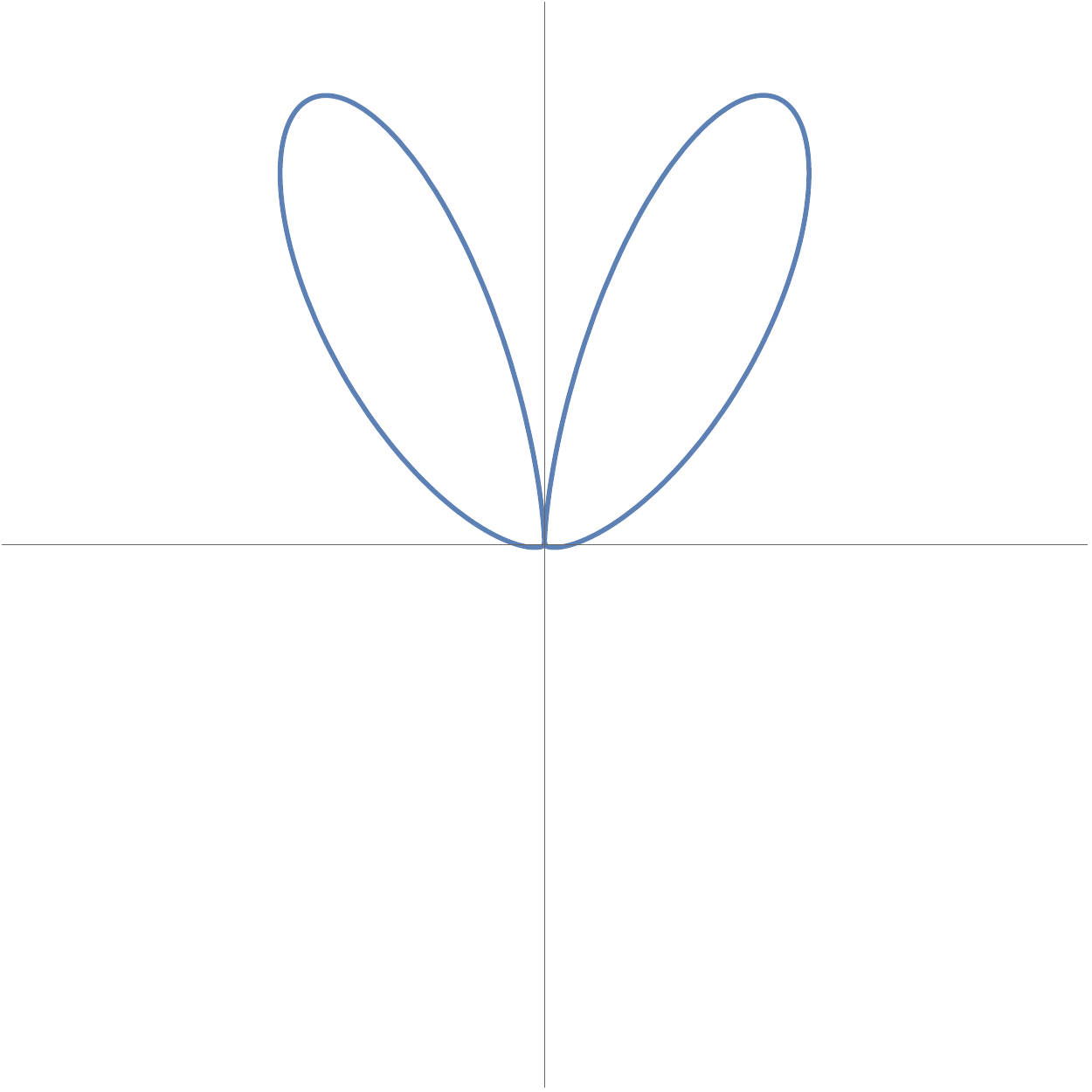}}\quad  \rotatebox{0}{\includegraphics[width=1.2in]{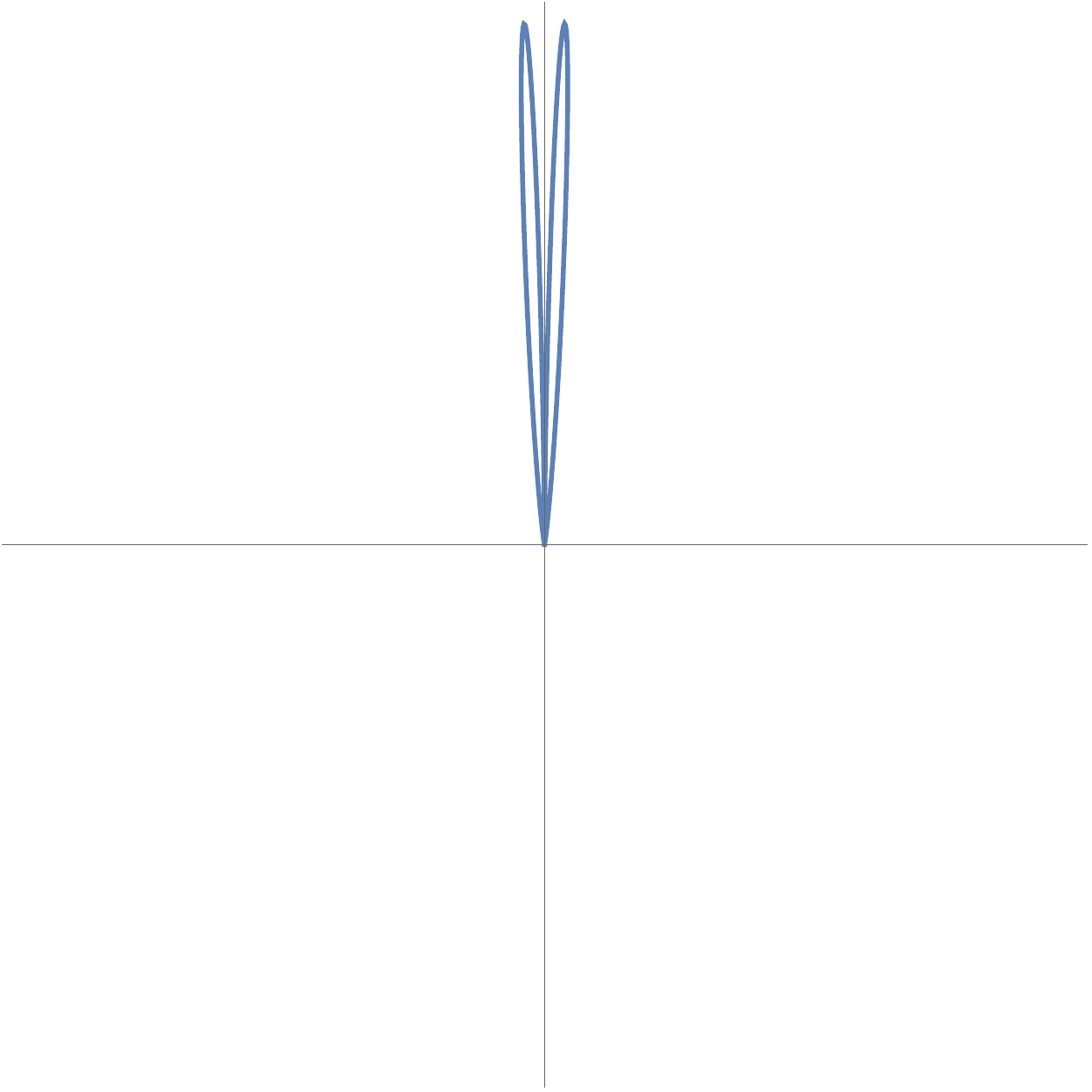}}} 

\caption{ Catenary Angular Distribution; The catenary hangs as a chain would with the vertical axis $x$ and horizontal axis as $z$.  A polar plot with $v_R=0.0001$ and $\beta = 0.001, 0.333,$ $0.665, 0.997$. The electron moves in both dimensions but at high speeds moves mostly in the $x$-direction (thus the reason for the beaming in the $x$-direction).  } 
\end{center}
\end{figure} 

\newpage

\begin{figure}[ht]
\begin{center}
{\rotatebox{0}{\includegraphics[width=1.2in]{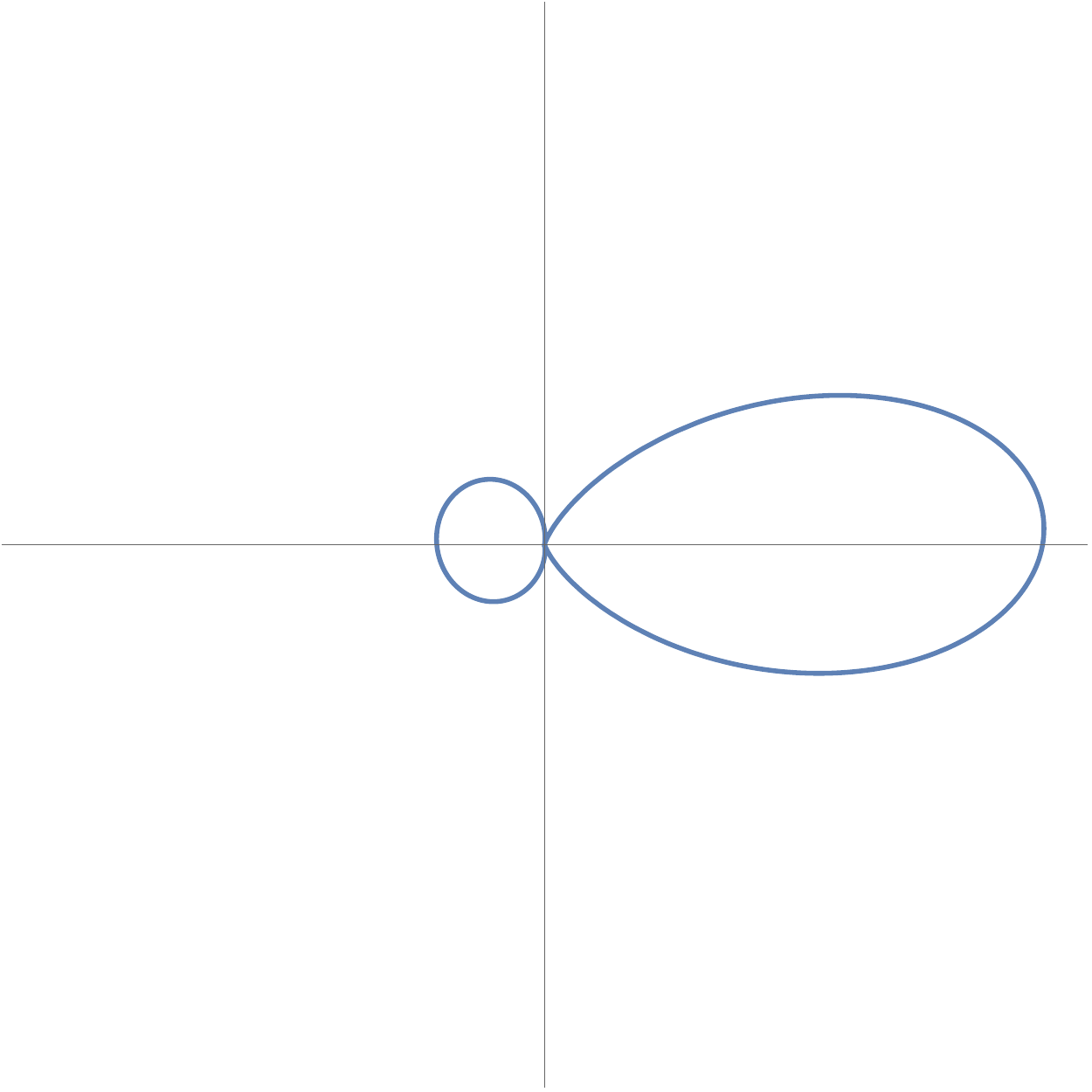}}\quad  \rotatebox{0}{\includegraphics[width=1.2in]{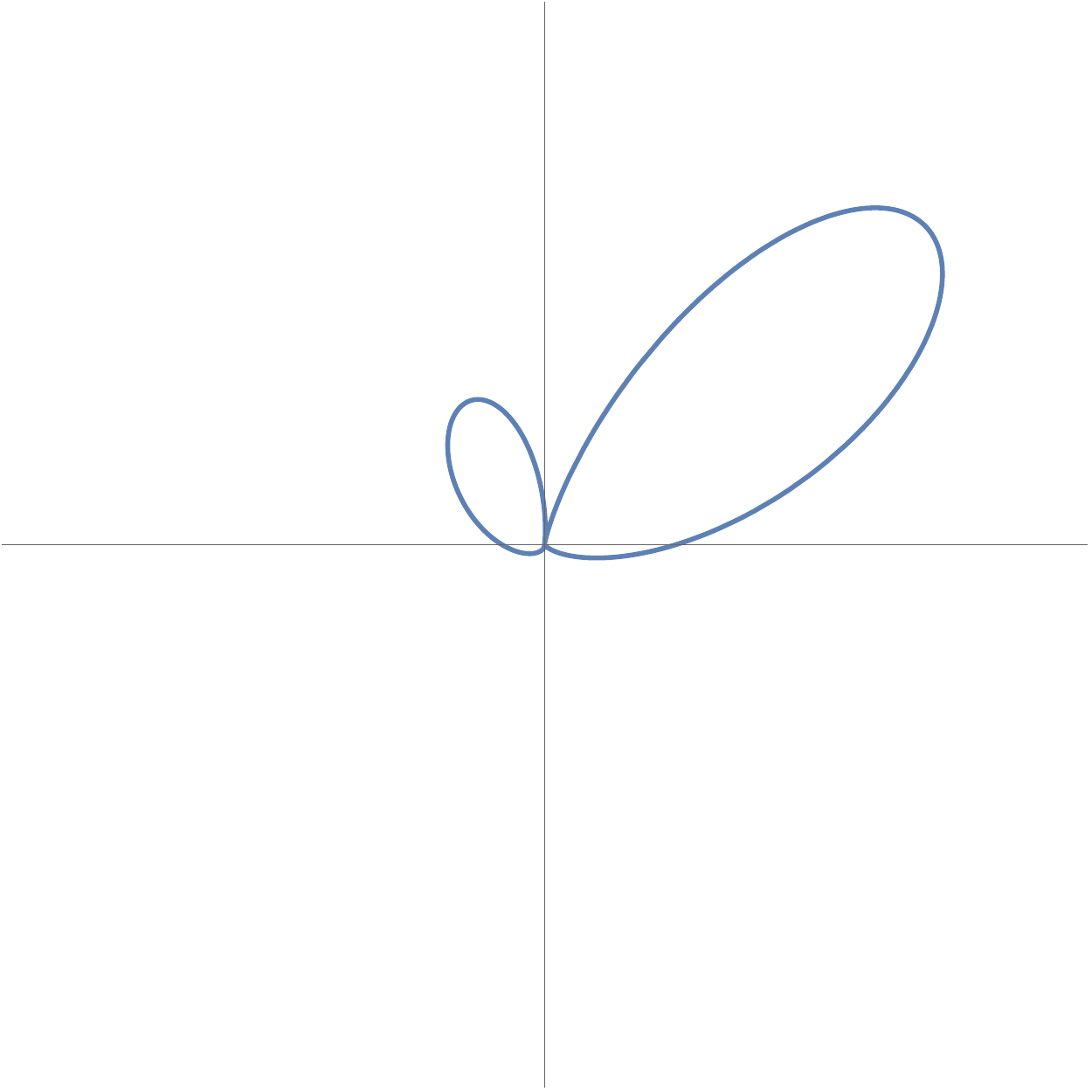}}\quad  \rotatebox{0}{\includegraphics[width=1.2in]{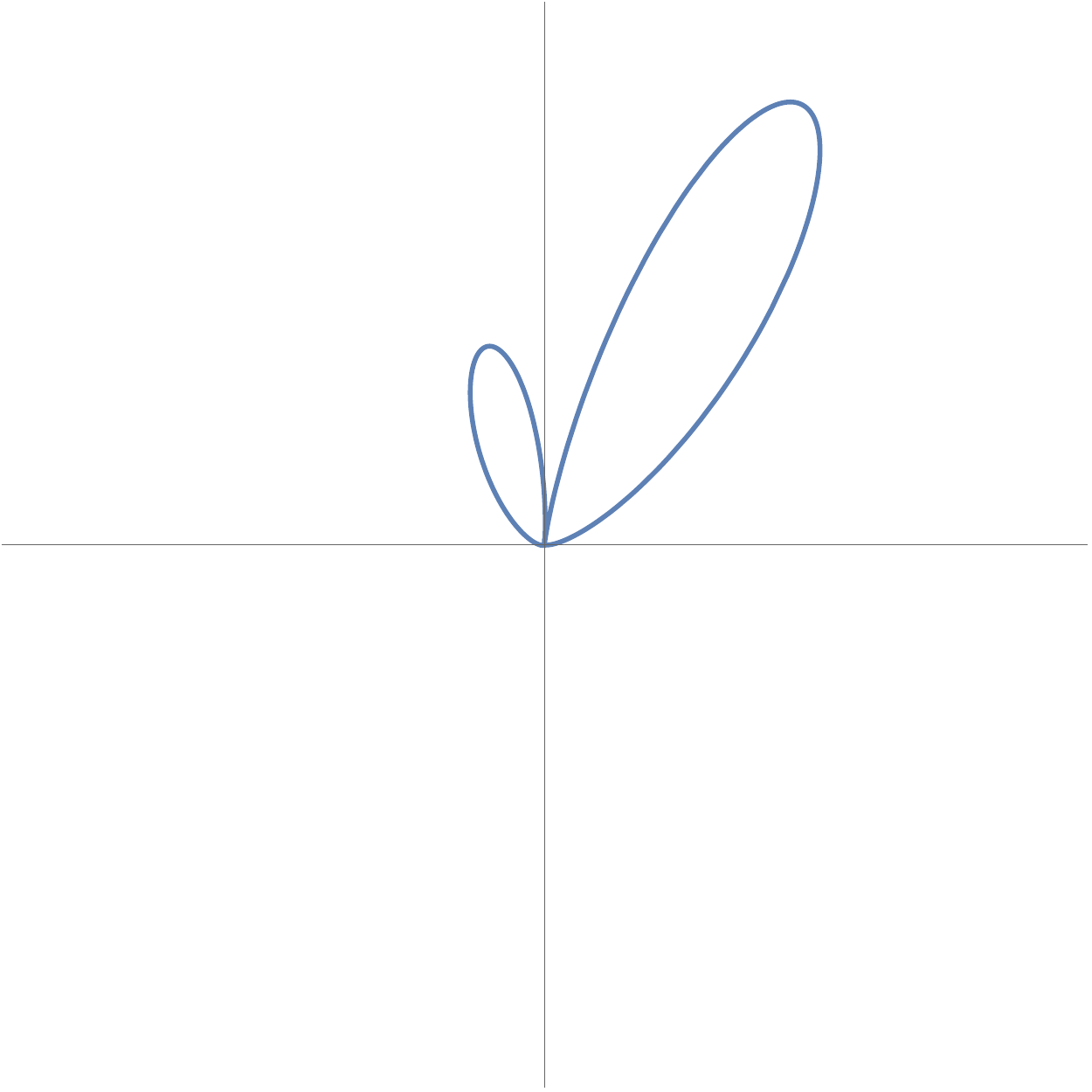}}\quad  \rotatebox{0}{\includegraphics[width=1.2in]{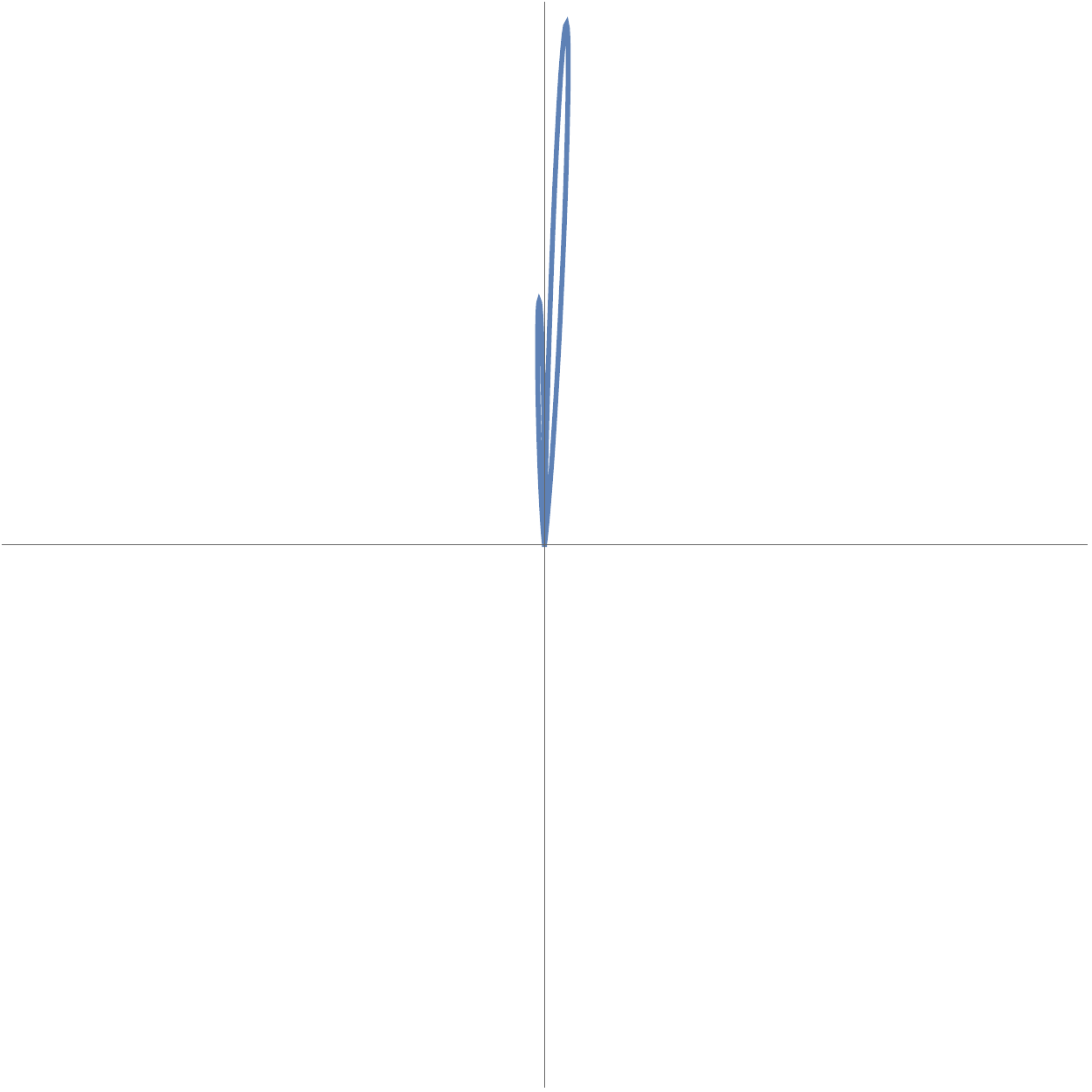}}} 

\caption{ Catenary Angular Distribution; Additional torsion is added with substantial Rindler drift: $v_R=0.25$.  Here $\beta = 0.251, 0.500,$ $0.749, 0.998$ for the four plots respectively.  } 
\end{center}
\end{figure} 
\begin{figure}[ht]
\begin{center}
{\rotatebox{0}{\includegraphics[width=1.2in]{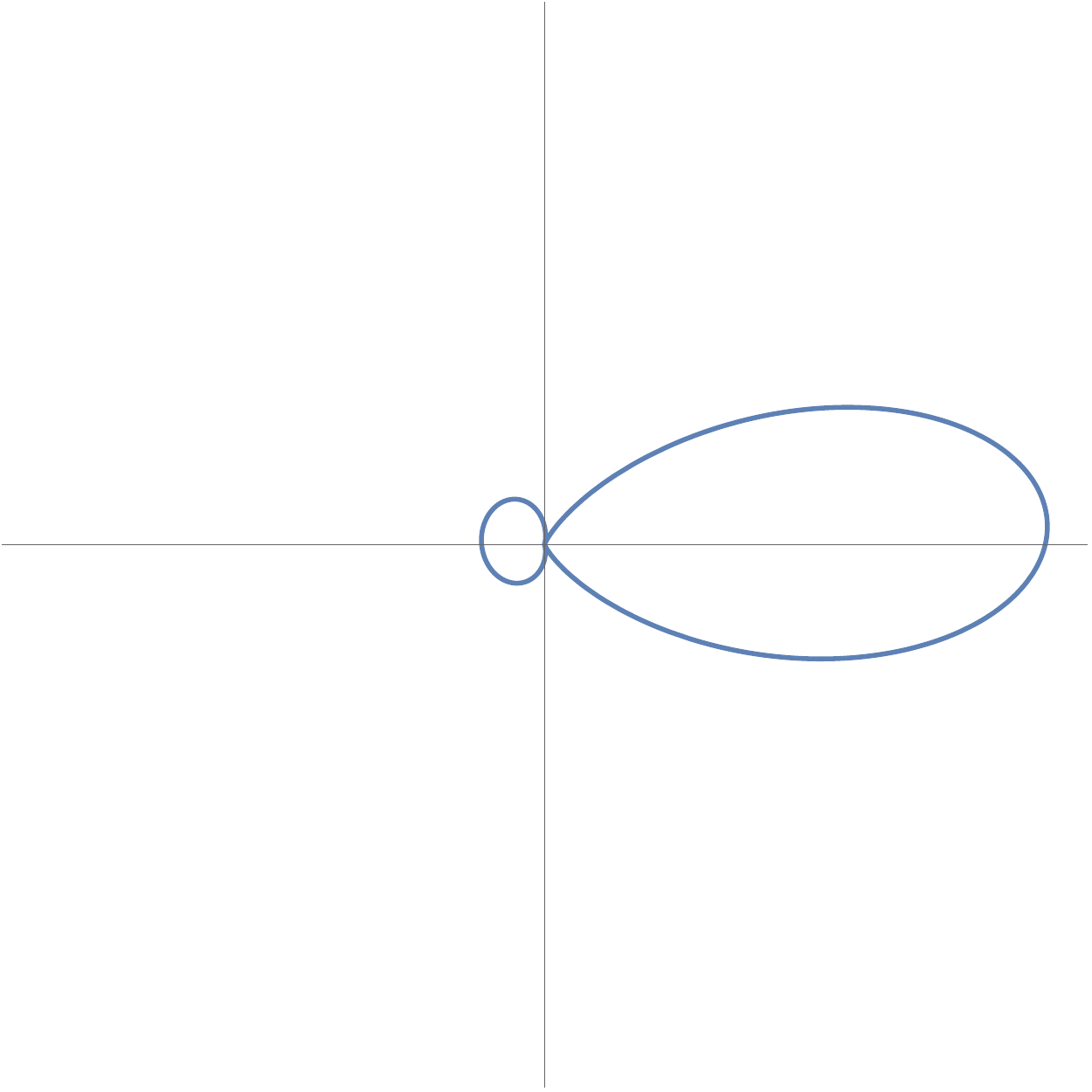}}\quad  \rotatebox{0}{\includegraphics[width=1.2in]{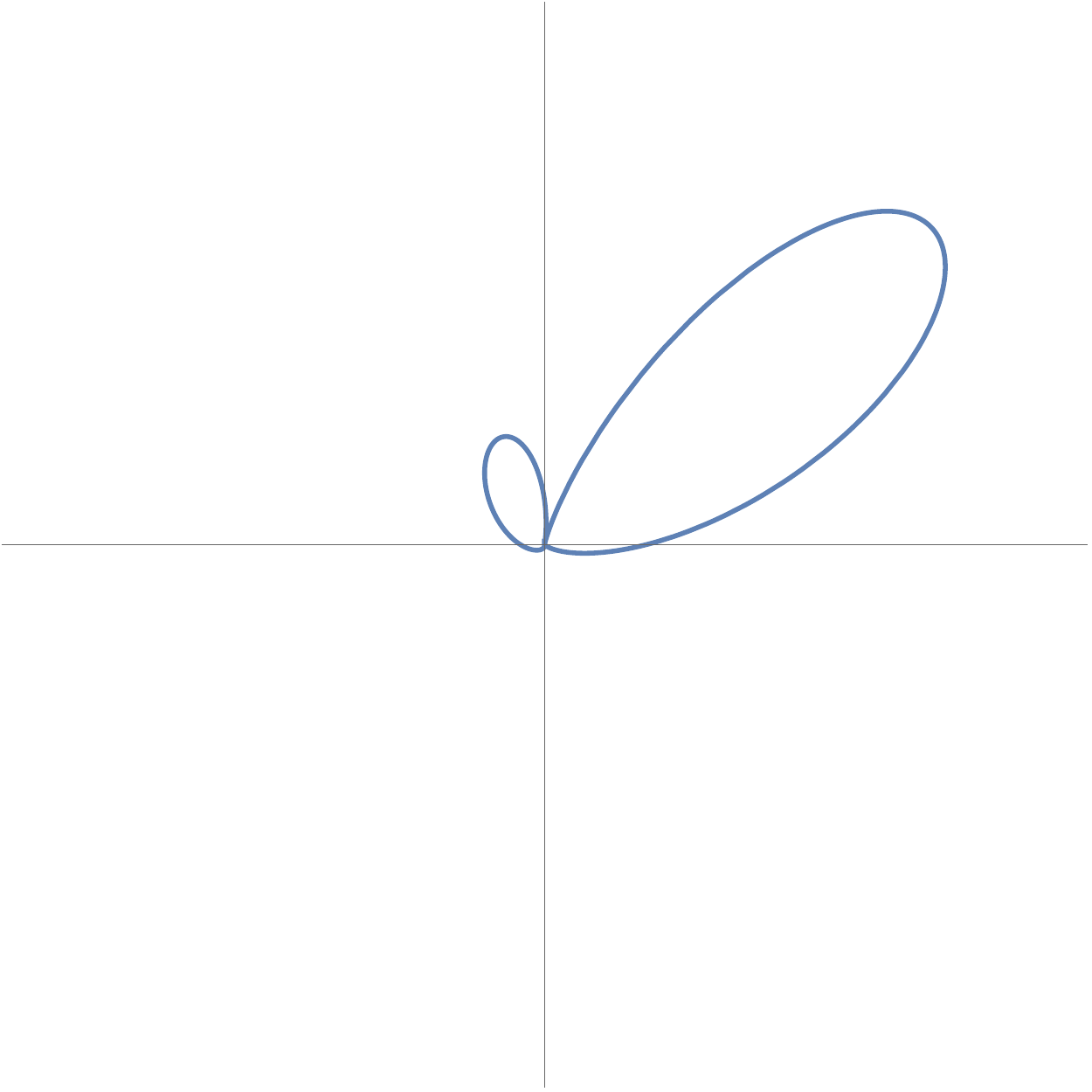}}\quad  \rotatebox{0}{\includegraphics[width=1.2in]{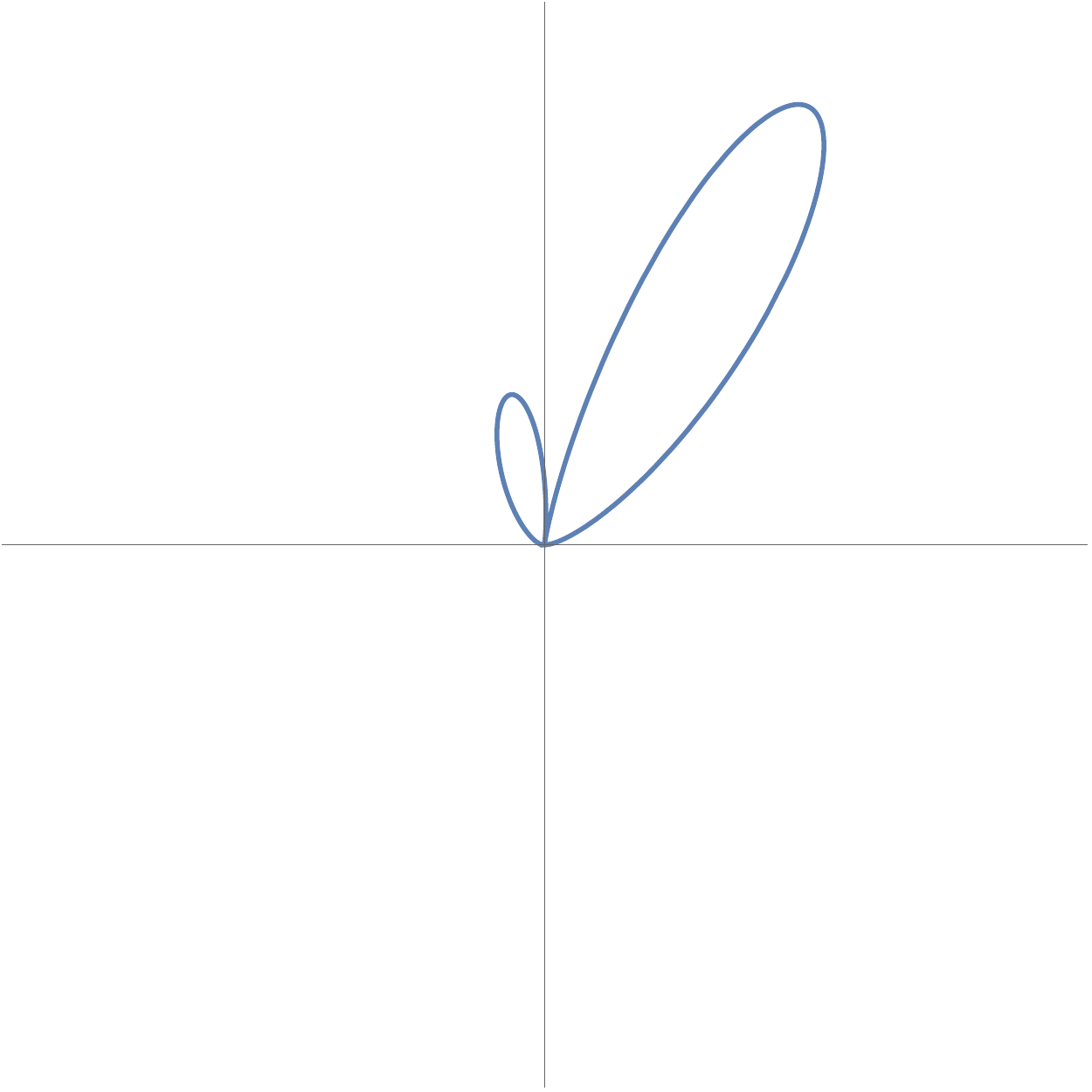}}\quad  \rotatebox{0}{\includegraphics[width=1.2in]{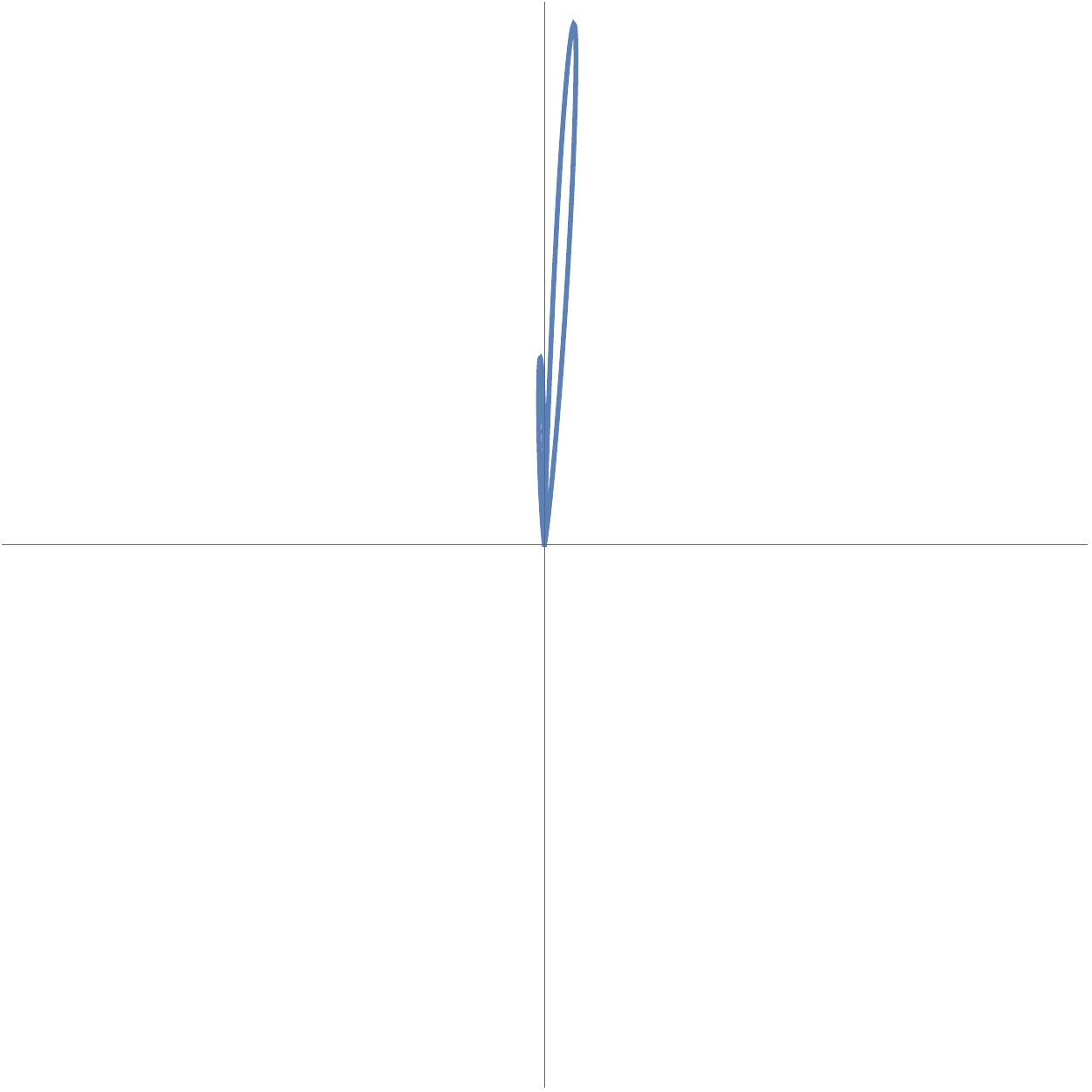}}} 
\caption{Catenary Angular Distribution; The Rindler drift requires an initial starting speed $\beta > v_R$.  Here $v_R=0.333$, and $\beta = 0.334$, $0.555$, $0.776$, $0.997$.} 
\end{center}
\end{figure}

\chapter{Power Distributions 3D Plots}\label{Appendix:3D}

\section*{Nulltor}
\begin{figure}[ht]
\begin{center}
{\rotatebox{0}{\includegraphics[width=1.5in]{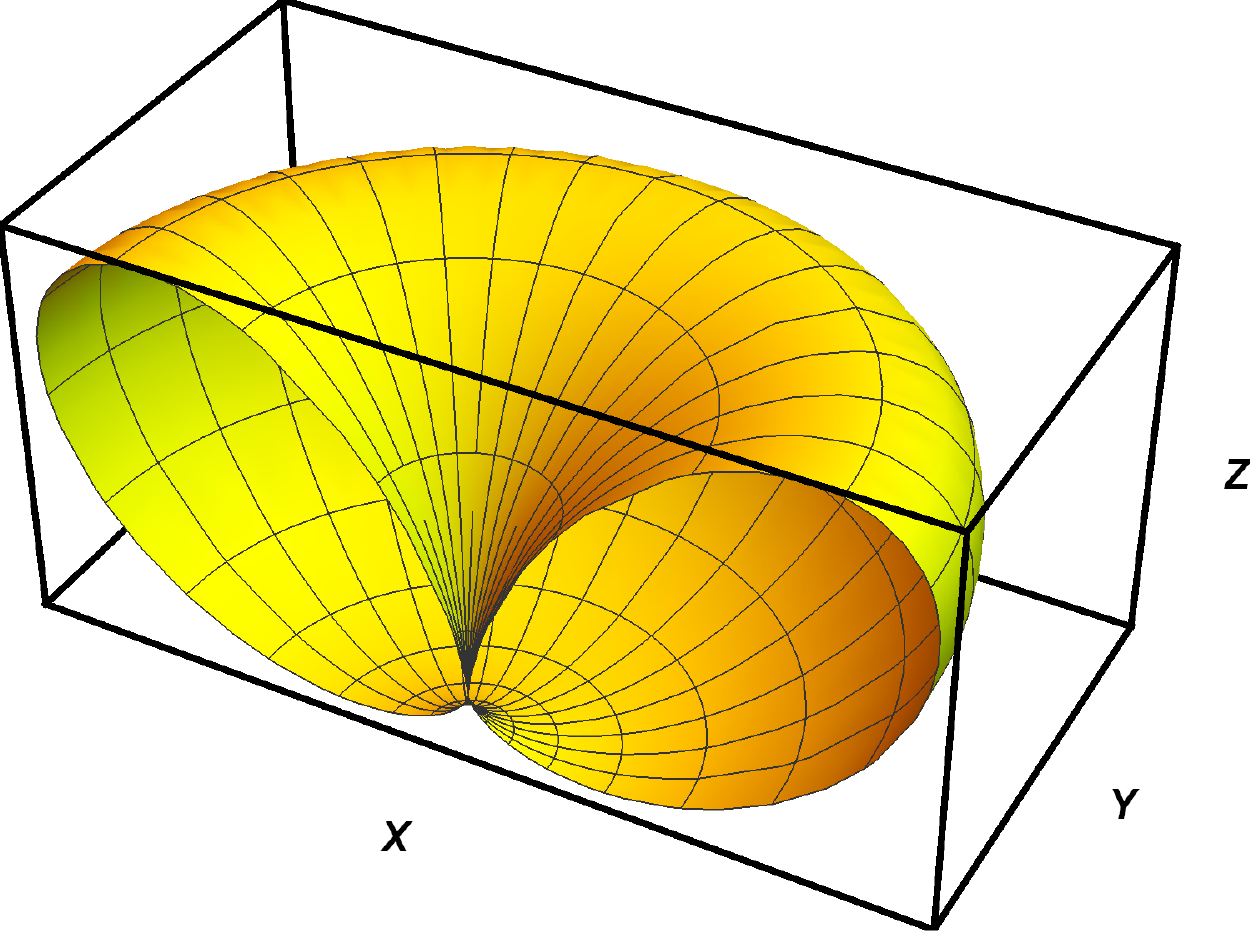}}\quad
\rotatebox{0}{\includegraphics[width=1.5in]{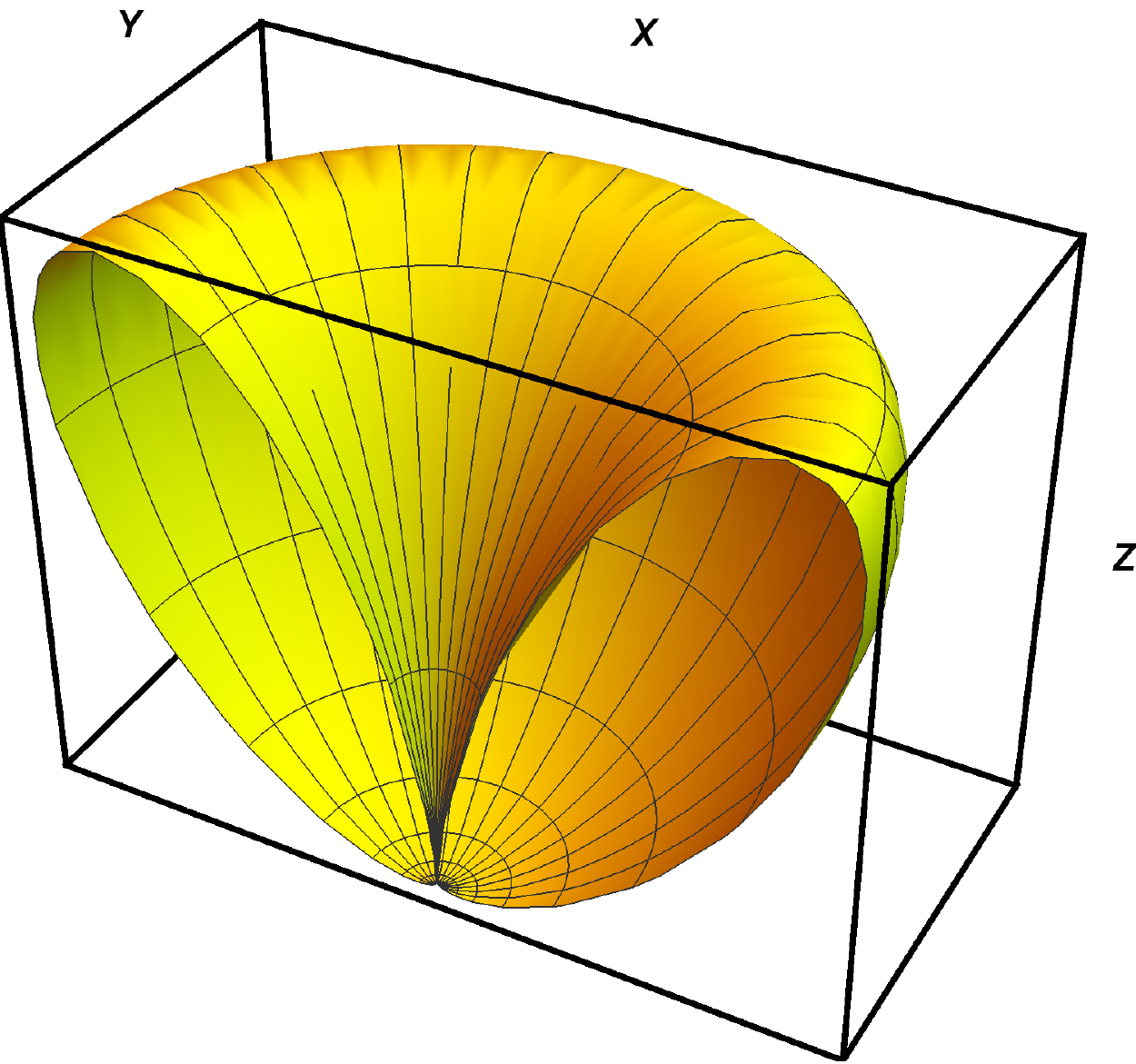}}\quad
\rotatebox{0}{\includegraphics[width=1.5in]{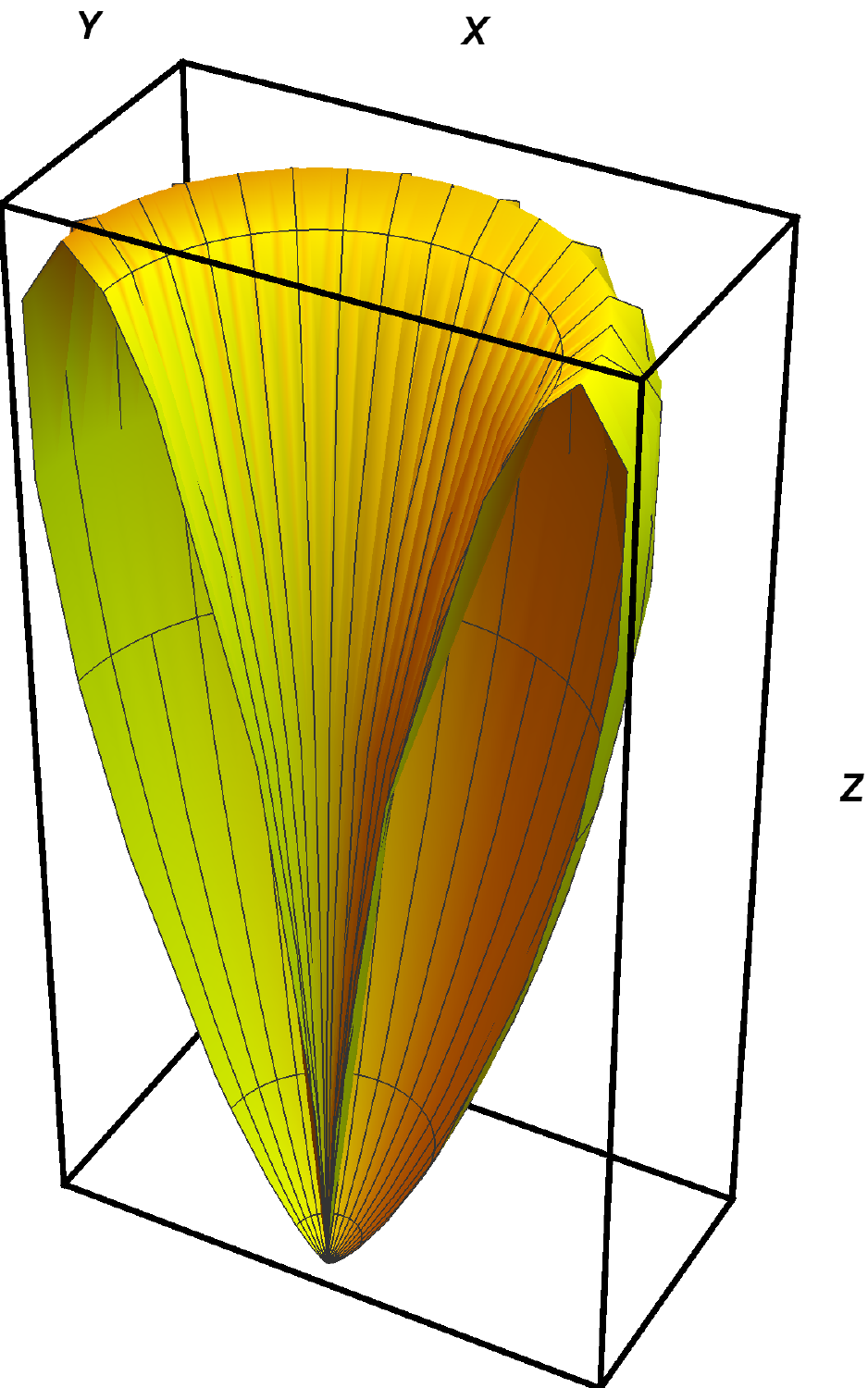} }}
\caption{3-Dimensional Angular Distribution; A polar plot with $\beta =  0.3,$ $0.6, 0.9$.  The electron moves forward along the $+z$  direction. The plot range is $\theta=[0,\pi]$ and $\phi=[0,\pi]$. } 
\end{center}
\end{figure}  
\newpage
\section*{Ultrator}
\begin{figure}[ht]
\begin{center}
{\rotatebox{0}{\includegraphics[width=1.2in]{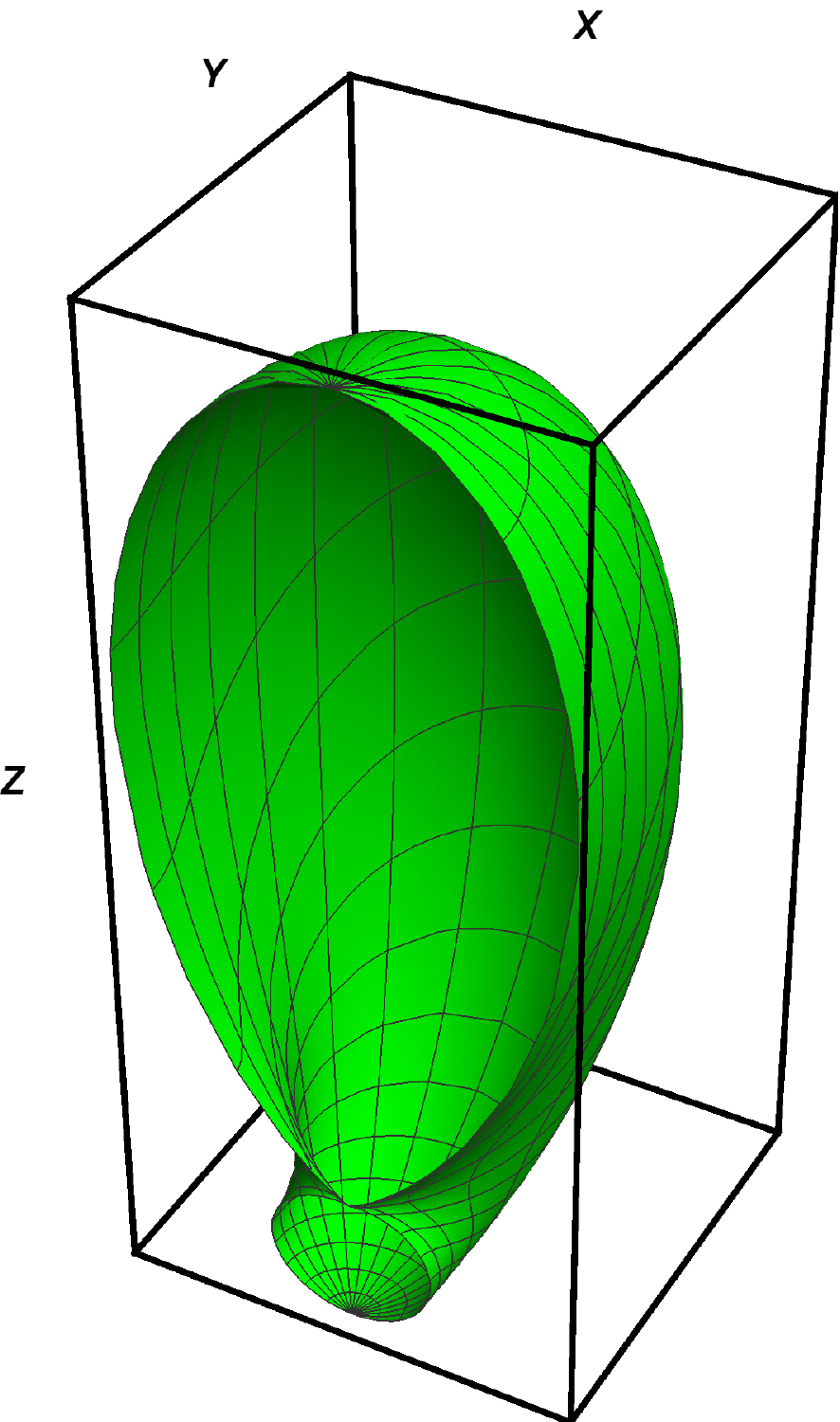}}\quad
\rotatebox{0}{\includegraphics[width=1.2in]{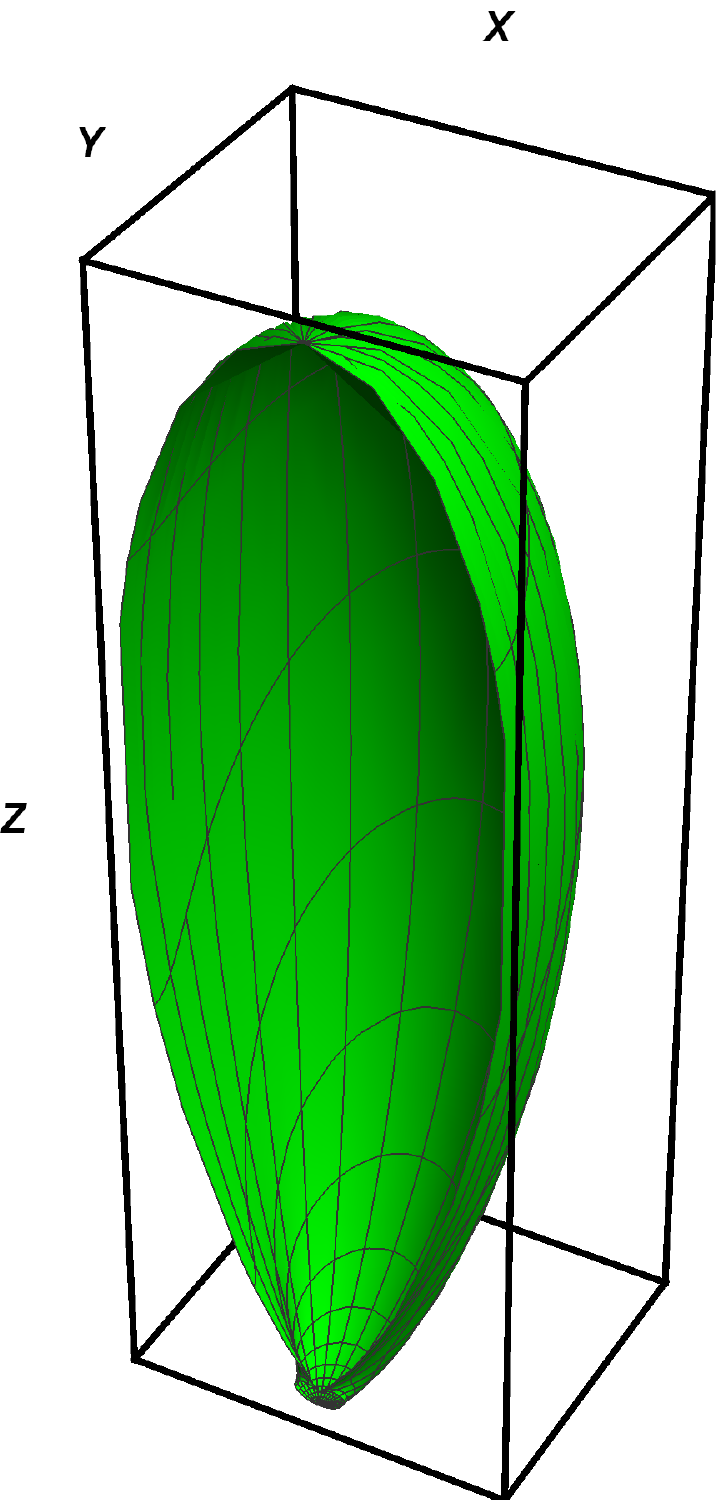}}\quad
\rotatebox{0}{\includegraphics[width=1.2in]{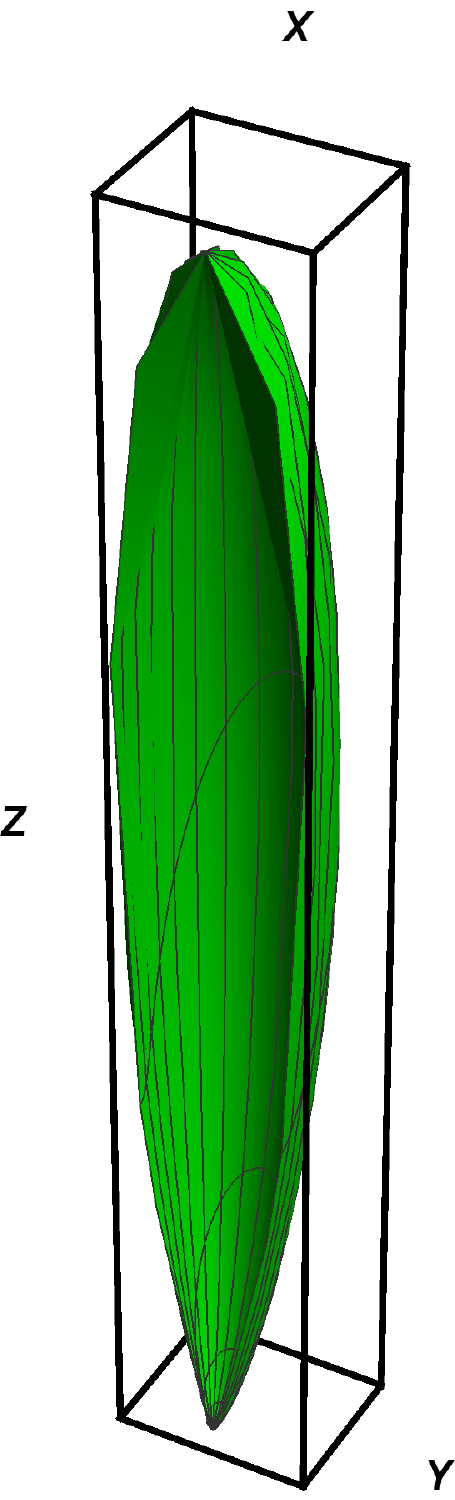}}}
\caption{ Synchrotron Angular Distribution in 3D; A spherical 3D plot with $\beta = 0.3,$ $0.6, 0.9$. The electron moves in the $z$ direction, but around the $x$ axis in circular motion.  The plot range is $\theta=[0,\pi]$ and $\phi=[0,\pi]$.} 
\end{center}
\end{figure} 

\newpage
\section*{Parator}
\begin{figure}[ht]
\begin{center}
{\rotatebox{0}{\includegraphics[width=1.4in]{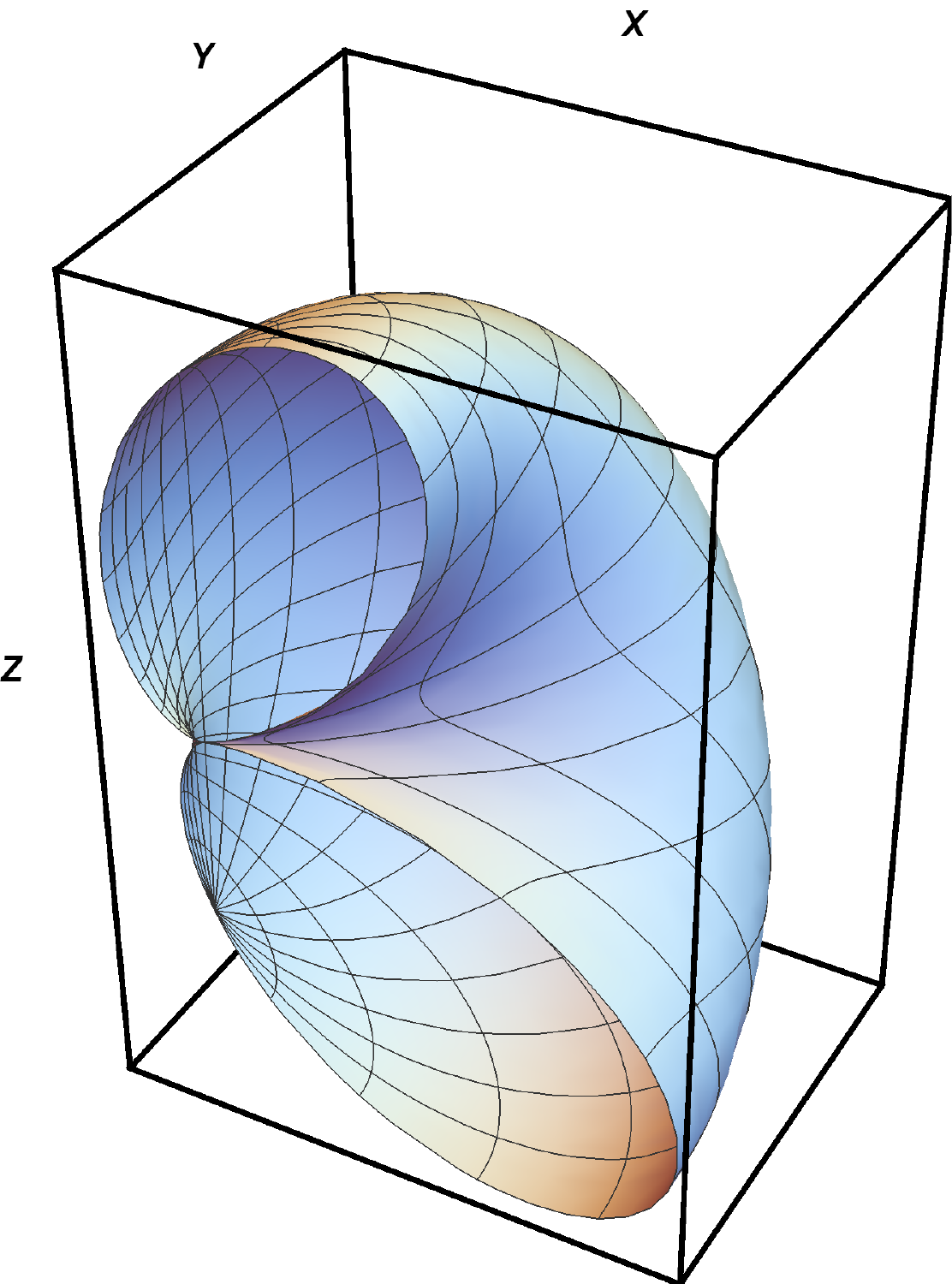}}\quad
\rotatebox{0}{\includegraphics[width=1.4in]{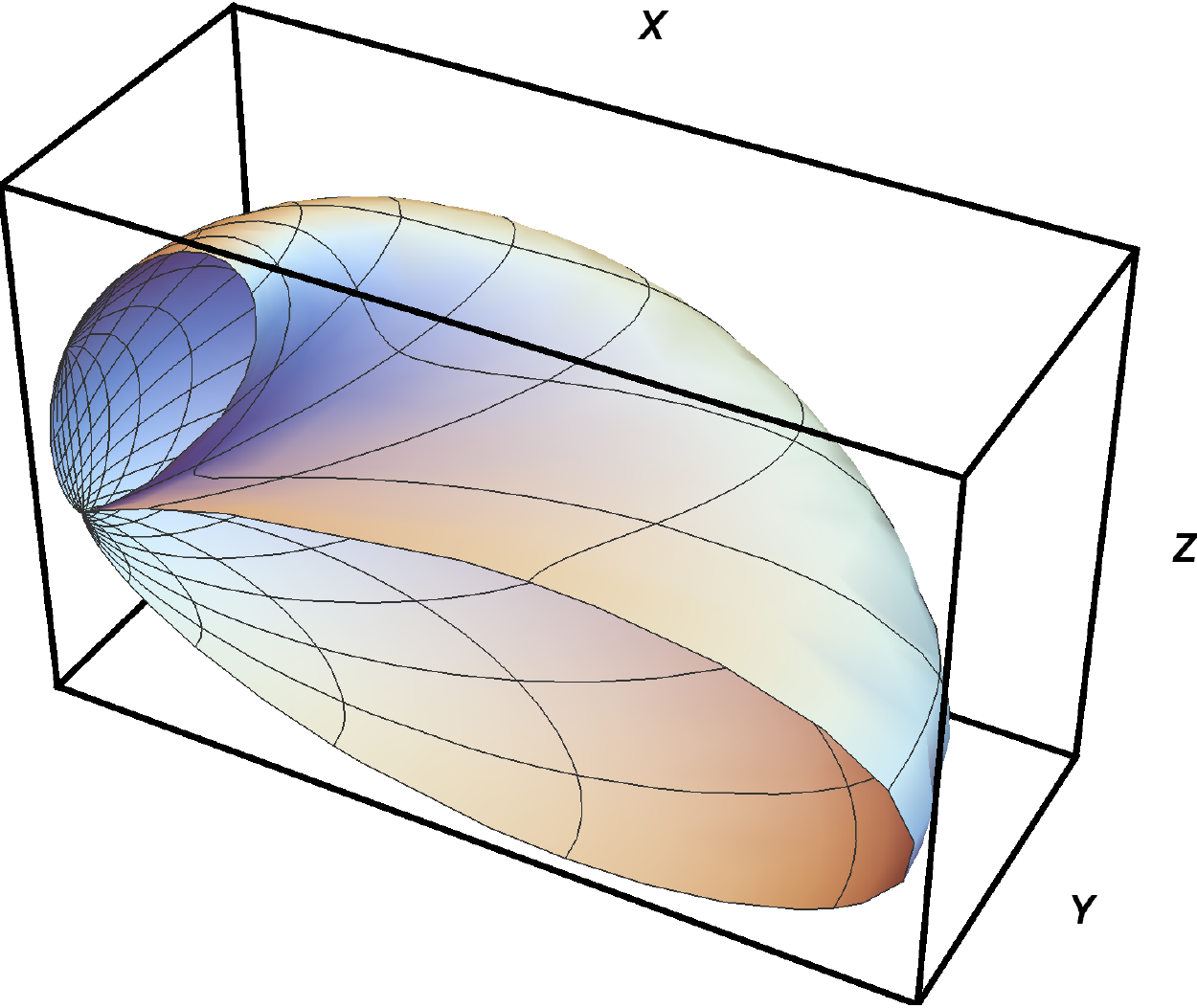}}\quad
\rotatebox{0}{\includegraphics[width=1.4in]{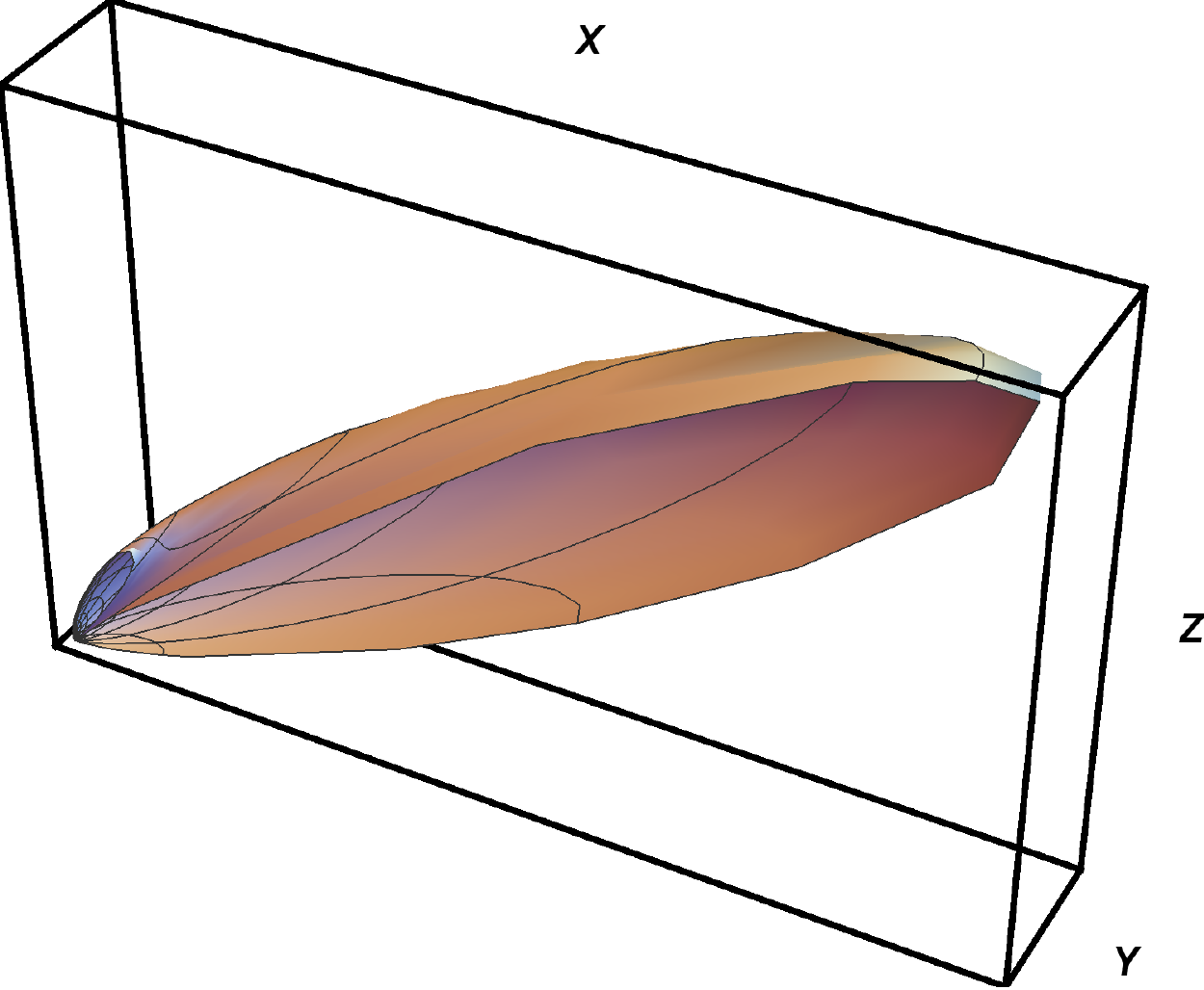} }}
\caption{ Cusp Angular Distribution; A polar plot with $\beta =  0.3,$ $0.6, 0.9$. The electron moves between $x$ and $z$ dimensions,directed as resulting vector.  The plot range is $\theta=[0,\pi]$ and $\phi=[0,\pi]$.  } 
\end{center}
\end{figure}

\section*{Infrator}
\begin{figure}[ht]
\begin{center}
{\rotatebox{0}{\includegraphics[width=1.4in]{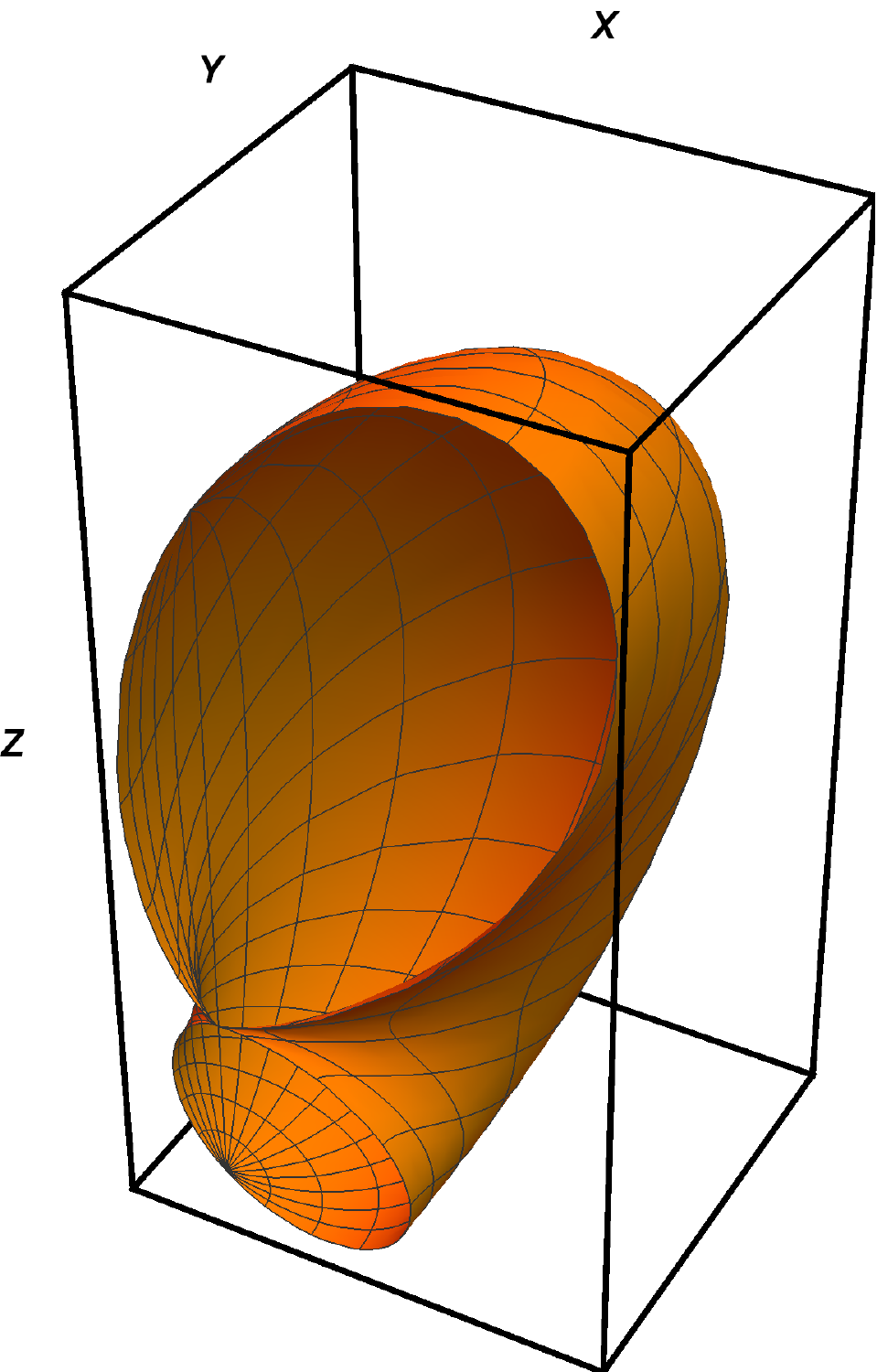}}\quad
\rotatebox{0}{\includegraphics[width=1.4in]{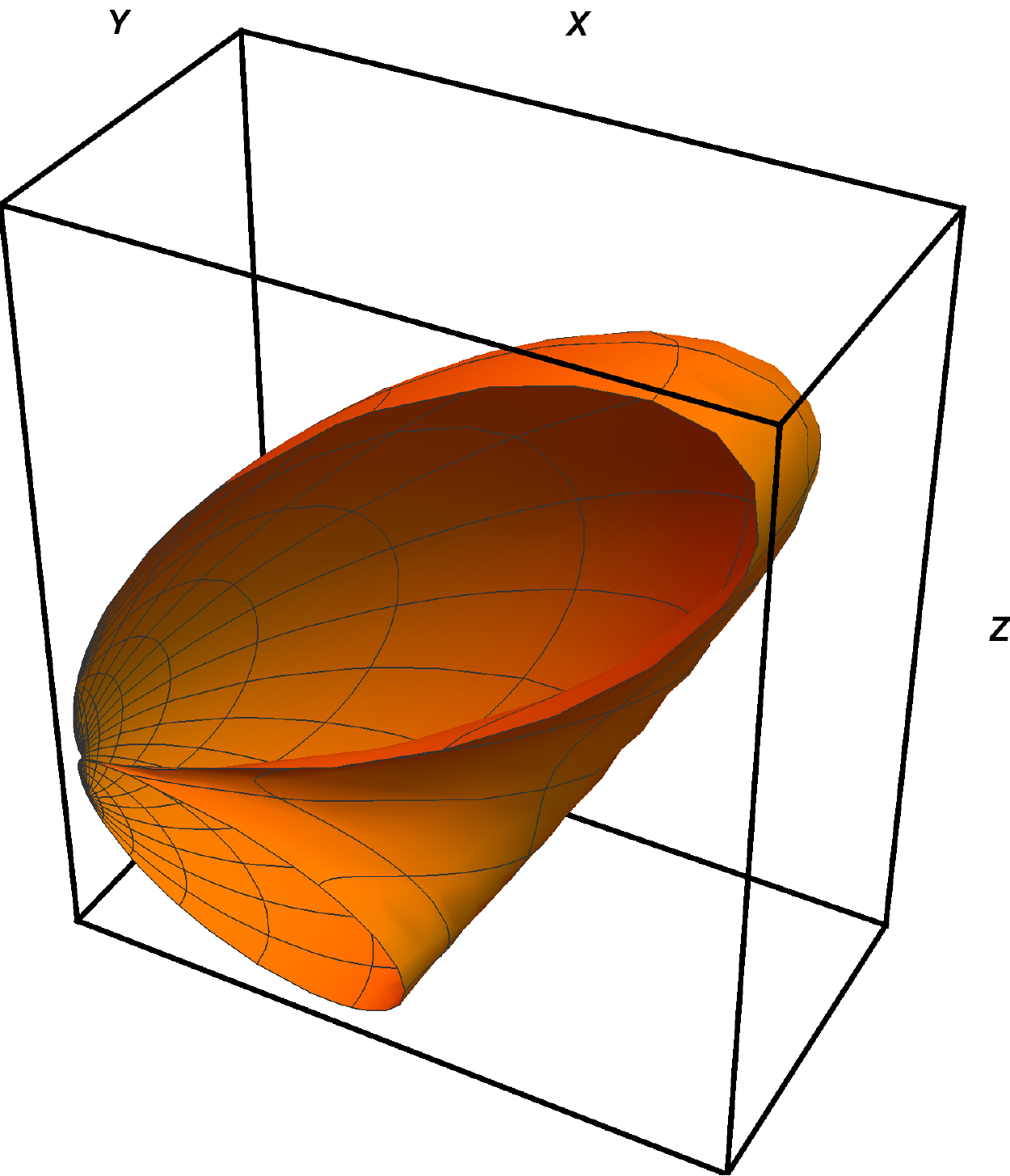}}\quad
\rotatebox{0}{\includegraphics[width=1.4in]{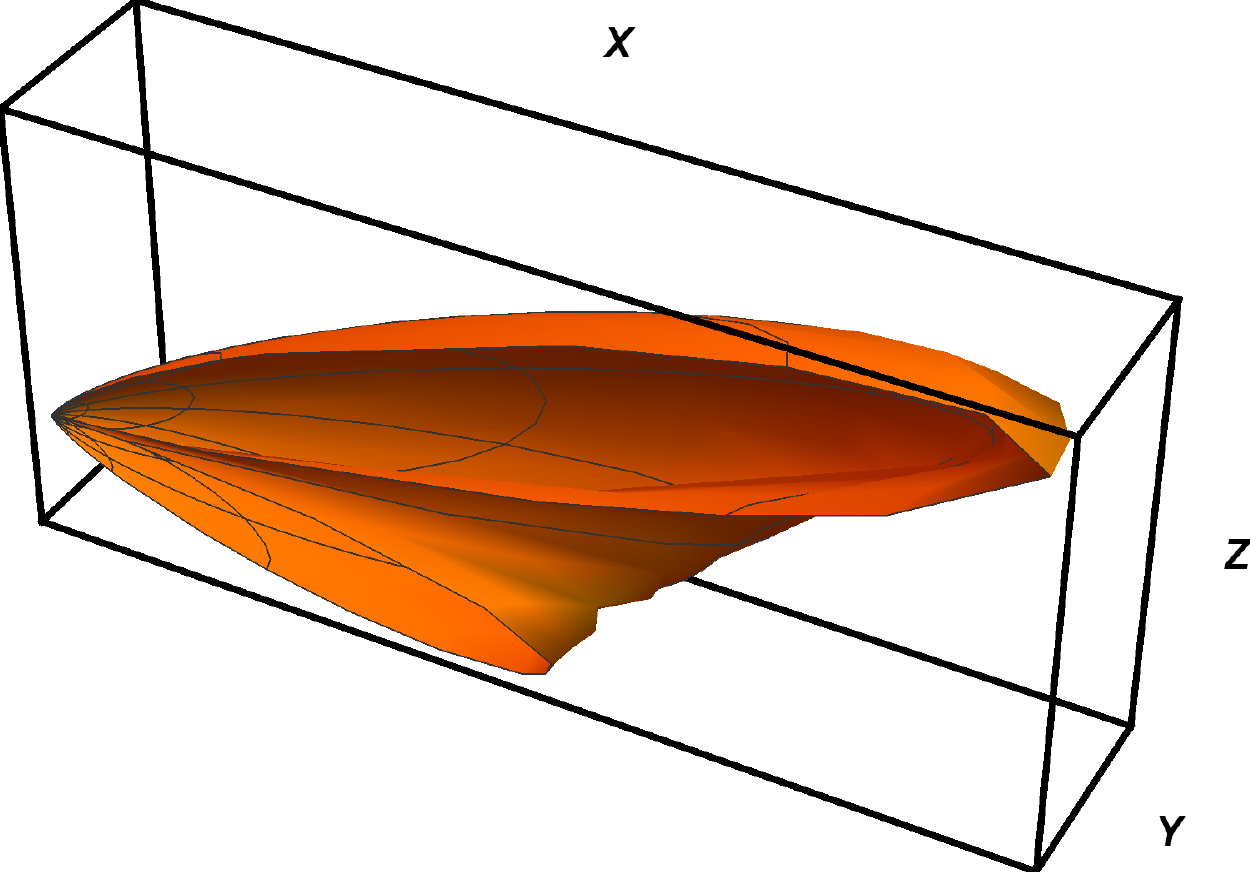} }}
\caption{ Catenary Angular Distribution; Additional torsion is added with substantial Rindler drift: $v_R=0.2$.  Here $\beta = 0.3,$ $0.6, 0.9$ . Direction of electron appear as resulting vector from components along $x$ and $z$ dimensions.  The plot range is $\theta=[0,\pi]$ and $\phi=[0,\pi]$.} 
\end{center}
\end{figure} 

\newpage
\section*{Hypertor}
\begin{figure}[ht]
\begin{center}
{\rotatebox{0}{\includegraphics[width=1.4in]{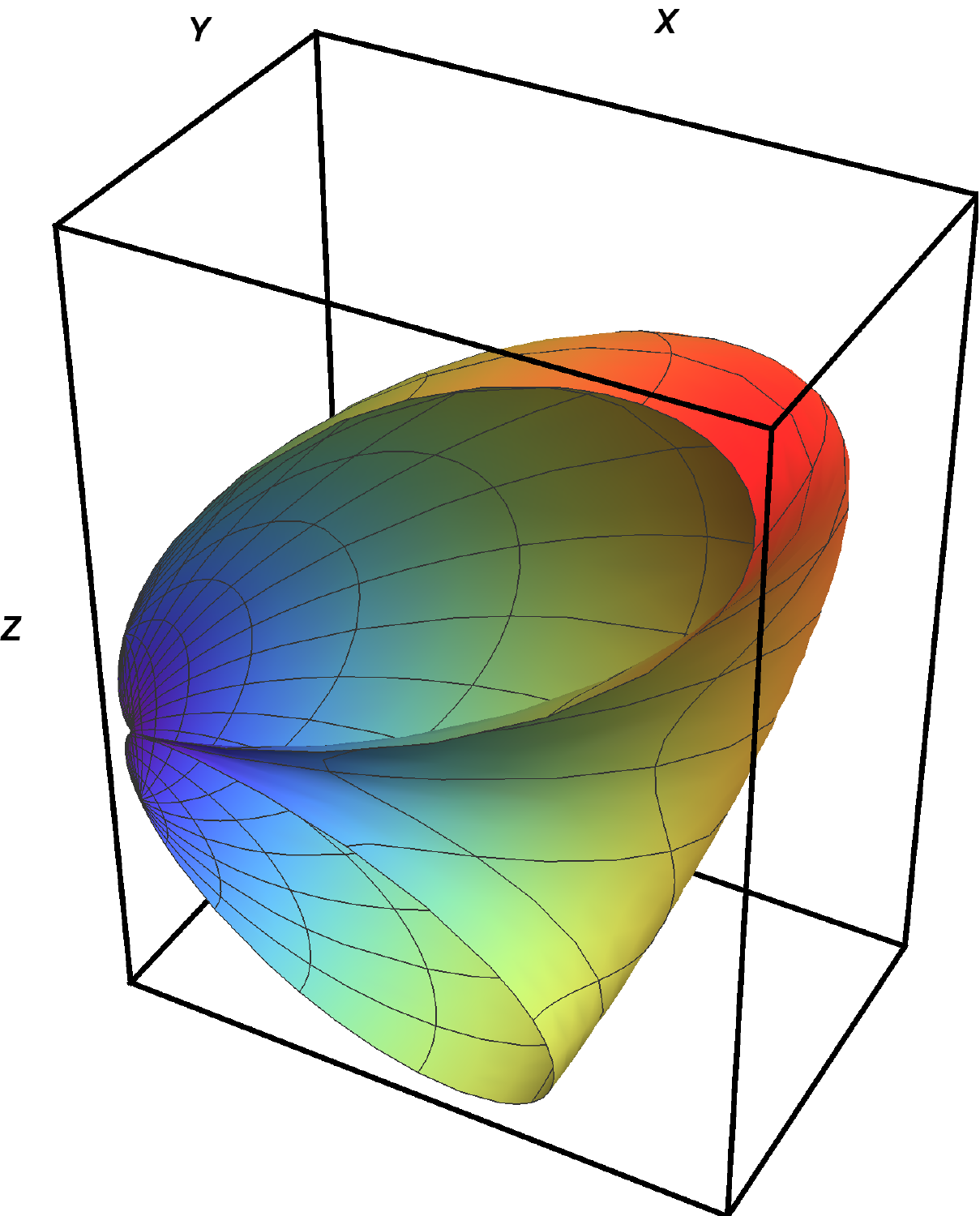}}\quad
\rotatebox{0}{\includegraphics[width=1.4in]{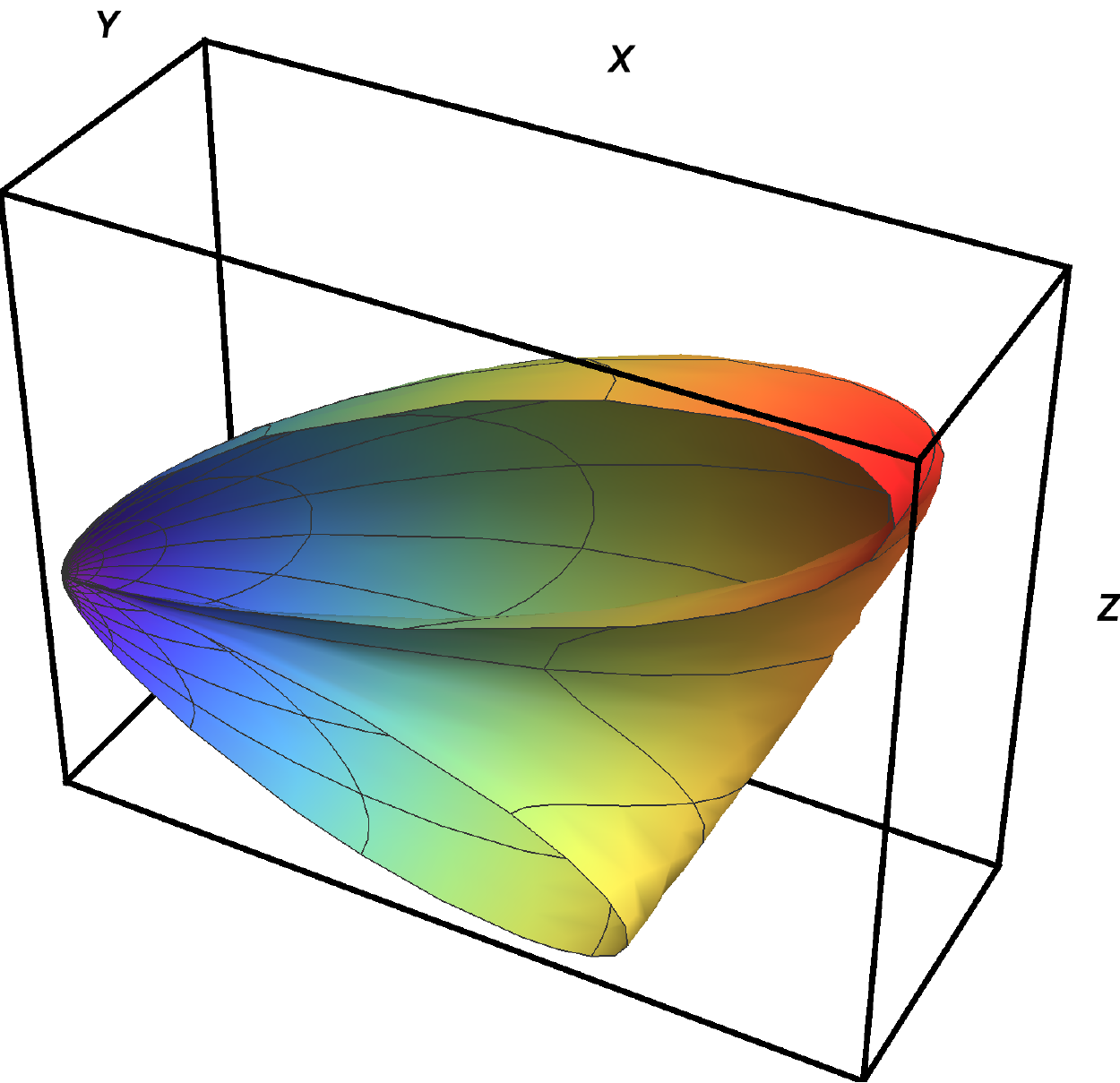}}\quad
\rotatebox{0}{\includegraphics[width=1.4in]{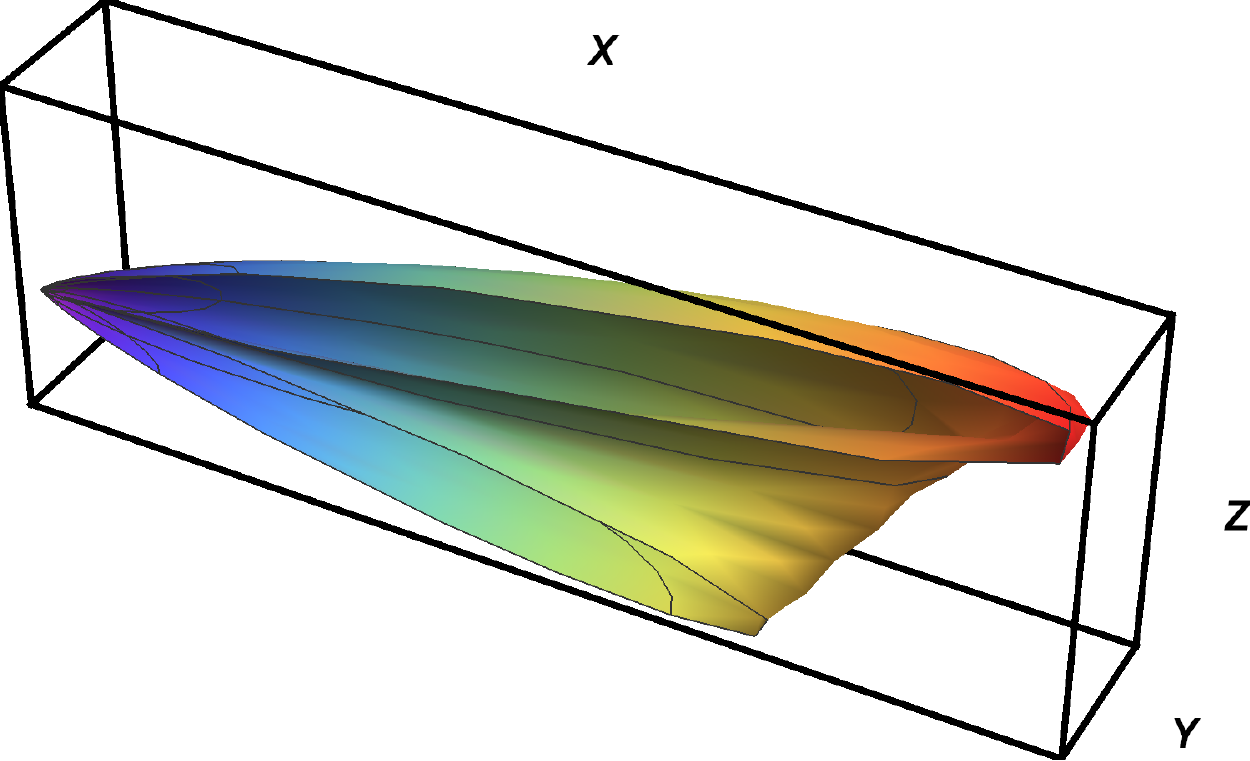} }}
\caption{ Helix Angular Distribution; $\kappa = 1$, $\tau=\nu=0.1$, $\beta = 0.3, 0.6,$ $0.9$ . The plot range is $\theta=[0,\pi]$ and $\phi=[0,\pi]$.} 
\end{center}
\end{figure} 

\begin{figure}[ht]
\begin{center}
{\rotatebox{0}{\includegraphics[width=1.4in]{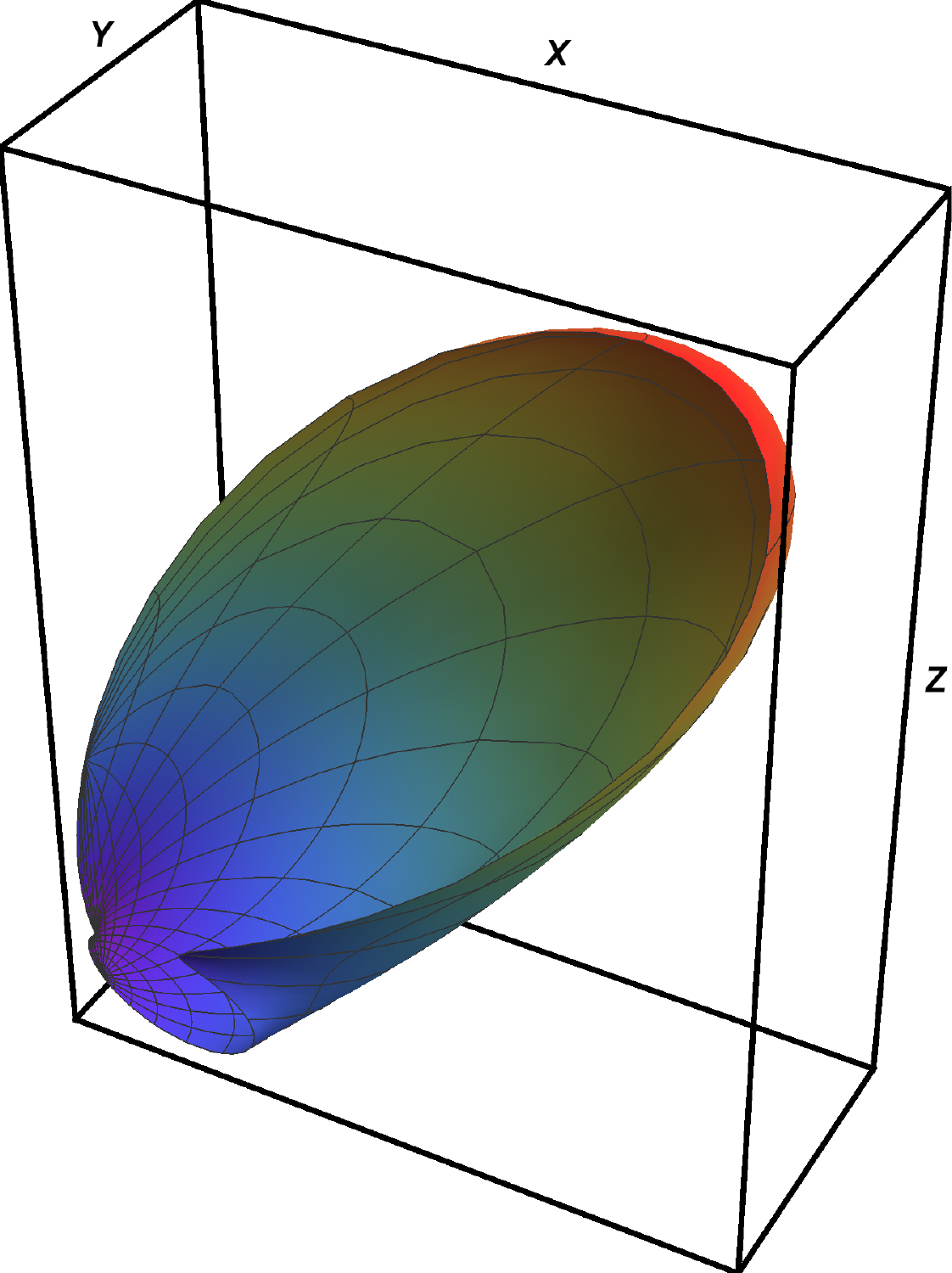}}\quad
\rotatebox{0}{\includegraphics[width=1.4in]{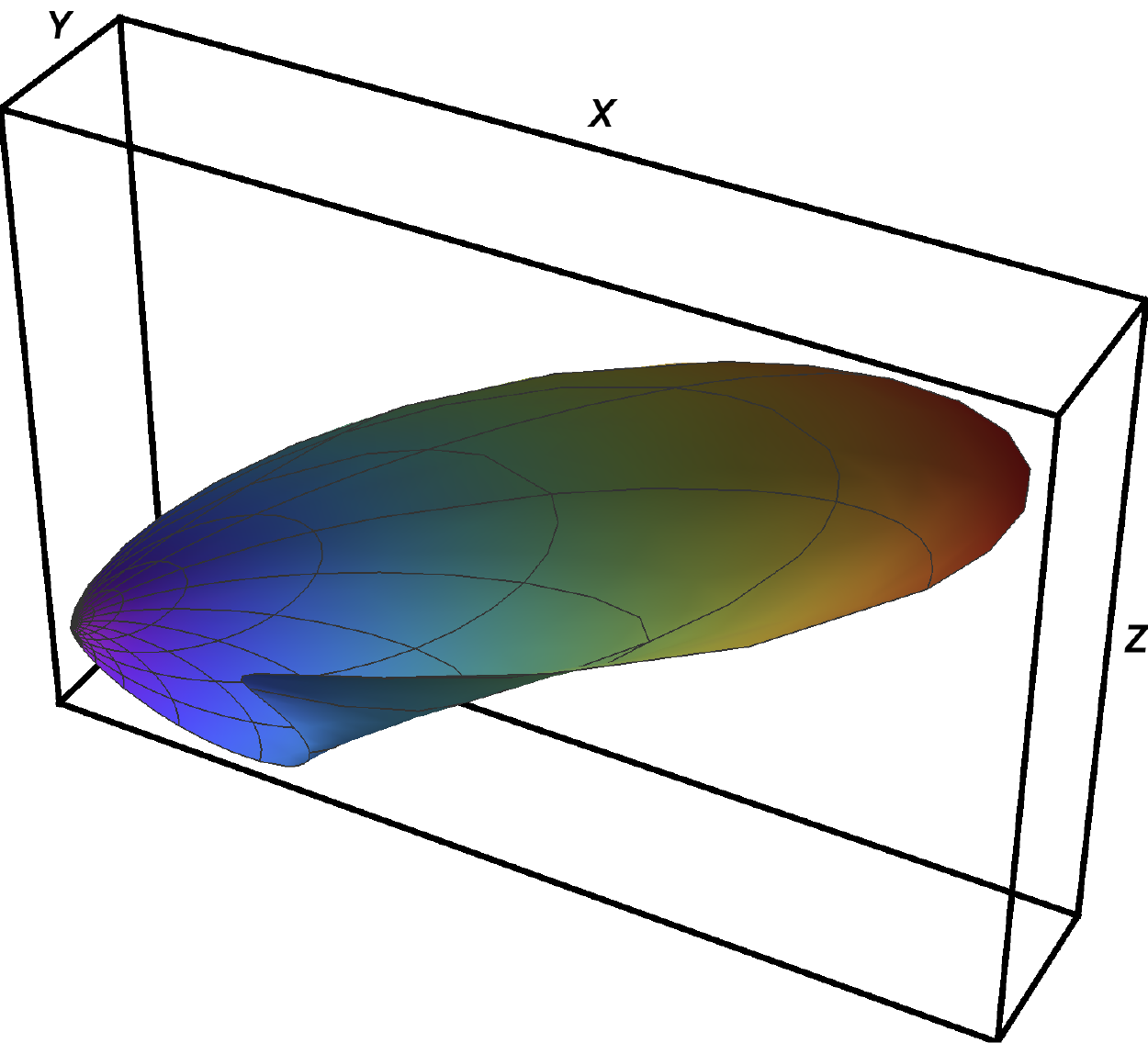}}\quad
\rotatebox{0}{\includegraphics[width=1.4in]{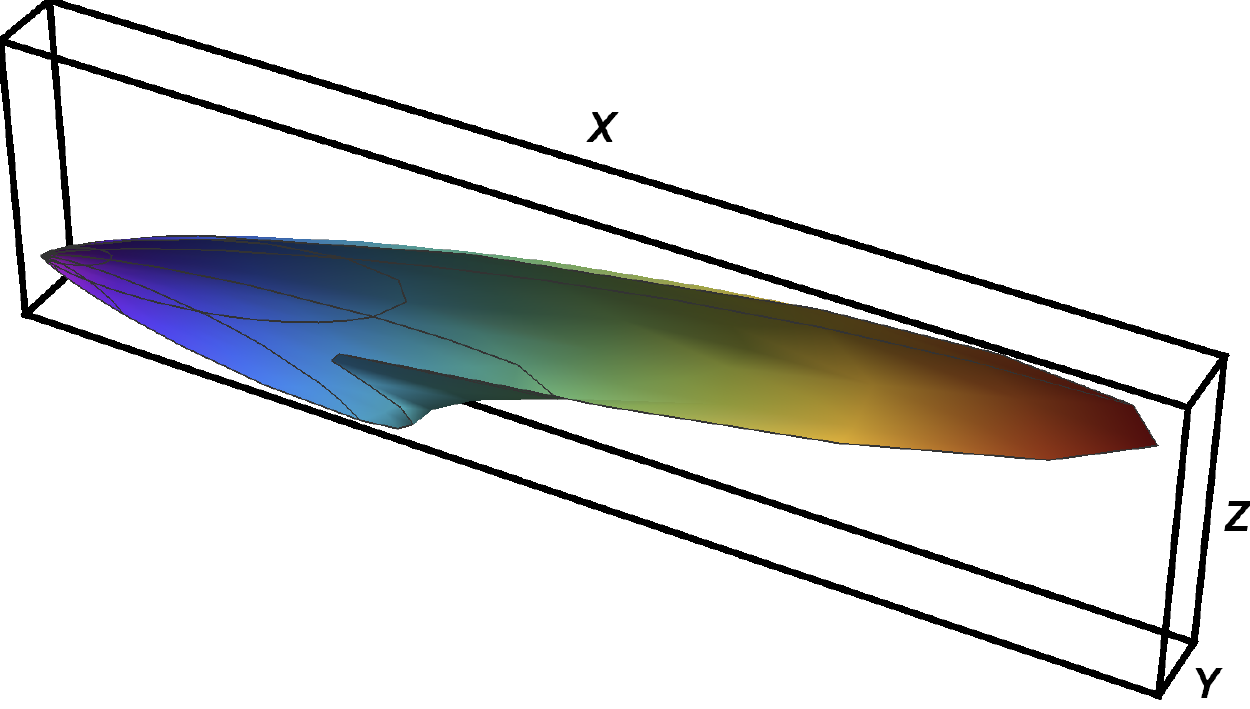} }}
\caption{ Hypertor Angular Distribution with  invariants  : $\kappa = 1,   \tau=\nu=0.5$, and speed  $\beta = 0.3 , 0.6 ,$ $0.9$ . The plot range is  $\theta=[0,\pi]$ and $\phi=[0,\pi]$.} 
\end{center}
\end{figure}







\printbibliography[title={References}]


\end{document}